\documentclass[letter,11pt]{article}
\pdfoutput=1
\usepackage{heppub} 
\allowdisplaybreaks
\usepackage{hyperref}
\usepackage{graphicx}
\usepackage{amsmath}
\usepackage{amssymb} 
\usepackage[normalem]{ulem}
\usepackage{xspace}
\usepackage{slashed}
\usepackage[utf8]{inputenc}
\usepackage[T1]{fontenc}
\usepackage{subcaption}
\usepackage{scalerel,stackengine}
\usepackage{xfrac}
\usepackage{braket}
\usepackage[cal=cm,scr=euler]{mathalpha}
\usepackage{enumitem}
\usepackage[usenames,dvipsnames]{xcolor}
\usepackage[export]{adjustbox}
\usepackage{mathrsfs}

\usepackage{bm}
\usepackage{slashed}
\usepackage{upgreek}

\usepackage{marginnote}
\usepackage[normalem]{ulem}

\preprint{CERN-TH-2026-128}

\title{Factorization of the triple-collinear $q \to qc\bar{c}$ splitting function at first order in opacity}

\author[a,b]{Shahin Iqbal}
\author[a]{and Urs Achim Wiedemann}

\affiliation[a]{Theoretical Physics Department, CERN, CH-1211 Geneva 23, Switzerland}
\affiliation[b]{National Centre for Physics, Shahdra Valley Road, Islamabad, 45320, Pakistan}
\emailAdd{shahin.iqbal@cern.ch}
\emailAdd{urs.wiedemann@cern.ch}

\abstract{
In strongly ordered collinear limits, triple-collinear $1\to 3$ splitting functions factorize into products of $1\to 2$ splitting functions. Here, we investigate whether, and under which conditions, this factorization persists in a finite QCD medium. To this end, we reformulate the opacity expansion as a medium-modification of the compact Catani-Grazzini representation of collinear $1\to n$ splitting functions. We compute the fully differential medium-modified triple-collinear $q\to q c\bar{c}$ splitting function to first order in opacity and leading order in $N_c$, and show that it can be expressed as a sum of momentum-shifted vacuum $q\to q c\bar{c}$ splitting functions multiplied by simple interference factors.
In vacuum, this splitting function has a single strongly ordered collinear limit, in which the invariant mass of the outgoing $c\bar{c}$ pair is much smaller than all other invariant masses. The medium-modified splitting function instead exhibits three distinct strongly ordered limits, depending on how the relevant invariant masses -- multiplied by a boost factor and interpretable as inverse formation times -- compare with the medium length. Remarkably, in all three cases the triple-collinear splitting factorizes also in the medium into a tensor product of $q\to qg$ and $g\to c\bar{c}$ splitting functions. In two of these limits, the latter are medium modified to first order in opacity. This first proof shows that factorization extends to medium-modified triple-collinear splitting functions. Lastly, we relate our results to the longstanding problem of overlapping formation times and discuss their implications for jet-quenching parton showers based on medium-modified $1\to 2$ splittings. We also outline further applications of our formalism and note that our results may inform future phenomenological studies of spin (de)correlations in final-state radiation in dense QCD matter.
}

\begin{document} 
\maketitle
\flushbottom

\section{Introduction}
\label{sec:intro}
The space-time picture of ultra-relativistic heavy ion collisions implies that partons entering a hard high-momentum-transfer process are unaffected by the quark-gluon-plasma (QGP) formed in these collisions. In contrast, outgoing partons propagate through the QGP over an extended timescale, during which their propagation and fragmentation can be significantly modified. Since the 1990s, this picture has motivated so-called "radiative parton energy loss" calculations, which describe medium-modified $1 \to 2$ parton splitting processes~\cite{Baier:1996kr,Baier:1996sk,Zakharov:1996fv,Wiedemann:2000za,Gyulassy:2000er,Gyulassy:1993hr,Wang:2001ifa,Arnold:2002zm}. In the early 2000's, radiative parton energy loss was identified as the dominant mechanism responsible for the suppression of high-transverse-momentum hadron spectra observed at RHIC~\cite{PHENIX:2004vcz,STAR:2005gfr}. With the greatly improved experimental access to high-$p_T$ phenomena at the LHC, research since the 2010s focusses on understanding medium-induced modifications of a broad range of jet- and jet-substructure observables~\cite{ALICE:2022wpn,CMS:2024krd}. The substantial progress achieved in this field over the last decades has been reviewed extensively, see~\cite{Wang:2025lct} and references therein.

In high-energy particle physics, jet physics phenomenology relies on parton showers that are rigorously motivated in certain strongly ordered limits, where collinear $1 \to n$ branchings simplify to Markovian sequences of successive $1\to 2$ splittings. In these limits, splittings lose memory of previous branchings as quantum interference between subsequent splittings becomes negligible. Although parton showers are commonly applied beyond the strict validity of these approximations, the strongly ordered limits remain essential for understanding and improving shower accuracy.  The accuracy with which parton showers resum large logarithmic enhancements from multiple soft or collinear radiations is typically classified by the logarithmic powers retained at each order in the strong coupling.  While leading -logarithmic (LL) accuracy was long the standard,  many modern parton showers aim for next-to-leading-logarithmic (NLL) accuracy~\cite{Forshaw:2020wrq,Nagy:2020rmk,Herren:2022jej,Preuss:2024vyu}, and recent developments~\cite{vanBeekveld:2024wws} have achieved next-to-next-to-leading logarithmic (NNLL) accuracy for specific event-shape observables. 

In high-energy nuclear physics, one may ask whether similarly rigorous arguments justify modeling the medium modification of parton showers traversing QGP as Markovian sequences of medium-modified $1\to 2$ splittings. Since such medium effects are power-suppressed higher-twist corrections, albeit geometrically enhanced, and since the logarithmic accuracy of vacuum parton showers does not systematically include power-suppressed contributions, a positive answer is far from obvious. Heuristic arguments in favor of a Markovian implementation typically start from the observation that known medium-modified $1\to 2$ splitting functions interpolate between totally coherent and incoherent limits governed by formation times. As formation times and the corresponding space-time separation  between splittings increase, this geometrical separation is argued to suppress quantum interference between subsequent splittings, thus allowing for a probabilistic description. In this context, Zapp et al.~\cite{Zapp:2008af} showed that the dominant quantum interference effects incorporated in the early energy-loss calculations of Refs.~\cite{Baier:1996kr,Baier:1996sk,Zakharov:1996fv,Wiedemann:2000za} can be implemented probabilistically through a formation-time constraint.  Caucal et al.\cite{Caucal:2018dla,Caucal:2019uvr} further used the concepts of formation time and angular broadening to argue, within a LL  approximation, how a vacuum-like probabilistic parton fragmentation history emerges in a dense QCD medium of finite size. In general, medium-modified parton showers (such as JEWEL~\cite{Zapp:2012ak,Zapp:2013vla}, Martini~\cite{Schenke:2009gb}, LBT~\cite{He:2015pra,Luo:2023nsi}, Jetscape~\cite{Putschke:2019yrg}, JetMed~\cite{Caucal:2020xad,Caucal:2020uic} or the Hybrid Model~\cite{Casalderrey-Solana:2014bpa}) must specify the space-time embedding of splittings in a QCD medium and thus make assumptions about the spatial separation of successive splittings, implicitly requiring this separation to be large enough for decoherence between them. 

In this work, we study medium-modified triple-collinear $1\to 3$ splitting functions, as they can provide insight into when overlaps between the formation times of successive splittings become negligible and a factorized probabilistic implementation is justified. Several existing calculations already capture aspects of medium-modified $1\to 3$ processes. In particular, studies of medium-induced gluon radiation from a QCD antenna~\cite{Armesto:2011ir,Casalderrey-Solana:2011ule,Mehtar-Tani:2010ebp,Mehtar-Tani:2011hma,Mehtar-Tani:2011lic,Mehtar-Tani:2012mfa,Barata:2021byj,Abreu:2024wka} have explored the medium-modified interference patterns associated with gluon emission from two color currents. Furthermore, at leading-log accuracy and in the limit where the longitudinal momentum fraction of the first emitted gluon is soft and that of the second emitted gluon is parametrically softer, it has been shown~\cite{Liou:2013qya,Blaizot:2014bha,Wu:2014nca,Casalderrey-Solana:2015bww} that radiative corrections to medium-induced $1\to 2$ splittings generate double logarithmic enhancements in the in-medium path length, which can be absorbed into a renormalisation of the quenching parameter. Over the past decade, Arnold et al.\cite{Arnold:2015qya,Arnold:2016kek,Arnold:2018yjd,Arnold:2020uzm,Arnold:2021pin,Arnold:2023qwi} have studied the problem of overlapping formation times in $1\to 3$ processes without restrictions on the longitudinal momentum fractions in a body of work that sets the state of the art. Among other results, these works clarify how to separate overlapping double bremsstrahlung from the case of consecutive independent bremsstrahlung, and provide evidence that genuine overlap effects are numerically small for transverse-momentum-integrated observables.

The starting point of this work differs both conceptually and technically from that of Refs.~\cite{Arnold:2015qya,Arnold:2016kek,Arnold:2018yjd,Arnold:2020uzm,Arnold:2021pin,Arnold:2023qwi}. Conceptually, those works discuss differential medium-induced $1\to 3$ splitting rates within a QCD transport framework, rather than medium modifications of vacuum splitting functions. 
As a result, their vacuum baseline vanishes, whereas ours is given by the leading-order triple-collinear vacuum splitting function.  Technically, they assume that partons propagating through the QGP undergo continuous colour evolution along their trajectories. Medium averaging then amounts to evaluating Wilson line correlators. As the complexity of these averages is greatly reduced when splitting rates are integrated over transverse momenta, fully transverse-momentum-differential expressions have not yet been obtained. However, such expressions are required to determine the invariant masses of partonic subsystems relevant for the study of collinear limits.  In contrast, we start from the collinear $1\to n$ vacuum splitting functions and systematically incorporate medium modifications in an opacity expansion, for which the color averages can be calculated straightforwardly.  Our calculations remain fully differential. 

More than a decade ago, Fickinger et al.~\cite{Fickinger:2013xwa} studied $1 \to 3$ splitting functions within soft-collinear effective theory (SCET), including medium corrections to first order in opacity and deriving fully differential expressions. They showed that their approach reproduces the triple-collinear vacuum splitting functions of Catani and Grazzini, and numerically investigated how medium effects modify the vacuum baseline in selected angular correlations. Despite this achievement, the complexity of the resulting expressions appears to have hindered further studies. To the best of our knowledge, no subsequent investigations of $1\to 3$ splitting processes within the opacity expansion have been reported.

The present work focuses on the $q \to qc\bar{c}$ splitting function for massless charm quarks, where the notation '$c$' serves only to indicate a flavour different from '$q$'. This case is somewhat simpler than that of other $1\to 3$ splitting functions because only a single branching history contributes to the vacuum amplitude. Technically, our starting point differs from that of Ref.\cite{Fickinger:2013xwa} in that we do not work within the SCET framework. Moreover, a number of technical choices -- most notably the separate treatment of contributions from longitudinally and transversely polarized gluons  -- allow us to expose a remarkably simple underlying structure hidden beneath the apparent complexity of the final result. In particular,  the medium-induced contributions can be understood as momentum-shifted vacuum expressions multiplied by simple interference factors. Conceptually, identifying these structures makes it straightforward to explore strongly ordered limits and separates the problem of overlapping formation times from the complications of the Dirac algebra. In the final section, we summarize the novel results obtained in this way and provide an outlook on questions that may now be within reach.

\subsection{The starting point of this work}
When a subset of $m$ partons becomes collinear, QCD $n$-parton scattering amplitudes $\vert {\cal M}_n\vert^2$ factorize into the product of an $(n-m+1)$-parton scattering amplitude and a universal $1 \to m$ collinear splitting function $P_{1\to 2 \dots m+1}$, which depends only on the kinematics and quantum numbers of the collinear subsystem: 
\begin{equation}
	\mathscr{C}_{2,\dots, m+1}
	\vert {\cal M}_n(p_2, \dots, p_{n+1})\vert^2 = \left( \frac{2\, (4\pi\alpha_s)}{p_1^2} \right)^m
	\vert {\cal M}_{n-m+1}(p_1; p_{m+2}, \dots, p_{n+1})\vert^2\, P_{1\to 2, \dots, m+1}\, .
	\label{eq1.0}
\end{equation}
Here, the symbol $\mathscr{C}_{2,\dots, m+1}$ denotes the limit in which the $m$ partons  become collinear. We label the partons in the collinear subsystem by $i = 2, \dots, m+1$. They originate from the splitting of a {\it parent} parton, labelled by $'1'$, which carries the four-momentum  
$p_1 \equiv p_2+\dots+p_{m+1}$ of the $m$ daughter partons. For notational convenience, we set 
\begin{equation}
	4 \pi \alpha_s = g_s^2 \overset{!}{\equiv } 1\, ,
\end{equation}
throughout the following, since the appropriate powers of the strong coupling constant $g_s$ can be restored trivially. 

The starting point of our work is the compact expression in which Catani and Grazzini have recast the spin-averaged quark and gluon $1\to m$ collinear splitting functions in terms of the dispersive parts $V^{(n)}_{a_2 \dots a_{m+1}}(p_2,\dots, p_{m+1})$ and $V^{\mu\nu\, (n)}_{a_2 \dots a_{m+1}}(p_2,\dots, p_{m+1})$ of the quark and gluon two-point functions,\cite{Catani:1999ss}
\begin{align}
	\hat{P}_{q\to a_2\dots a_{m+1}} &= {\rm Tr} \left[ \frac{\not{n} V^{(n)}_{a_2 \dots a_{m+1}}(p_2,\dots, p_{m+1}) }{4\, n\cdot p_1} \right]
	\, , \label{eq1.1}\\
	\hat{P}_{g\to a_2\dots a_{m+1}} &= -\frac{1}{2} d^{\rho}_{\mu}(p_1)\, V^{\mu\nu\, (n)}_{a_2 \dots a_{m+1}}(p_2,\dots, p_{m+1})\, d_{\nu\rho}(p_1)\, .
	\label{eq1.2}
\end{align}
The precise form of the gluon polarization tensor $d_{\nu\rho}(p_1)$ and the diagrammatic representation of the dispersive parts will be explained in the main text. The normalization of the $1\to m$ collinear splitting functions \eqref{eq1.1}, \eqref{eq1.2} includes the explicit powers  
$\left( \tfrac{2\, (4\pi\alpha_s)}{p_1^2} \right)^m$ of the squared invariant mass of the parent parton that appear in \eqref{eq1.0}. 

The most general parametrization of the collinear momenta is
\begin{equation}
	p_i^{\mu} = \xi_i p^\mu + {\bf p}_i^\mu + \frac{{\bf p}_i ^2}{\xi_i} \frac{n^\mu}{2 p\cdot n}\, , \quad i= 2, \dots m+1\, .\label{eq1.3}
\end{equation}
The collinear four-momentum $p^\mu$ combined with the auxiliary light-like vector $n^{\mu}$ specifies what is meant by "transverse". Here, our metric is mostly negative, boldface momenta without Lorentz index are two-dimensional transverse vectors with scalar product ${\bf p}_i^2 > 0$, and the transverse four vector ${\bf p}_i^\mu $ in \eqref{eq1.3}  satisfies ${\bf p}_i^\mu {\bf p}_{i\, \mu}  = -  {\bf p}_i^2 < 0$ and ${\bf p}_i \cdot p = {\bf p}_i  \cdot n = 0$. We take the longitudinal momentum fractions $\xi_i$ to be normalized, $\sum_{i=2}^{m+1} \xi_i = 1$.  The invariant mass of any two-parton final state 
\begin{equation}
	\tilde{s}_{ij} \equiv (p_i + p_j)^2 = 2 {p_i}_\mu p_j^\mu = \frac{\left({\bf p}_j \xi_i - {\bf p}_i \xi_j \right)^2 }{\xi_i\, \xi_j}   \label{eq1.4}
\end{equation}
can be expressed in terms of characteristic combinations of transverse momenta $\left({\bf p}_j \xi_i - {\bf p}_i \xi_j \right)$. These relative transverse momenta are sometimes referred to as boost-invariant transverse momenta since they are invariant under a common transverse boost ${\bf p}_i \to {\bf p}_i + \xi_i {\bf r} $ and therefore characterize the internal kinematics of the collinear system. 
%
%
\begin{figure}[!t]
    \centering
    \includegraphics[width=0.46\textwidth]{figures/Dv}\\
    \includegraphics[width=0.46\textwidth,trim=85bp 350bp 100bp 300bp,clip]{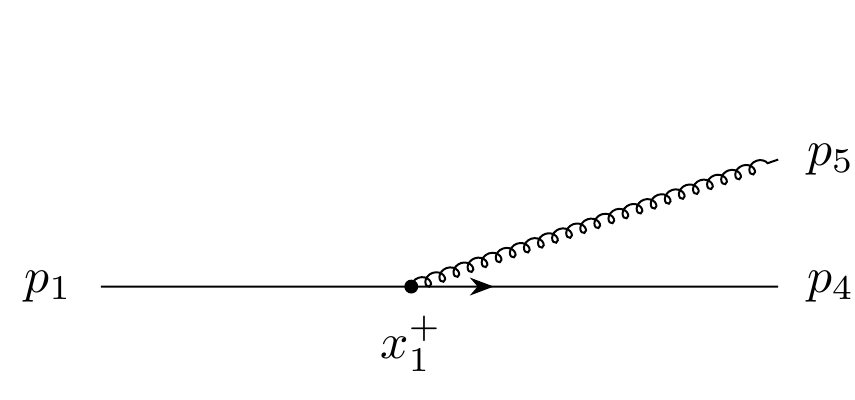}\hfill
    \includegraphics[width=0.46\textwidth,trim=85bp 358bp 100bp 300bp,clip]{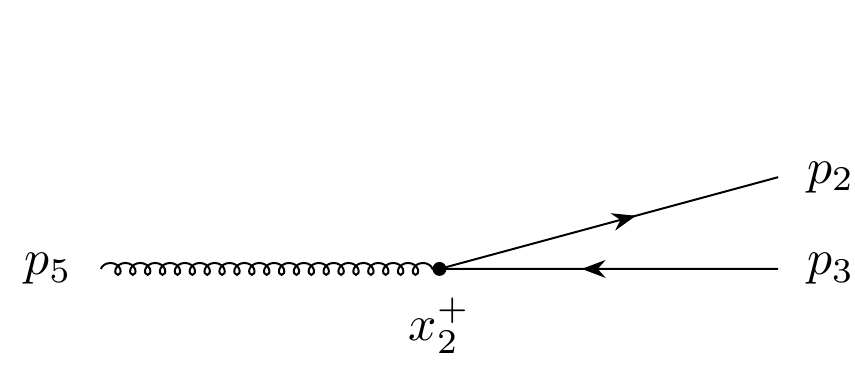}
    \caption{First row: Amplitude associated to the triple-collinear vacuum splitting function $q(1) \to q(4)\, c(2)\, \bar{c}(3)$, with the index labeling used throughout this work. Second row: the diagrams for the subprocesses $q(1) \to q(4)\, g(5)$ and  $g(5) \to c(2)\, \bar{c}(3)$, which form the building blocks of the $q(1) \to q(4)\, c(2)\, \bar{c}(3)$ splitting.}
    \label{fig0}
\end{figure}
%

Figure ~\ref{fig0} shows the vacuum amplitude of the triple-collinear splitting function $q\to q\, c\, \bar{c}$ in the notation used throughout this paper. It will be convenient to work with longitudinal momentum fractions $\zeta$ for the first and $z$ for the second splitting, such that the normalization $\sum_{i=2}^{4} \xi_i = 1$ is automatically implemented, 
\begin{equation}
	p_2^+ = p_1^+ \zeta z\, , \qquad p_3^+ = p_1^+ \zeta (1-z)\, , \qquad p_4^+ = p_1^+(1-\zeta)\, .
	\label{eq1.5}
\end{equation}
The boost-invariant transverse momenta associated to the first and second splitting in Figure ~\ref{fig0} read 
\begin{align}
	{\bm \kappa}_1 &=  \zeta {\bf p}_4 -(1-\zeta)({\bf p}_2+{\bf p}_3) \, ,\label{eq1.6}\\
	{\bm \kappa}_2 &=  (1-z) {\bf p}_2  - z {\bf p}_3 \, . \label{eq1.7}
\end{align}
\subsection{Outline of this work}
To extend the compact representations \eqref{eq1.1}, \eqref{eq1.2} of collinear splitting functions to a finite QCD medium, they must be supplemented by information about the space-time embedding of the parton splittings. Section~\ref{sec2} develops the necessary formalism by reformulating eqs.~\eqref{eq1.1} and \eqref{eq1.2} in time-ordered perturbation theory (section~\ref{sec2.2}) and organizing the opacity expansion (section~\ref{sec2.4}) as a sum over products of color terms, Dirac terms which contain the spinor structure but no space-time information, and longitudinal phase integrals which are simple analytical integrals encoding the full space-time dependence. In the remainder of section~\ref{sec2}, we rederive the first order opacity corrections to the $g\to c\bar{c}$ and $q \to qg$ splitting functions. Besides illustrating the formalism in its simplest applications, this allows us to explain
the absence of instantaneous contributions that can complicate time-ordered perturbation theory, to rewrite the Dirac terms in a form without temporal derivatives that is amenable to symbolic algebra packages such as FeynCalc, and  to introduce, in a consistent notation, the building blocks needed for the analysis of the collinear limits of $q\to qc\bar{c}$ in section~\ref{sec5}. 
 
 In section~\ref{sec3}, we revisit the triple-collinear vacuum splitting function $q\to qc\bar{c}$. Although its derivation in time-ordered perturbation theory is more involved than the standard one, it provides the baseline for the opacity expansion developed in section~\ref{sec4}. Unlike the $1\to 2$ splitting functions discussed in section~\ref{sec2}, the intermediate gluon propagator in the vacuum $q\to qc\bar{c}$ splitting contains a non-vanishing instantaneous contribution. In section ~\ref{sec3.1}, we show how this contribution cancels and why it does not affect interactions with the medium. Section ~\ref{sec3.2} derives the longitudinal phase integrals appearing in time-ordered perturbation theory, relates them to invariant masses in propagator denominators, and introduces a decomposition of the vacuum $q\to qc\bar{c}$ splitting function into contributions with longitudinally and transversely polarized intermediate gluons. This completes the technical groundwork for the calculation of medium-modifications. 
 
 Section~\ref{sec4} derives the medium-modification of the $q\to qc\bar{c}$ splitting function to first order in opacity. It presents the corresponding color factors (section~\ref{sec4.1}), the longitudinal phase integrals (section~\ref{sec4.2}) and Dirac spinor structures (section~\ref{sec4.3}), before summarizing the complete result in section~\ref{sec4.4}. A central technical result is that all Dirac structures can be expressed as simple momentum shifts of compact vacuum expressions. 
 
 Section~\ref{sec5} is dedicated to the strongly ordered collinear limits of the medium-modified $q\to qc\bar{c}$ splitting function. The decomposition of the opacity expansion into products of color factors, longitudinal phase integrals and Dirac terms is ideally suited for analyzing factorization. This is so, since (the purely transverse contributions to) all medium-modified Dirac terms remain parametrically of the same order in the collinear limit, and since they inherit their structure from the vacuum contribution and can thus be expressed, under certain conditions, as products of spin-dependent $1\to 2$ splitting tensors. As a result, the simplifications of strongly ordered collinear limits arise entirely from studying the analytically simple longitudinal phase integrals. This enables us to prove the main results summarized in the abstract. 
 
 Our derivation in section~\ref{sec4} provides a compact representation of medium modifications to a fully differential triple-collinear splitting function. The result is suited for numerical studies and may therefore help clarify the issue of overlapping formation times beyond analytically accessible collinear limits. Although the present work focuses on the $q\to qc\bar{c}$ splitting, and other triple-collinear splitting functions are more involved combinatorially as they involve more than one vacuum amplitude, we expect that the techniques developed here can be extended to all medium-modified triple-collinear splitting functions. In the outlook section, we discuss these and other directions for future work, including applications to refined jet quenching parton showers and to  the study of spin-(de)correlations in medium-modified jets. 

\section{The splitting function $g \to c\bar{c}$}
\label{sec2}
The main aim of this work is to calculate a fully differential, medium-modified $1 \to 3$ splitting function to first order in opacity (see section~\ref{sec4}).  To this end, we shall re-organize the opacity expansion in a formalism in which color, spatio-temporal information and complications arising from the Dirac algebra factorize, and can be cast separately into particularly compact expressions. In the present preparatory section, we introduce the main elements of this re-organization of the opacity expansion, and we illustrate its use by rederiving the known $N=1$ opacity corrections to $1 \to 2$ splitting functions. 

\subsection{Propagators in axial gauge and in time-ordered perturbation theory}
\label{sec2.1}
We begin with some technical remarks. Physical observables are, of course, gauge-invariant. However, within QCD collinear perturbation theory, splitting functions constitute only one ingredient in the calculation of gauge-invariant quantities. In particular, it is in the axial gauge, $n\cdot A = 0$, that interference between the $1\to m$ collinear splitting functions and the remainder of the hadronic collision is known to be negligible. For this reason, Eqs.~\eqref{eq1.1} and \eqref{eq1.2} should be evaluated in axial gauge \cite{Catani:1999ss}. 

In contrast to vacuum splittings, the presence of a medium introduces genuine  space-time dependence into the splitting process. To account for this, we employ time-ordered perturbation theory, expressing propagators in a mixed representation in which the small light-cone momentum component is traded for light-cone time. In axial gauge, however, this leads to technical complications, as certain components of the quark and gluon propagators develop instantaneous terms. We make this issue explicit here, as this will allow us to show later how the resulting complications can be circumvented. 

We work  in light-cone coordinates, with light-cone time 
\begin{equation}
x^+ \equiv \frac{x^0+x^3}{\sqrt{2}}
\label{eq2.1}
\end{equation}
 being the Fourier conjugate of light-cone energy $p^- = p_+$. We denote transverse momenta in bold face, so that the squared four-momentum reads $p^2 = 2 p^+ p^- - {\bf p}^2$. For $\mu, \nu = +, -, {\mathbf \perp}$, the light-cone metric is 
\begin{equation}
	g^{\mu\nu} = g_{\mu\nu} = \begin{bmatrix} 0 & 1 & 0 & 0 \\
	 1 & 0 & 0 & 0 \\
	  0 & 0 & -1 & 0 \\
	   0 & 0 & 0 & -1 
	\end{bmatrix}\, .
	\label{eq2.2}
\end{equation}
Choosing the auxiliary vector to point along the light-cone minus direction, the axial (light-cone) gauge is fixed by the condition 
\begin{equation}
	A^+ = n \cdot A \overset{!}{=} 0\, ,\qquad   n^{\mu} = \left(0,1,{\bf 0}\right)\, .
	\label{eq2.3}
\end{equation}
In this gauge, the gluon propagator takes the form 
\begin{equation}
	G^{\mu\nu}(p) = \frac{i}{p^2 + i\epsilon} d^{\mu\nu}(p)\, ,
	\label{eq2.5}
\end{equation}
where the gluon polarization tensor is
\begin{equation}
	d^{\mu\nu}(p)\ = - g^{\mu\nu} + \frac{p^\mu n^\nu + p^\nu n^\mu}{p\cdot n }\, .
	\label{eq2.5}
\end{equation}
For the following discussion, it will be important that all $+$ -components of the gluon polarization tensor vanish, and that only $d^{--}(p)$ depends on $p^-$. Using $p\cdot n = p^+$, one finds
\begin{align}
d^{++}(p) &=  d^{+-}(p) =  d^{-+}(p) =  d^{+\perp}(p)=0 \, ,   \nonumber \\
 d^{--}(p) &= \frac{2 p^-}{p^+} \, ,\qquad  
d^{\perp -}(p) = d^{-\perp }(p) = \frac{{\bf p}}{p^+}\, ,\nonumber \\ 
d^{ij}(q) &= - g^{ij}\, , \qquad \hbox{with ${i,j \in \perp}$ }\, .
\label{eq2.6}
\end{align}
We now write the gluon propagator in the mixed representation $G^{\mu\nu} \left( p^+, {\bf p}; x^+ \right)$, in which the small  $p^-$-component is traded for the light-cone time $x^+$,
\begin{align}
	G^{\mu\nu} \left( p^+, {\bf p}; x^+ \right) &= \int \frac{dp^-}{2\pi} e^{- i p^- x^+}\, G^{\mu\nu}\left(p^+,p^-,{\bf p}\right)\, ,
	\nonumber \\
	&= d^{\mu\nu} \left(p^+, p^- \to i \frac{\partial}{\partial x^+}, {\bf p} \right) \int \frac{dp^-}{2\pi}  \frac{i\, e^{- i p^- x^+}} {2 p^+ p^- - {\bf p}^2 + i\epsilon }
	\nonumber \\
	&=  d^{\mu\nu} \left(p^+, p^- \to i \frac{\partial}{\partial x^+}, {\bf p} \right)  \Theta(x^+)\, \frac{1}{2 p^+}
	\exp\left[ - i \frac{{\bf p}^2}{2p^+}  x^+ \right]\, .
	\label{eq2.7}
\end{align}
Here, in the second line, replacing $p^-$ by $ i \tfrac{\partial}{\partial x^+}$ in the gluon polarization tensor allows one to pull $d^{\mu\nu}$ out of the $p^-$-integral. The third line then follows from evaluating the $q^-$-contour integral via the pole at $p^- = \tfrac{{\bf p}^2}{2p^+} - i \epsilon$. As only the component $d^{--}(p)$ depends on $p^-$, the derivative $ i \tfrac{\partial}{\partial x^+}$ appears only in this component
\begin{align}
	G^{--} \left(p^+, {\bf p}; x^+ \right)  &=  \frac{ i \tfrac{\partial}{\partial x^+}} { \left(p^+\right)^2}  \left(  \Theta(x^+)\, 
	\exp\left[ - i \frac{{\bf p}^2}{2p^+}  x^+ \right] \right) \nonumber \\
	&=  \frac{ i } { \left(p^+\right)^2} \delta(x^+)\, 
	+  \frac{ 1} { \left(p^+\right)^2} \frac{{\bf p}^2}{2p^+}   \Theta(x^+)\, e^{ - i \frac{{\bf p}^2}{2p^+}  x^+ }
	\, .\qquad  \label{eq2.8}
\end{align}
This is the only component of the gluon propagator \eqref{eq2.7} that contains an {\it instantaneous} contribution $\propto  \delta(x^+)$. 

In close analogy, the quark propagator can be written in mixed representation as 
\begin{align}
	G^F \left( p^+, {\bf p}; x^+ \right) &= \int \frac{dp^-}{2\pi} e^{- i p^- x^+}\, G^F\left(p^+,p^-,{\bf p}\right)\, ,
	\nonumber \\
	&=  d^F \left(p^+, p^- \to i \frac{\partial}{\partial x^+}, {\bf p} \right)  \Theta(x^+)\, \frac{1}{2 p^+}
	\exp\left[ - i \frac{{\bf p}^2}{2p^+}  x^+ \right]\, ,
	\label{eq2.9}
\end{align}
where
\begin{equation}
	d^F \left(p^+, p^- \to i \frac{\partial}{\partial x^+}, {\bf p} \right)  = i \gamma^-  p^+  -  \gamma^+  \frac{\partial}{\partial x^+}  - i \gamma^\perp  p^\perp \, .
	\label{eq2.10}
\end{equation}
Therefore, only the $\gamma^+$-term of the quark propagator carries an {\it instantaneous} contribution. 

%
\begin{figure}[!htbp]
    \centering
      \includegraphics[width=0.8\textwidth]{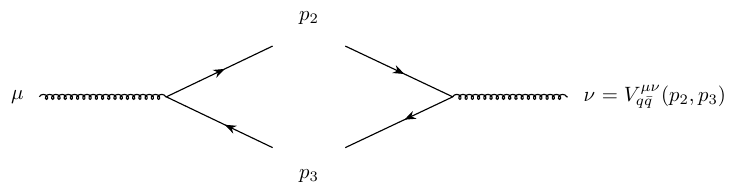}
    \vspace{-0.5em}
    \caption{The dispersive part of the gluon two-point function to lowest order.}
    \label{fig1}
\end{figure}
\subsection{The vacuum splitting $g\to q\bar{q}$ in time-ordered perturbation theory }
\label{sec2.2}
Some generic features of time-ordered perturbation theory can be illustrated using vacuum splitting functions alone. According to the dispersive expression \eqref{eq1.2}, depicted diagrammatically in Fig.~\ref{fig1}, the leading-order helicity-averaged $g \to q\bar{q}$ vacuum splitting reads
\begin{align}
	\hat{P}_{g\to q\bar{q}} &= -\frac{1}{2} \frac{d^{\rho\mu}(p_5)}{p_5^2 + i\epsilon} {\rm Tr}\left[ \gamma_\mu \not{p}_3 \gamma_\nu \not{p}_2 \right]
	\frac{d^{\nu}_{\, \rho}(p_5)  }{p_5^2 - i\epsilon} {\cal N}_{g \to q\bar{q}} \nonumber \\
	&=  \frac{2 }{p_5^2 }  T_R \left( z^2 + (1-z)^2 \right)\, ,
	\label{eq2.11}
\end{align} 
where  $p_5 = p_2 + p_3$. As explained in the text following \eqref{eq1.2}, this expressions consists of the normalization $\tfrac{2}{p_5^2}$ involving the squared invariant mass $p_5^2$ of the parent gluon, multiplied by the standard LO $g \to c\bar{c}$ splitting function. The corresponding color factor, 
\begin{equation}
{\cal N}_{g \to q\bar{q}}\ = \frac{1}{N_c^2 -1} {\rm Tr}\left[ T^a\, T^a \right] = \frac{1}{2}=T_R \, ,
	\label{eq2.12}
\end{equation}
 is obtained by averaging over the color of the incoming gluon.

The only component $G^{--}$ of the gluon propagator capable of generating an instantaneous term \eqref{eq2.8} does not contribute to the vacuum splitting function \eqref{eq2.11}. To see this,  consider  \eqref{eq2.11} with $d^{\rho\mu} \to d^{--}$ and contract with the second gluon polarization tensor 
\begin{equation}
	d^{\rho\mu}  \, d^{\nu}_{\, \rho} \to d^{--} d^{\nu}_{\, -} = d^{--} d^{\nu +} = 0\, .
	\label{eq2.13}
\end{equation}
Here, the last equality follows from the absence of a non-zero $+$-component of the gluon polarization tensor ~\eqref{eq2.6}. Since $G^{--}$ does not contribute, 
we can set  the $p^-$-argument of the gluon polarization tensor to zero and write \eqref{eq2.11} with gluon propagators in mixed representation \eqref{eq2.7}
\begin{align}
	\hat{P}_{g\to c\bar{c}} =& \frac{-1}{2} G^{\rho\mu}(p_5) {\rm Tr}\left[ \gamma_\mu \not{p}_3 \gamma_\nu \not{p}_2 \right]
	{G^{\nu}_{\, \rho}}^*(p_5)  {\cal N}_{g \to q\bar{q}} \nonumber \\
	=& {\cal N}_{g \to q\bar{q}} \frac{-1}{2} \int dx^+ e^{i\, (p_2^- + p_3^-) x^+}\,   \int d\bar{x}^+ e^{-i\, (p_2^- + p_3^-) \bar{x}^+}\, \nonumber \\
	& \times G^{\rho\mu}(p_5^+,{\bf p}_5; x^+) {\rm Tr}\left[ \gamma_\mu \not{p}_3 \gamma_\nu \not{p}_2 \right] \, {G^{\nu}_{\, \rho}}^*(p_5^+,{\bf p}_5; \bar{x}^+)  
	\nonumber \\
	=& \underbrace{{\cal N}_{g \to q\bar{q}} }_{\hbox{Color}} \,
	\underbrace{ \int_0^\infty dx^+ e^{i\, \left( p_2^- +  p_3^- -\tfrac{{\bf p}_5^2}{2p_5^+}  \right) x^+ - \epsilon x^+}\,  
	 \int_0^\infty d\bar{x}^+ e^{-i\, \left( p_2^- +  p_3^- -\tfrac{{\bf p}_5^2}{2p_5^+}  \right) \bar{x}^+ - \epsilon \bar{x}^+}}_{\hbox{Longitudinal phase integrals}}\, \nonumber \\
	& \times 
	 \underbrace{\frac{-1}{2} \frac{1}{4 (p_5^+)^2} 
	d^{\rho\mu}(p_5^+,0,{\bf p}_5) {\rm Tr}\left[ \gamma_\mu \not{p}_3 \gamma_\nu \not{p}_2 \right] \, d^{\nu}_{\, \rho}(p_5^+,0,{\bf p}_5) }_{\equiv D^{g\to c\bar{c}}_{\rm vac} \, ,\,\, \, \hbox{Dirac term}} \, .
	\label{eq2.14}
\end{align} 
This expression is more involved than the momentum-space representation \eqref{eq2.11} and is therefore of little use for evaluating vacuum splittings. However, it highlights several features that persist in the presence of in-medium interactions and are relevant for the calculation of medium-modified splitting functions:
\begin{enumerate}
	\item {\it Factorization into factors for color,  longitudinal phase and Dirac structure.}\\
	Eq.\eqref{eq2.14}  factorizes as indicated. Longitudinal phase integrals are readily done analytically. The Dirac terms can be evaluated using standard symbolic computation packages for the Dirac algebra. 
	\item {\it Longitudinal phase factors such as $\left( p_2^- +  p_3^- -\tfrac{{\bf p}_5^2}{2p_5^+}  \right)$ in \eqref{eq2.14}}
	\begin{itemize}
		\item can be determined diagrammatically as the sum of transverse energies flowing out of a vertex (here $\sum_{i=2,3} p_i^- = \sum_{i=2,3} \tfrac{{\bf p}_i^2}{2 p_i^+}$) minus the  transverse energy flowing into the vertex (here: $\tfrac{{\bf p}_5^2}{2p_5^+} $). 
				\item arise as the difference between on-shell values and pole values. For instance, in \eqref{eq2.14}, the phase $\exp\left[{- i\, \left(p_2^- + p_3^-  \right) x^+}\right]$ originates from Fourier-transforming with-respect to on-shell final-state momenta, 	 $p_i^- = \tfrac{{\bf p}_i^2}{2 p_i^+}$ since $ 2 p_i^+\, p_i^- - {\bf p}_i^2 = 0$. By contrast, the transverse energy $\tfrac{{\bf p}^2}{2p^+}$  is the off-shell pole value of the gluon propagator \eqref{eq2.7}.  
		\item can be viewed as inverse formation times. For instance
		\begin{equation}
	C_5 \equiv p_2^- + p_3^- - \frac{{\bf p}_5^2}{2p_5^+}  = \frac{(p_2+p_3)^2}{2 n.p_5} =  \frac{\left( (1-z){\bf p}_2 - z {\bf p}_3 \right)^2}{2 z (1-z)\, p_5^+ } 
	 = \frac{\bm{\upkappa}_2  \cdot  \bm{\upkappa}_2}{2z(1-z) p_5^+}\, ,
	\label{eq2.15}
\end{equation}
where $p_2^+ = z p_5^+$, $p_3^+ = (1-z)\, p_5^+$ and ${\bf p}_5 = {\bf p}_2 + {\bf p}_3$. Regarding the inverse of the gluon's off-shellness (invariant mass) $1/ \sqrt{p_2+p_3)^2}$ as measuring the lifetime in its rest frame, and boosting with factor $n \cdot p_5/ \sqrt{p_2+p_3)^2}$
to the medium rest-frame in which the gluon carries energy $n \cdot p_5$, one may interprete $C_5 = \frac{(p_2+p_3)^2}{2 n.p_5}$ in \eqref{eq2.15} as inverse formation time of the splitting. 
	\end{itemize}
\item {\it Diagrammatic rules for Dirac terms}\\
The diagrammatic rules for writing the Dirac terms are the same as those used in Eqs.~\eqref{eq1.1} and \eqref{eq1.2}, with the exception that each propagator denominator $\tfrac{1}{2p_i^+ p_i^- - {\bf p}_i^2 + i \epsilon}$ is replaced by the factor $1/(2\, p_i^+)$. This replacement accounts for the fact that, in mixed representation, the denominators of propagators become the product of a factor $1/(2\, p_i^+)$ times a phase factor, see Eqs.~\eqref{eq2.7} and \eqref{eq2.9}. 
\end{enumerate}

The vacuum splitting function $\hat{P}_{g\to q\bar{q}}$ in \eqref{eq2.14} can then be written in the form 
\begin{equation}
	\hat{P}_{g\to c\bar{c}} = T_R \, \frac{1}{C_5^2}\, D_{\rm vac}^{g\to c\bar{c}} \, ,
	\label{eq2.16}
\end{equation} 
where $\tfrac{1}{C_5^2}$ is the result of the longitudinal phase integrals in \eqref{eq2.14} and the Dirac term reads,
\begin{equation}
	D_{\rm vac}^{g\to c\bar{c}} = \frac{1}{(p_5^+)^2} \left( z^2 + (1-z)^2 \right)\, \frac{\bm{\upkappa}_2  \cdot  \bm{\upkappa}_2}{2z(1-z)} \, .
	\label{eq2.17}
\end{equation}
Both the Dirac term $D_{\rm vac} $ and the inverse formation time $C_5$ depend on the boost-invariant transverse momentum $\bm{\upkappa}_2$ introduced in  \eqref{eq1.7}. One easily checks that \eqref{eq2.16} reproduces \eqref{eq2.11}. 
\subsection{In-medium interactions in the Gyulassy-Wang model}
\label{sec2.3}
We work in light-cone gauge \eqref{eq2.3} with $A^+=0$. Partonic projectiles with large light-cone energy $p^+$ interact with a medium-induced color field that is predominantly $A^-$. We parametrize this field in the Gyulassy-Wang model as a sum of individual scattering potentials located at positions $\check{y}_i$, 
\begin{align}
	A^-({\bf y},y^+) &=  \sum_i   T^{d_i} \int \frac{d^3q}{(2\pi)^3} a({\bf q}) \exp\left[i ({\bf y}-\check{\bf y}_i).{\bf q}  + i(y^+ -\check{y}^+_i)q^- \right]
	\nonumber \\
	&=  \sum_i   T^{d_i}  \delta\left(y^+ -\check{y}^+_i\right) \int \frac{d{\bf q}}{(2\pi)^2} a({\bf q}) \exp\left[i ({\bf y}-\check{\bf y}_i).{\bf q}  \right]
	\label{eq2.18}
\end{align}
Here, the scattering potential $a({\bf q})$ transfers only transverse momentum ${\bf q}$, implying that the interaction occurs at a definite light-cone  time $y^+ = \check{y}^+_i$. We further assume that the positions $\check{y}_i$ of scattering centers are distributed homogenous with density 
\begin{equation}
	n(\check{\bf y},\check{y}^+) = n_0\, \theta(\check{y}^+) \, \theta(L-\check{y}^+)\, .
	\label{eq2.19}
\end{equation}
An expansion  of medium-induced interactions to first order in opacity is then defined by keeping terms linear in $n_0 L \int \tfrac{d^3q}{(2\pi)^3} \vert a({\bf q}) \vert^2 (...)$ .

\begin{figure}
    \centering
      \includegraphics[width=\textwidth]{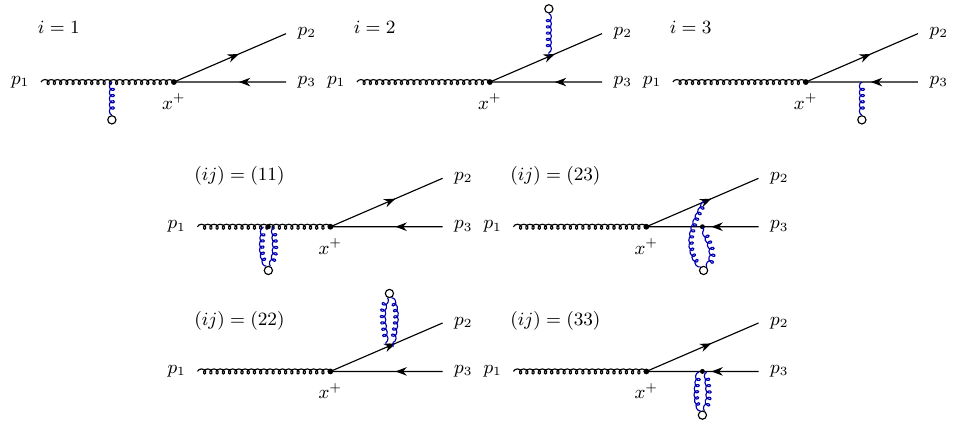}
    \caption{First row: the three real diagrams contributing to the medium modification of the $g\to q\bar{q}$ splitting function at first order in the opacity expansion. Second and third row: the four virtual diagrams contributing to the medium-modification of the $g\to q\bar{q}$ splitting function at first order in the opacity expansion. Although these virtual contributions are of order $O\left( \vert a({\bf q})\vert^2 \right)$ in the amplitude, they contribute at first order in opacity through their interference with the zeroth-order complex-conjugate amplitude.}
    \label{fig2}
\end{figure}
%
\subsection{The medium-modification of $g\to c\bar{c}$ to first order in opacity}
\label{sec2.4}
The first-order opacity expansion of the medium modification of the  $g\to c\bar{c}$ splitting function in Fig.~\ref{fig1} corresponds to second order in the Gyulassy--Wang scattering potential in \eqref{eq2.18}. We organize these medium-modifications in the form 
\begin{equation}
\hat{P}_{g\to c\bar{c}}^{(N=1)} = \int\frac{d{\bf q}}{(2\pi)^2} \vert a({\bf q}) \vert^2  
 \sum_{\substack{ (i,j) \\ i\leq j} } \left(
 \underbrace{  C^{g\to c\bar{c}}_{(ij)} D^{g\to c\bar{c}}_{(ij)} {\cal P}^{g\to c\bar{c}}_{(ij)} }_{\it real}  + 
 \underbrace{  \bar{C}^{g\to c\bar{c}}_{(ij)} \bar{D}^{g\to c\bar{c}}_{(ij)} \bar{\cal P}^{g\to c\bar{c}}_{(ij)}  }_{\it virtual}   \right)\, ,
 \label{eq2.20}
\end{equation}
Here, the momentum ${\bf q}$ exchanged between the medium and the splitting process is distributed according to the elastic cross section $ \vert a({\bf q}) \vert^2 $ associated with the color field in the Gyulassy-Wang model \eqref{eq2.18}. The index pair $(ij)$, with $i,j = 1,2,3$,  specifies whether the two interactions with the medium occur with the parent gluon $(1)$, the charm $(2)$ or the anti-charm quark $(3)$, respectively. The sum over pairs $(ij)$ is restricted to $i\leq j$ because the contribution $(ij)$ is symmetrized, i.e., it includes both the case in which the medium interacts with parton $i$ in the amplitude and the parton $j$ in its complex conjugate amplitude, and the case obtained under the interchange $i \leftrightarrow j$. If one were to consider unsymmetrized pairs $(ij)_{\rm unsym}$ instead, the color factors and Dirac terms would remain invariant under the interchange $i \leftrightarrow j$, while the phase factors would satisfy  ${\cal P}_{(ij)_{\rm unsym}} = {\cal P}_{(ji)_{\rm unsym}}^*$. Therefore, the symmetrization condition will be made explicit in the definition of the phase factors given below.  

The contributions to the cedium-modified splitting function \eqref{eq2.20}  can be classified as {\it real} or {\it virtual}. We refer to contributions as {\it real} when the medium exchanges one gluon with the amplitude in Fig.\ref{fig1} and one with its complex conjugate. In real contributions, momentum is transferred from the medium to the final state. The first row of Figure~\ref{fig2} shows the three ways in which the medium exchanges a gluon through the scattering potential \eqref{eq2.18}. In \eqref{eq2.20}, this yields to nine real contributions obtained from squaring the sum of the three real amplitudes. 

In addition, {\it virtual} contributions arise when the vector potential appears at second order in the amplitude and at zeroth order in its complex conjugate. These contributions ensure probabillity conservation. Since the initial and final state momenta of the amplitude and its complex conjugate must match, virtual contributions do not transfer net momentum between the medium and the splitting process. Instead, one of the two gluon exchanges carries transverse momentum ${\bf q}$ while the other carries $-{\bf q}$. Moreover, both insertions of the Gyulassy--Wang potential must occur at the same light-cone time $\check{y}^+$. We therefore have only four virtual contributions entering \eqref{eq2.20}.

\subsubsection{A first illustration: the contribution $\hat{P}^{(N=1)}_{g\to c \bar{c}\, (33)}$  }
\label{sec2.4.1}
\begin{figure}
    \centering
      \includegraphics[width=0.8\textwidth]{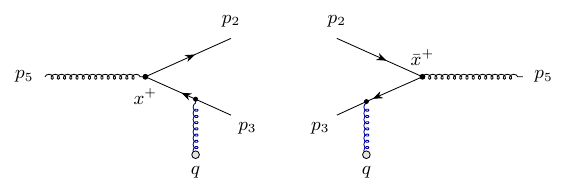}
    \caption{Diagram of the real scattering contribution $(33)$ to the medium-modified $g\to c\bar{c}$ splitting function \eqref{eq2.2}.  }
    \label{fig3}
\end{figure}

We illustrate how the structure in \eqref{eq2.20} arises by presenting an explicit calculation of one of the terms contributing to this sum, namely  the term 
$\hat{P}^{(N=1)}_{g\to c \bar{c}\, (33)}$, depicted in Fig.~\ref{fig3}, in which the medium exchanges a gluon with the anti-charm quark in both the amplitude and its complex conjugate. The color factor of this contribution corresponds to that of a fermion loop with four insertions, averaged over the color states of the incoming gluon:
\begin{equation}
	C^{g\to c\bar{c}}_{(33)} =  \frac{1}{N_c^2-1}  {\rm Tr}\left[ T^a\, T^{b}\, T^{b}\, T^a\right]  =  \frac{1}{N_c^2-1}   C_F^2 
	{\rm Tr}\left[ \mathbf{1}  \right] =  \frac{1}{2} C_F \, .
	\label{eq2.21}
\end{equation}
%
%
%
The contribution $\hat{P}^{(N=1)}_{g\to c \bar{c}\, (33)}$ is obtained from the vacuum expression \eqref{eq2.14} by inserting a fermion propagator $D^F(p_3+q)$ together with an adjacent medium-induced quark--gluon vertex $i \gamma^+ A^-$ in both the amplitude and the complex conjugate amplitude, see Fig.~\ref{fig3}.  The resulting anti-charm propagators are then expressed in the mixed representation.  This  yields phase factors proportional to $\tfrac{1}{2 p_3^+}\,  \exp\left[{i\, \left(p_3^- - \tfrac{({\bf p}_3 + {\bf q})^2}{2(1-z)p^+}  \right) \check{y}^+ } \right] $, where the light-cone time is fixed at the position $\check{y}^+$ of the scattering center due to the $\delta$-function in \eqref{eq2.18}. We find
\begin{align}
	\hat{P}^{(N=1)}_{g\to c \bar{c}\, (33)}
	&= C^{g\to c\bar{c}}_{(33)}\, n_0  \int\frac{d{\bf q}}{(2\pi)^2} \vert a({\bf q}) \vert^2   \int_0^L d\check{y}^+  \nonumber \\
	& \times \frac{1}{2 p_5^+} \int_0^\infty dx^+ e^{i\, \left(p_2^- + \tfrac{({\bf p}_3 + {\bf q})^2}{2(1-z)p_5^+} - \tfrac{({\bf p}_5+{\bf q})^2}{2p_5^+}  \right) x^+ - \epsilon x^+}  
	\frac{1}{2 p_3^+} e^{i\, \left(p_3^- - \tfrac{({\bf p}_3 + {\bf q})^2}{2(1-z)p_5^+}  \right) \check{y}^+ } \Theta(\check{y}^+ - x^+)
	 \nonumber \\
	& \times \frac{1}{2 p_5^+} 
	 \int_0^\infty d\bar{x}^+ e^{-i\, \left(p_2^- + \tfrac{({\bf p}_3 + {\bf q})^2}{2(1-z)p_5^+}  -\tfrac{({\bf p_5}+{\bf q})^2}{2p_5^+}  \right) \bar{x}^+ - \epsilon \bar{x}^+}
	 \frac{1}{2 p_3^+}
	 e^{-i\, \left(p_3^- - \tfrac{({\bf p}_3 + {\bf q})^2}{2(1-z)p_5^+}  \right) \check{y}^+ } \Theta(\check{y}^+ - \bar{x}^+) \nonumber \\
	& \times 
	d^{\rho\mu}(p_5^+,0,{\bf p}_5) {\rm Tr}\left[ \gamma_\mu \left( \slashed{p}_3 + \slashed{q}\right) \gamma^+ \slashed{p}_3 \gamma^+ \left( \slashed{p}_3 + \slashed{q}\right) \gamma_\nu \slashed{p}_2 \right] \, d^{\nu}_{\, \rho}(p_5^+,0,{\bf p}_5)  \, .
	\label{eq2.22}
\end{align} 
In this expression, we have replaced $d^F \left( p_3^+, p_3^- \to i \frac{\partial}{\partial x^+}, {\bf p}_3 + {\bf q} \right) \gamma^+$ $ \to i\, \left( \slashed{p}_3 + \slashed{q}\right) \gamma^+$ since the $\gamma^+$-term of $D^F$ in \eqref{eq2.10} does not contribute when $D^F$ appears adjacent to $i \gamma^+ A^-$. As a consequence, no instantaneous contributions associated to fermion propagators arise in our calculation. We note a few additional technical points:
\begin{itemize}
\item 
Eq.~\eqref{eq2.22} averages over the transverse position $\check{\bf y}$ of the in-medium interaction using the distribution \eqref{eq2.19}, and it averages over the color exchanged with the medium. Color flow ensures that the same scattering center is probed in both the amplitude and its complex conjugate. The integration over $\check{\bf y}_i$ then enforces that the same momentum ${\bf q}$ is exchanged in both. As a result, the vector potentials $a({\bf q})$ in the amplitude and complex conjugate amplitude  combine into $\int\frac{d{\bf q}}{(2\pi)^2} \vert a({\bf q}) \vert^2 (\hbox{... integrand...}) $. 
\item
Five integrals over light-cone times enter \eqref{eq2.22}: one from averaging over the distribution of scattering centers and four from propagators written in mixed representations. However, each in-medium interaction with the scattering potential \eqref{eq2.18} introduces a $\delta$-function, so that only three light-cone time integrals  remain.
\item 
 The gluon polarisation tensors in \eqref{eq2.22} can be replaced by their momentum space representation since the component $d^{--}(p)$ does not contribute. This generalizes to all in-medium interactions. Specifically, connecting the three-gluon vertex $V^{\mu\lambda\nu}$ to $d^{\rho\mu}(p) = d^{--}(p)$ and to $A^\lambda = A^-$ requires the component $V^{++\nu}$, which is non-vanishing only for $\nu = -$. However,  the gluon propagator $G^{\rho\nu}$ contracted with $V^{++\nu}$ would need a non-vanishing $+$-component, which does not exist, as seen from \eqref{eq2.6}.  
 \item 
  In mixed representation, each propagator \eqref{eq2.7}, \eqref{eq2.9} contributes a factor $\tfrac{1}{2p^+}$, where $p^+$ is the light-cone energy carried by the propagator. For the diagram Fig.~\ref{fig3}, there are two gluon and two anti-quark propagators, and this leads to the factors $\tfrac{1}{(2p_5^+)^2} \tfrac{1}{(2p_3^+)^2}$
 in \eqref{eq2.22}.
\end{itemize}
These remarks explain why \eqref{eq2.22} takes a form consistent with the diagrammatic rules for longitudinal phase factors and  Dirac terms stated in section~\ref{sec2.2}. In particular, all phases in \eqref{eq2.22} are written as sums of transverse energies flowing out of a vertex minus the transverse energy flowing into it, multiplied by the light-cone time at which the vertex is located. Moreover, due to the absence of instantaneous contributions, the Dirac terms follow the diagrammatic rules of writing \eqref{eq1.1} in momentum space, except that propagator denominators are replaced by a factor $1/(2p_i^+)$ for each propagator. Expression \eqref{eq2.22} takes the form
\begin{align}
	\hat{P}_{g\to c\bar{c}\, (33)}^{(N=1)} 
	&= C^{g\to c\bar{c}}_{(33)}\, \int\frac{d{\bf q}}{(2\pi)^2} \vert a({\bf q}) \vert^2  
	\underbrace{ \int_0^L d\check{y}^+  P_3(\check{y}^+) \,   n(\check{y}^+) \, P_3^*(\check{y}^+)}_{\equiv {\cal P}^{g\to c\bar{c}}_{(33)}} \nonumber \\
	& \times 
	\underbrace{\frac{1}{(2p_5^+)^2} \frac{1}{(2p_3^+)^2} 
	d^{\rho\mu}(p_5) {\rm Tr}\left[ \gamma_\mu \left( \not{p}_3 + \not{q}\right) \gamma^+ \not{p}_3 \gamma^+ \left( \not{p}_3 + \not{q}\right) \gamma_\nu \not{p}_2 \right] \, d^{\nu}_{\, \rho}(p_5) }_{\equiv D^{g\to c\bar{c}}_{(33)}} \, .
	\label{eq2.23}
\end{align}

\subsection{The $N=1$ medium-modified $g\to c\bar{c}$ splitting function rederived}
\label{sec2.5}
Following the steps outlined in the previous subsection, we now present the complete set of contributions \eqref{eq2.20} to the $g\to c\bar{c}$ splitting function at first order in the opacity expansion.  Although this is a rederivation of a known result, it serves to illustrate our formalism in a simple setting and allows us to state explicitly several results that will be needed in the analysis of the opacity corrections to  the $1\to 3$ collinear splitting functions presented in section~\ref{sec5}.

\subsubsection{Color factors $C_{(ij)}^{g\to c\bar{c}}$ for $\hat{P}_{g\to c\bar{c}}^{(N=1)}$}
Useful relations for the evaluation of color factors are summarized in Refs.~\cite{Haber:2019sgz,Peigne:2024srm}. We use the label $a$ for the color of the incoming parent gluon and $b$ for the color of the gluon exchanged with the medium. We average over incoming color, i.e. we divide by $\tfrac{1}{N_c^2-1}$. For scattering on the external gluon field $b$, the $SU(3)$ generator of index $ b$ appear then in the representation of the scattered partons, i.e. $T^b$ for a quark, $-T^b$ for an anti-quark and $-i\, f_{abc}$ for the  incoming gluon. In this way, we find 
\begin{align}
	C_{(55)}^{g\to c\bar{c}} &= \frac{1}{N_c^2-1}  (-i\, f_{abc}) (-i\, f_{c'b a}) Tr\left[ T^c T^{c'}\right] 
		=   \frac{N_c}{2}   \, ,		
		\label{eq2.24} \\
		C_{(22)}^{g\to c\bar{c}} = C_{(33)}^{g\to c\bar{c}}
		 &= \frac{1}{N_c^2-1}   Tr\left[ T^a\, T^{b}\, T^{b}\, T^a\right]  
		=  \frac{1}{2} C_F \approx \frac{N_c}{4} \, ,
		\label{eq2.25} \\
		C_{(25)}^{g\to c\bar{c}} = C_{(35)}^{g\to c\bar{c}}  
		&=  \frac{1}{N_c^2-1}   (-i\, f_{abc}) Tr\left[ T^a\, T^{b}\, T^{c}\right]  
		= \frac{N_c}{4}  \, ,
			\label{eq2.26}\\
			C_{(23)}^{g\to c\bar{c}}  &= \frac{1}{N_c^2-1}   Tr\left[ T^a\, T^{b}\, T^{a}\, T^{b}\right]  =  - \frac{1}{4\, N_c}    \approx  0 \, .
				\label{eq2.27}	
\end{align}
The expressions given after the $\approx$ signs correspond to the leading-order approximation in $N_c$. Five of the six terms contribute at leading order in $N_c$. 

The probability-conserving virtual contributions take the form
\begin{equation}
	\bar{C}_{(55)}^{g\to c\bar{c}}= - \frac{N_c}{2}  \, ,\quad \bar{C}^{g\to c\bar{c}}_{(22)}=  \bar{C}^{g\to c\bar{c}}_{(33)} = - \frac{N_c}{4}  \, ,\quad
	\bar{C}_{(23)}^{g\to c\bar{c}} =  - \frac{1}{4\, N_c}    \approx  0\, .
					\label{eq2.28}	
\end{equation}
There are no virtual contributions of the form $(25)$ and $(35)$, since a scattering center located at $\check{y}^+$ cannot exchange one gluon before the splitting and another after it. 

%
%
\subsubsection{The Dirac terms $D_{(ij)}^{g\to c\bar{c}}$  for $\hat{P}_{g\to c\bar{c}}^{(N=1)}$}
\label{sec2.5.3}
Except for the modified treatment of propagator denominators explained in section~\ref{sec2.2}, the diagrammatic rules for the
Dirac terms are the same as those used to evaluate \eqref{eq1.1} and \eqref{eq1.2}. Therefore, the Dirac terms can be written down straightforwardly  and simplified using standard symbolic computation packages. Using FeynCalc~\cite{Shtabovenko:2016sxi,Shtabovenko:2020gxv}, we establish that all Dirac terms take the form
\begin{equation}
	D_{(ij)}^{g\to c\bar{c}} =  \frac{1}{\left(p_5^+\right)^2} \left( z^2 + (1-z)^2 \right)\, \frac{\bm{\upkappa}_2^{(i)}  \cdot  \bm{\upkappa}_2^{(j)*}  }{2z(1-z)} \, .
	\label{eq2.29}
\end{equation}
These terms can be obtained from the vacuum Dirac structure \eqref{eq2.17} by simple momentum shifts of the boost-invariant
transverse momentum $\bm{\upkappa}_2$ 
\begin{equation}
\bm{\upkappa}_2^{(i)} = \bm{\upkappa}_2\vert_{{\bf p}_i \to {\bf p}_i + {\bf q}}\, .
\label{eq2.30}
\end{equation} 
Diagrammatically, these transverse momenta are naturally associated with the $g\to c\bar{c}$ vertex in the three amplitudes shown in Fig.~\ref{fig2}, where the medium exchanges transverse momentum ${\bf q}$ with the parton $i$. The  superscript $*$ in $\bm{\upkappa}_2^{(i)*} $  indicates that the corresponding shifted momentum is associated with the vertex in the complex conjugate amplitude. For instance, an interaction of the medium with the charm quark (second diagram in Fig.~\ref{fig2}, $i=2$) implies, in our sign convention, that the charm-quark leaves the $g\to c\bar{c}$-vertex with transverse momentum ${\bf p}_2 + {\bf q}$, which amounts to shifting $\bm{\upkappa}_2$ to $\bm{\upkappa}_2^{(2)}$. Likewise 
\begin{equation}
\bm{\upkappa}_2^{(1)} = \bm{\upkappa}_2\, ,\qquad 
\bm{\upkappa}_2^{(2)} = \bm{\upkappa}_2 +  (1-z) {\bf q}\, ,\qquad
\bm{\upkappa}_2^{(3)} = \bm{\upkappa} _2 -z {\bf q}\, .
\label{eq2.31}
\end{equation}

We next turn to the virtual contributions where the two gluon exchanges with the amplitude carry transverse momenta ${\bf q}$ and $-{\bf q}$, respectively, while the complex conjugate amplitude is the vacuum one. The corresponding Dirac terms are of the form 
\begin{equation}
	\bar{D}_{(ij)}^{g\to c\bar{c}} =  \frac{1}{\left(p_5^+\right)^2}
	 \left( z^2 + (1-z)^2 \right)\, \frac{\bm{\upkappa}_2^{(ij)}  \cdot  \bm{\upkappa}_2^*}{2z(1-z)} \, ,
	\label{eq2.32}
\end{equation}
where the transverse momentum $\bm{\upkappa}_2^{(ij)}$ is associated with the amplitude that interacts twice with the medium while 
$\bm{\upkappa}_2^*$ is associated with the vacuum splitting vertex in the complex amplitude. In particular, for the only off-diagonal virtual 
contribution (23), the transverse momenta leaving the $g\to c\bar{c}$ vertex are ${\bf p}_2 + {\bf q}$ and ${\bf p}_3 - {\bf q}$ and
one finds by explicit calculation
\begin{align}
	 \bar{D}_{(23)}^{g\to c\bar{c}}
	&=\frac{1}{(2p_5^+)^2} \frac{1}{(2p_2^+)\, (2p_3^+)} 
	d^{\rho\mu}(p) {\rm Tr}\left[ \gamma_\mu \left( \not{p}_3 + \not{q}\right) \gamma^+ \not{p}_3  \gamma_\nu  \not{p}_2     \gamma^+ \left( \not{p}_2 - \not{q}\right)  \right] \, d^{\nu}_{\, \rho}(p)  
	\nonumber \\
	& = \frac{1}{\left(p_5^+\right)^2} 
	\left( z^2 + (1-z)^2 \right)\, \frac{  \bm{\upkappa}_2^{(23)}  \cdot  \bm{\upkappa}_2  }{2z(1-z)}\, , 
	\label{eq2.33}
\end{align} 
where
\begin{equation}
	\bm{\upkappa}_2^{(23)}  = \bm{\upkappa}_2\Big\vert_{{\bf p}_2 \to {\bf p}_2 + {\bf q}\, , \,  {\bf p}_3 \to {\bf p}_3 - {\bf q}}
	= \bm{\upkappa}_2 + {\bf q}\, .
	\label{eq2.34}
\end{equation}
For virtual contributions $(jj)$, both gluons are exchanged with the same parton in the splitting process. The transverse momentum flow therefore remains unchanged relative to the vacuum diagram, and consequently
\begin{equation}
\bm{\upkappa}_2^{(jj)}  = \bm{\upkappa}_2\quad \hbox{for}\, j=1,2,3\, .
\label{eq2.35}
\end{equation}
Since the $N=1$ opacity modification of $g\to c\bar{c}$ is relatively simple and known, rewriting it as a sum over many contributions $(ij)$ in \eqref{eq2.20} may appear to be of limited practical value. 
For the $N=1$ opacity modifications to $q \to qc\bar{c}$, however, the Dirac terms are significantly more involved, and understanding how they can be constructed from simple momentum shifts applied to the vacuum expressions will facilitate the further physics analysis significantly. It is for this reason that we have exposed the use of these momentum shifts here in detail.  

\subsubsection{The longitudinal phase factors for $\hat{P}_{g\to c\bar{c}}^{(N=1)}$}
As emphasized previously, all phase factors are built from sums of transverse energies. To formalize this structure further,  we introduce short-hand notations for the transverse energies,
\begin{align}
	\Gamma_{23} &= \frac{({\bf p}_2 + {\bf p}_3)^2}{2 p_5^+}\, , \quad 
	\Gamma_2 = \frac{ {\bf p}_2^2}{2 z\, p_5^+}\, , \quad 
	\Gamma_3 = \frac{  {\bf p}_3^2}{2 (1-z)\, p_5^+}\, , \label{eq2.36} \\
	\bar{\Gamma}_{23} &= \frac{({\bf p}_2 + {\bf p}_3 + {\bf q})^2}{2 p_5^+}\, , \quad 
	\bar{\Gamma}_2 = \frac{ ({\bf p}_2   + {\bf q})^2}{2 z\, p_5^+}\, , \quad 
	\bar{\Gamma}_3 = \frac{  ({\bf p}_3 + {\bf q})^2}{2 (1-z)\, p_5^+}\, , \label{eq2.37}
\end{align}
and we associate the following phases to the three diagrams in Fig.\ref{fig2} 
\begin{align}
	P_5(\check{y}^+) &= \int_0^\infty dx^+ \theta\left(x^+-\check{y}^+\right)
	\exp\left[ i \check{y}^+ \underbrace{\left( \Gamma_{23} -  \bar{\Gamma}_{23} \right)}_{\equiv A_5}  + \, i x^+ \underbrace{\left(  \Gamma_2  +  \Gamma_3 - \Gamma_{23}   \right)}_{\equiv C_5}  \right]\, ,
	\label{eq2.38}\\
	P_2(\check{y}^+) &= \int_0^\infty dx^+ \theta\left(\check{y}^+ - x^+\right) 
	\exp\left[  i \check{y}^+ \underbrace{\left(\Gamma_2 - \bar{\Gamma}_2  \right)}_{\equiv A_2}  + \, i x^+ \underbrace{  \left( \bar{\Gamma}_2 + \Gamma_3  - \bar{\Gamma}_{23} \right)}_{\equiv C_2}  \right],
	\label{eq2.39}\\
	P_3(\check{y}^+) &= \int_0^\infty dx^+  \theta\left(\check{y}^+ - x^+\right) 
	\exp\left[ i \check{y}^+ \underbrace{\left( \Gamma_3 -  \bar{\Gamma}_3 \right)}_{\equiv A_3}  + \, i x^+ \underbrace{\left( \Gamma_2 + \bar{\Gamma}_3 -  \bar{\Gamma}_{23}  \right)}_{\equiv C_3}  \right]\, .
	\label{eq2.40}
\end{align}
The integration variables $x^+$ and $\check{y}^+$ denote the light-cone times at which the $g \to c \bar{c}$ splitting and the interaction with the medium occur, respectively. The $\theta$-functions in these phases enforce the relative ordering of the two vertices in the diagrams of Fig.\ref{fig2}.
All  light-cone time integrals are understood to include an $\epsilon$-regulator, $e^{-\epsilon x^+}$, whenever the integration extends to infinity; the limit $\epsilon \to 0$ is taken only after the upper integration limit is sent to infinity.  

In the phases \eqref{eq2.38}-\eqref{eq2.40},  we have grouped the sums of transverse energies $\Gamma_{23}$, $\Gamma_2$, $\Gamma_3$, $\bar{\Gamma}_{23}$, $\bar{\Gamma}_2$, $\bar{\Gamma}_3$ into six combinations  $A_5$, $C_5$, $A_2$, $C_2$, $A_3$, $C_3$ that satisfy three linear relations
\begin{align}
	C_2 - C_5 &= A_5 - A_2 \, ,
	\label{eq2.41}\\
	C_5 - C_3 &= A_3 - A_5 \, , 
	\label{eq2.42} \\
	C_2 - C_3 &= A_3 - A_2 \, .
	\label{eq2.43}
\end{align}
This allows us to express all phases $P_i(\check{y}^+)$ in terms of  $C_i$.  The combination $C_5$ of transverse energies  has appeared already in the vacuum splitting, see \eqref{eq2.15}. In terms of the boost-invariant transverse momentum $ \bm{\upkappa}_2 $  and the corresponding shifted transverse momenta \eqref{eq2.31}, we parametrize the combination of transverse momenta in \eqref{eq2.15} in the form
\begin{align}
	 C_5 &\equiv 	 \left(\Gamma_2 + \Gamma_3 - \Gamma_{23} \right) =  \frac{ \bm{\upkappa}_2^2}{2p_5^+ z (1-z)} =  \frac{ \bm{\upkappa}_2^{(1)} \cdot  \bm{\upkappa}_2^{(1)}  }{2p_5^+ z (1-z)}  \, ,
	 	\label{eq2.44} \\
	C_2 &\equiv 	 \left(\bar{\Gamma}_2 + \Gamma_3 - \bar{\Gamma}_{23} \right) =  \frac{\left( \bm{\upkappa}_2 + {\bf q}(1-z)\right)^2}{2p_5^+ z (1-z)}
	=  \frac{ \bm{\upkappa}_2^{(2)} \cdot  \bm{\upkappa}_2^{(2)}  }{2p_5^+ z (1-z)}  \, ,
		\label{eq2.45}\\	
	C_3 &\equiv	 \left(\Gamma_2 + \bar{\Gamma}_3 - \bar{\Gamma}_{23} \right) = \frac{\left( \bm{\upkappa}_2 - {\bf q} z\right)^2}{2p_5^+ z (1-z)} 
	= \frac{ \bm{\upkappa}_2^{(3)} \cdot  \bm{\upkappa}_2^{(3)}  }{2p_5^+ z (1-z)}  \, .
		\label{eq2.46}
\end{align}

The longitudinal phase integrals ${\cal P}_{(ij)}$ appearing in the medium-modified splitting function \eqref{eq2.20}  are then obtained from simple integrals over products of phases \eqref{eq2.38}--\eqref{eq2.40}, 
\begin{equation}
{\cal P}_{(ij)}  \equiv   \int_0^\infty d\check{y}^+ \, n(\check{y}^+)  \frac{P_i(\check{y}^+) \,  P_j^*(\check{y}^+) + P_j(\check{y}^+) \,  P_i^*(\check{y}^+)}{1 + \delta_{ij}} \, .
\label{eq2.47}
\end{equation}
In deriving \eqref{eq2.23}, we already encountered one specific example of this general form.  As explained in the text following \eqref{eq2.20}, we define these longitudinal phase integrals for symmetrized pairs $(ij)$. To write ${\cal P}_{(ij)}$  in explicit analytic form, it is useful to parametrize interference terms using the shorthand 
\begin{equation}
	{\cal S}(C) \equiv \left(1 - \frac{\sin C L}{C L} \right)\, ,
	\label{eq2.48}
\end{equation}
where $C$ denotes (linear combinations of) the transverse energies. The interference factor ${\cal S}(C) $ can be thought of as comparing the inverse formation time $C$ to the in-medium pathlength $L$, interpolating between the totally coherent ($L \ll 1/C$) and totally incoherent ($L \gg 1/C$) limits
\begin{equation}
{\cal S}(C) =
\begin{cases}
0\quad \text{for}  & C L \to 0 \\
1 \quad \text{for}  & C L  \to \infty
\end{cases}\, .
\label{eq2.49}
\end{equation}
For the homogeneous distribution \eqref{eq2.19} of scattering centers,  evaluating \eqref{eq2.47} yields
\begin{align}
	{\cal P}_{(55)}^{g\to c\bar{c}} &= 
	\frac{n_0\, L}{C_5^2}\, ,
	\label{eq2.50} \\
	{\cal P}_{(22)}^{g\to c\bar{c}} &= \frac{2\, n_0\, L}{C_2^2} {\cal S}(C_2) \, ,
	\label{eq2.51} \\
	{\cal P}_{(33)}^{g\to c\bar{c}} &= \frac{2\, n_0\, L}{C_3^2} {\cal S}(C_3) \, ,
	\label{eq2.52} \\
	{\cal P}_{(25)}^{g\to c\bar{c}} &=   - \frac{2\, n_0\, L}{C_5\, C_2} {\cal S}(C_2) \, ,
	\label{eq2.53}\\
	{\cal P}_{(35)}^{g\to c\bar{c}} &=   - \frac{2\, n_0\, L}{C_5\, C_3} {\cal S}(C_3) \, ,
	\label{eq2.54}\\
	{\cal P}_{(23)}^{g\to c\bar{c}} &=  \frac{ 2\, n_0 L}{ C_2\,  C_3} 
		 \left({\cal S}(C_2) + {\cal S}(C_3) - {\cal S}(C_2- C_3)   \right) \, .
		 \label{eq2.55}
\end{align}
For the virtual contributions, the phases associated with the amplitudes are
\begin{align}
	\bar{P}_{55}^{g\to c\bar{c}}(\check{y}^+) &= \int_{\check{y}^+}^\infty dx^+ \, e^{  i x^+\,  C_5 }\, ,\label{eq2.56} \\
	\bar{P}_{22}^{g\to c\bar{c}}(\check{y}^+) &= \bar{P}_{33}^{g\to c\bar{c}}(\check{y}^+) = \int_0^{\check{y}^+} dx^+  \, e^{ i x^+\,  C_5 }\, ,
	\label{eq2.57}\\
	\bar{P}_{23}^{g\to c\bar{c}}(\check{y}^+) &= \int_0^{\check{y}^+} dx^+ \, 
	\exp\left[  i \check{y}^+ \underbrace{\left(  \Gamma_2 -  \bar{\Gamma}_2 + \Gamma_3 -  \tilde{\Gamma}_3  \right)}_{= C_5 - \tilde{C}}  + \, i x^+ \underbrace{\left(   \bar{\Gamma}_2 + \tilde{\Gamma}_3 - \Gamma_{23}  \right)}_{\equiv \tilde{C}}  \right]\, .
	\label{eq2.58}
\end{align}
For the virtual $(23)$ contribution, the transverse momenta leaving the $g\to c\bar{c}$ vertex are ${\bf p}_2 + {\bf q}$ and ${\bf p}_3 - {\bf q}$, respectively. 
In addition to the shifted transverse energies $\bar{\Gamma}_2 = \tfrac{\left( {\bf p}_2+ {\bf q} \right)^2}{2 z p_5^+}$ and 
$\bar{\Gamma}_3 = \tfrac{\left( {\bf p}_3+ {\bf q} \right)^2}{2 (1-z) p_5^+}$, introduced in \eqref{eq2.37}, this leads to a transverse energy $\tilde{\Gamma}_3$ for which the transverse momentum ${\bf q}$ flows in the opposite direction,
\begin{equation}
\tilde{\Gamma}_3 = \frac{\left( {\bf p}_3 - {\bf q} \right)^2}{2 (1-z) p_5^+}\, .
\label{eq2.59}
\end{equation}
As indicated in \eqref{eq2.58}, the phase $\bar{P}_{23}(\check{y}^+) $ can then be expressed in terms of $C_5$ and 
\begin{equation}
	\tilde{C} = \frac{ \left( \bm{\upkappa}_2 + {\bf q} \right)^2 }{2 z (1-z) p_5^+}\, .
	\label{eq2.60}
\end{equation}

With these ingredients, we can now calculate the symmetrized longitudinal phase integrals associated with the virtual contributions,~\footnote{We comment on a subtelty about the symmetrization factors $1 + \delta_{ij}$.  For the phases \eqref{eq2.47} associated with the real contributions, ${\cal P}_{(jj)}$, there is only one diagrammatic contribution, and dividing by a factor $2 = 1 + \delta_{ij} \vert_{i=j}$ simply corrects the overcounting arising from the symmetrization condition. This is not the case for the symmetrized longitudional phase integrals  \eqref{eq2.61} associated with virtual contributions, where two diagrams contribute depending on whether the two gluons are exchanged in the amplitude or its complex conjugate. However, when both gluons are exchanged with the same parton at the same light-cone time $\check{y}$, a factor $\tfrac{1}{2}$ arises from the rules of time-ordered perturbation theory. Thus, although the symmetrization conditons for real and imaginary contributions are identical,  they have different origins.}
\begin{equation}
	\bar{\cal P}_{(ij)} =   \int_0^\infty d\check{y}^+ \, n(\check{y}^+)  \frac{ \bar{P}_{ij}(\check{y}^+) \,  P^*_{\rm vac} + 
	P_{\rm vac}\, \bar{P}_{ij}^*(\check{y}^+) }{1 + \delta_{ij}}\, ,
	\label{eq2.61}
\end{equation}
where the phase associated to the vacuum splitting amplitude
\begin{equation} 
P_{\rm vac} = \int_0^\infty dx^+\, e^{i x^+ C_5 - \epsilon x^+} = \frac{i}{ C_5}\, .
\label{eq2.62}
\end{equation} 
We find
\begin{align}
	\bar{\cal P}_{(55)}^{g\to c\bar{c}}  &= 
	\frac{n_0\, L}{C_5^2} \frac{\sin \left(C_5 L\right)}{C_5 L}
	\label{eq2.63} \\
	\bar{\cal P}_{(22)}^{g\to c\bar{c}} &= \bar{\cal P}_{(33)}^{g\to c\bar{c}} = \frac{n_0\, L}{C_5^2} {\cal S}(C_5) \, ,
	\label{eq2.64} \\
	\bar{\cal P}_{(23)}^{g\to c\bar{c}} &=  \frac{ 2\, n_0 L}{ C_5\,  \tilde{C}} 
	\left({\cal S}(C_5)  - {\cal S}(C_5- \tilde{C})   \right) \, .
	\label{eq2.65}
\end{align}
%

\subsection{The final result for $\hat{P}_{g\to c\bar{c}}^{(N=1)}$}
\label{sec2.6}
To write the complete result for the $g\to c\bar{c}$ splitting function \eqref{eq2.20} to first order in opacity and at leading order in $N_c$, 
we now combine the longitudinal phase integrals  \eqref{eq2.50}-\eqref{eq2.55} and \eqref{eq2.63}-\eqref{eq2.65}, the color factors \eqref{eq2.24}-\eqref{eq2.28} and the Dirac terms \eqref{eq2.29} and \eqref{eq2.32}.  

To this end, we note first that the color factor $C_{(23)}^{g\to c\bar{c}}$ and $\bar{C}_{(23)}^{g\to c\bar{c}}$ are subleading in $N_c$ and that 
$\bar{C}_{(55)}^{g\to c\bar{c}} = \bar{C}_{(22)}^{g\to c\bar{c}} + \bar{C}_{(33)}^{g\to c\bar{c}} = - N_c\, T_R$. 
The sum over all virtual contributions in \eqref{eq2.20} takes therefore to leading order in $N_c$ the form 
\begin{align}
 \sum_{\substack{ (i,j) \\ i\leq j} } \left( 
  \bar{C}_{(ij)}^{g\to c\bar{c}} \bar{D}_{(ij)}^{g\to c\bar{c}} \bar{\cal P}_{(ij)}^{g\to c\bar{c}}   \right) &= \sum_{j=5,2,3} D^{g\to c\bar{c}}_{\rm vac} \bar{C}^{g\to c\bar{c}}_{(jj)} \bar{\cal P}_{(jj)}^{g\to c\bar{c}} 
  = - N_c\, T_R\, D_{\rm vac} \, \frac{n_0\, L}{C_5^2} \nonumber \\
  &= - {C}_{(55)}^{g\to c\bar{c}} {D}_{(55)}^{g\to c\bar{c}} {\cal P}_{(55)}^{g\to c\bar{c}}   
   \label{eq2.66}
\end{align}
The sum over virtual contributions thus  cancels the real (55) term and the final result is the sum over the real contributions
(22), (33), (25) and (35),
\begin{equation}
\hat{P}_{g\to c\bar{c}}^{(N=1)} = n_0 L \frac{N_c}{2} \negmedspace\int\negmedspace\frac{d{\bf q}}{(2\pi)^2} \vert a({\bf q}) \vert^2 
 \left(\negmedspace\left( \frac{D_{(22)}^{g\to c\bar{c}} }{C_2^2} -  \frac{D^{g\to c\bar{c}}_{(25)}}{C_2 C_5} \right) {\cal S}(C_2) 
 +  \left( \frac{D^{g\to c\bar{c}}_{(33)}}{C_3^2} -  \frac{D^{g\to c\bar{c}}_{(35)}}{C_3 C_5} \right) {\cal S}(C_3) \negmedspace\right) \, ,
 \label{eq2.67}
\end{equation}
To check consistency with results in the literature, we introduce 
\begin{equation}
	\vert a_3({\bf q},z) \vert^2  \equiv \frac{N_c}{2 z^2} \vert a({\bf q}/z) \vert^2  + \frac{N_c}{2 (1-z)^2}    \vert a({\bf q}/(1-z)) \vert^2 \, ,
	\label{eq2.68}
\end{equation}
Rescaling the integration variable ${\bf q}$ in \eqref{eq2.67} by a factor $1/z$ in the first and a factor $1/(1-z)$ in the second term of \eqref{eq2.67}, 
one finds
\begin{equation}
\hat{P}_{g\to c\bar{c}}^{(N=1)} =  
	\frac{1}{2\left( p_5^+\right)^2}
\frac{z^2 + (1-z)^2}{z(1-z)}\, n_0 L  \int\frac{d{\bf q}}{(2\pi)^2} \vert a_3({\bf q},z) \vert^2 \, 
	{\cal S}(\tilde{C}) \left( 
	\frac{\left( \bm{\upkappa}_2  +  {\bf q}\right)^2}{\tilde{C}^2}  - \frac{\bm{\upkappa}_2  \cdot \left( \bm{\upkappa}_2   +  {\bf q}\right)}{C_5\, \tilde{C}} \right)\, ,
 \label{eq2.69}
\end{equation}
where $\tilde{C}$ is defined in \eqref{eq2.60}. This expression is consistent with the terms (A.1) and (A.4) in ~\cite{Attems:2022ubu}.  

\begin{figure}
   \centering
   \newcommand{\diagraminclude}[1]{\includegraphics[width=0.30\textwidth]{#1}}
   \diagraminclude{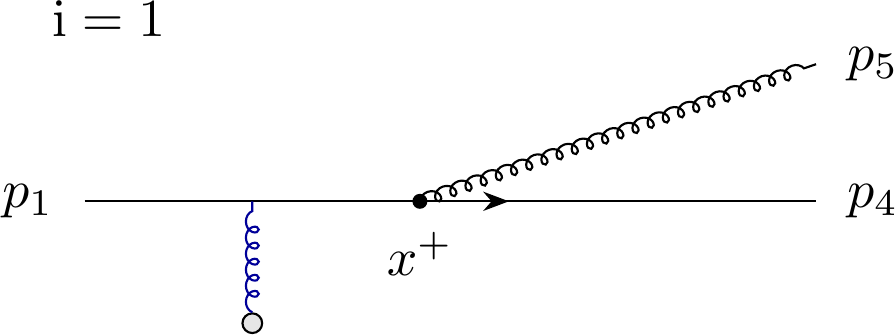}\hspace{0.8cm}
   \diagraminclude{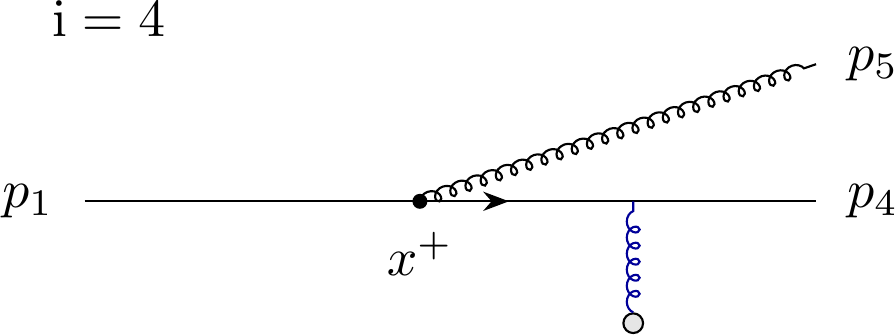}\\[0.2em]
   \diagraminclude{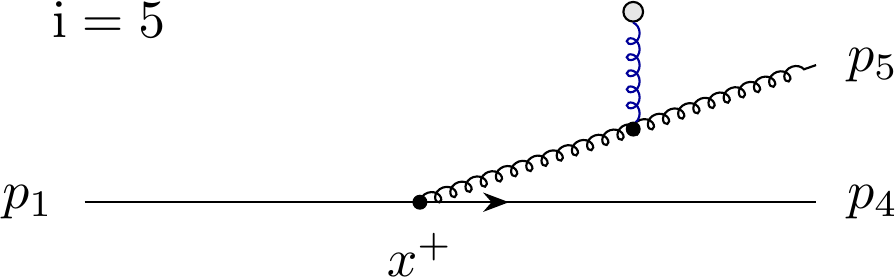}\hspace{0.8cm}
   \diagraminclude{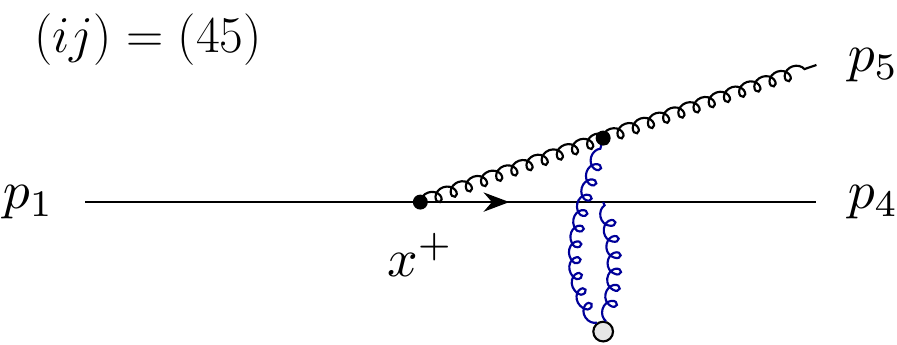}\\[0.2em]
   \diagraminclude{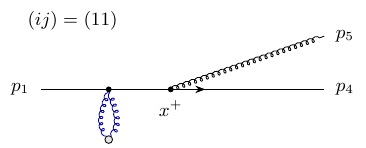}\hspace{0.8cm}
   \diagraminclude{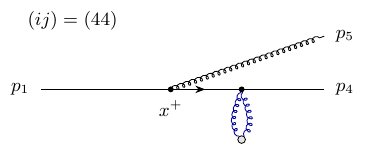}\\[0.2em]
   \diagraminclude{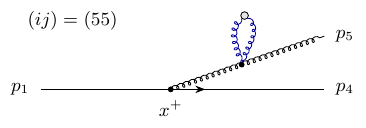}
   \caption{The three real scattering amplitudes $i = 1,4,5$ that contribute to the $q(1) \to q(4)\, g(5)$ splitting function at first order in opacity, the off-diagonal virtual amplitude $(ij)=(45)$,  and the diagonal virtual amplitudes $(ij)=(11)$, $(ij)=(44)$ and $(ij)=(55)$.}
   \label{fig6}
\end{figure}
\subsection{The result for $\hat{P}_{q\to qg}^{(N=1)}$}
In section~\ref{sec5}, we identify the $q\to qg$ splitting function at first order in opacity and leading order in $N_c$ as a building block of the collinear limit of  the $q\to qc\bar{c}$ splitting function. To ease the discussion in section~\ref{sec5}, we summarize here the relevant results for $q\to qg$, indexing the incoming quark as "1", the outgoing quark as "4" and the outgoing gluon as "5". 

The vacuum splitting function takes the form
\begin{align}
	\hat{P}_{q \to q g} &= \frac{d_{\mu\nu}(p_5)}{4 n\cdot \left(p_4+p_5\right)} {\rm Tr}\left[\slashed{n}\, \left(\slashed{p}_4 + \slashed{p}_5\right)\, \gamma^\mu\, \slashed{p}_4\, \gamma^\nu \,  \left(\slashed{p}_4 + \slashed{p}_5\right) \right] \nonumber \\
	&= \frac{2}{p_1^2} C_F\, \frac{1 + (1-\zeta)^2}{\zeta}
	\label{eq2.70}
\end{align}
In close analogy with \eqref{eq2.11}, this is the LO $q \to q g$ splitting function multiplied by the normalization $\tfrac{2}{p_1^2} $ involving the squared invariant mass of the parent parton. Likewise, the calculation of $q \to qg$ at first order in opacity proceeds in close analogy with that of $g\to c\bar{c}$, presented in the last subsection. The corresponding diagrammatic contributions are shown in Fig.~\ref{fig6}. and the result can be cast in the form
\begin{align}
\hat{P}_{q\to qg}^{(N=1)} &= \int\frac{d{\bf q}}{(2\pi)^2} \vert a({\bf q}) \vert^2  
  \sum_{\substack{ (i,j),\,  i\leq j   \\  i,j \in 1,4,5} }  \left(
  C^{q\to qg}_{(ij)} D^{q\to qg}_{(ij)} {\cal P}^{q\to qg}_{(ij)}  + 
  \bar{C}^{q\to qg}_{(ij)} \bar{D}^{q\to qg}_{(ij)} \bar{\cal P}^{q\to qg}_{(ij)}  \right)\, .
 \label{eq2.71}
\end{align}
To leading order in $N_c$, the color factors read 
\begin{equation}
	C_{11}^{q\to qg} = C_{44}^{q\to qg} = C_{15}^{q\to qg} = C_{45}^{q\to qg} = \frac{N_c^2}{4}\, , \quad C_{55}^{q\to qg} = \frac{N_c^2}{2}\, ,\quad C_{14}^{q\to qg} = 0\, .
	\label{eq2.72}
\end{equation}
For the probability-conserving virtual contributions, one finds the color factors
\begin{equation}
	\bar{C}_{11}^{q\to qg} = \bar{C}_{44}^{q\to qg} =  \bar{C}_{45}^{q\to qg} = - \frac{N_c^2}{4}\, , \quad \bar{C}_{55}^{q\to qg} = - \frac{N_c^2}{2}\, .
	\label{eq2.73}
\end{equation}
%
The Dirac term associated with the vacuum splitting can be expressed in terms of the boost-invariant transverse momentum $\bm{\upkappa}_1$ defined in \eqref{eq1.7} 
\begin{align}
	D^{q\to qg}_{\rm vac} &= \frac{1}{2^2 (p_1^+)^2}  \frac{1}{4 n\cdot p_1} d^{\mu\nu}(p_5)\, {\rm Tr}\left[ \slashed{n} \left( \slashed{p}_4 + \slashed{p}_5 \right) \gamma_\mu \slashed{p}_4\, \gamma_\nu \left( \slashed{p}_4 + \slashed{p}_5 \right)  \right] \nonumber \\
	&= \frac{1}{(p_1^+)^2}  \frac{(1-\zeta)^2 + 1}{\zeta}\,  \frac{ {\bm{\upkappa}_1} \cdot  {\bm{\upkappa}_1}  }{2 \zeta (1 - \zeta) }\, .
	\label{eq2.74}
\end{align}
so that the vacuum splitting function \eqref{eq2.70} can be written in close analogy with \eqref{eq2.16} as the product of a color factor $C_F$, a longitudinal phase factor $1/B_1^2$ defined in \eqref{eq2.82} below, and this Dirac term
\begin{align}
	\hat{P}_{q \to q g} &= C_F\, \frac{1}{B_1^2} D^{q\to qg}_{\rm vac} \, .
	\label{eq2.74b}
\end{align}
To first order in opacity, the Dirac terms for real medium-modifications read
\begin{equation}
	D^{q\to qg}_{(ij)} =  \frac{1}{(p_1^+)^2}  \frac{(1-\zeta)^2 + 1}{\zeta}\,  \frac{ {\bm{\upkappa}_1^{(i)} } \cdot  {\bm{\upkappa}_1^{(j)} }  }{2 \zeta (1 - \zeta) }\, ,
	\quad \hbox{for}\quad i,j = 1,4,5, \label{eq2.75}
\end{equation}
where we have introduced the shifted momenta
\begin{align}
	{\bm{\upkappa}_1^{(1)} } &= \bm{\upkappa}_1 \, ,\label{eq2.76} \\
	{\bm{\upkappa}_1^{(4)} } &= \bm{\upkappa}_1 \Big\vert_{{\bf p}_4 \to {\bf p}_4 + {\bf q}}  = \bm{\upkappa}_1 + \zeta \, {\bf q}\, ,\label{eq2.77} \\
	{\bm{\upkappa}_1^{(5)} } &= \bm{\upkappa}_1 \Big\vert_{{\bf p}_5 \to {\bf p}_5 + {\bf q}}  = \bm{\upkappa}_1 - (1-\zeta) \, {\bf q}\, .\label{eq2.78}
\end{align}
For virtual contributions, the diagonal terms are essentially the vacuum contribution,
\begin{equation}
	\bar{D}^{q\to qg}_{(jj)} =   D^{q\to qg}_{\rm vac} 
	\quad \hbox{for}\quad i,j = 1,4,5. \label{eq2.79}
\end{equation}
In addition, there is one off-diagonal virtual contribution, 
\begin{equation}
	\bar{D}^{q\to qg}_{(45)} = \frac{1}{(p_1^+)^2} \frac{(1-\zeta)^2 + 1}{\zeta}\,  \frac{ {\bar{\bm{\upkappa}}_1} \cdot  {\bm{\upkappa}_1}  }{2 \zeta (1 - \zeta) } \, ,
		\label{eq2.80}
\end{equation}
where the shifted transverse momentum in the amplitude is
\begin{equation}
{\bar{\bm{\upkappa}}_1} = {\bm{\upkappa}}_1 \Big\vert_{{\bf p}_4 \to {\bf p}_4 + {\bf q},\, \,{\bf p}_5 \to {\bf p}_5 - {\bf q}} = 
{\bm{\upkappa}}_1 + {\bf q}\, .
\label{eq2.81}
\end{equation}

Finally, the structure of the longitudinal phase factors does not depend on whether a quark or gluon scatters. These factors can therefore be inferred from our evaluation of phase factors \eqref{eq2.47} for $g \to c\bar{c}$. Changing  in all expressions \eqref{eq2.44} - \eqref{eq2.55} the subscripts 
$2 \to 4$ and $3 \to 5$, and introducing transverse energies
\begin{equation}
	B_i = \frac{  {\bm{\upkappa}_1}^{(i)} \cdot  {\bm{\upkappa}_1}^{(i)}   }{2 p_1^+ \zeta (1-\zeta)}\, ,\qquad i=1,4,5\, ,
	\label{eq2.82}
\end{equation} 
related to the shifted transverse momenta defined in \eqref{eq2.76}--\eqref{eq2.78}. The longitudinal phase integrals of real contributions read then
\begin{align}
	{\cal P}_{(11)}^{q\to qg} &= 
	\frac{n_0\, L}{B_1^2}\, ,
	\label{eq2.83} \\
	{\cal P}_{(44)}^{q\to qg} &= \frac{2\, n_0\, L}{B_4^2} {\cal S}(B_4) \, ,
	\label{eq2.84} \\
	{\cal P}_{(55)}^{q\to qg} &= \frac{2\, n_0\, L}{B_5^2} {\cal S}(B_5) \, ,
	\label{eq2.85} \\
	{\cal P}_{(14)}^{q\to qg} &=   - \frac{2\, n_0\, L}{B_1\, B_4} {\cal S}(B_4) \, ,
	\label{eq2.86}\\
	{\cal P}_{(15)}^{q\to qg} &=   - \frac{2\, n_0\, L}{B_1\, B_5} {\cal S}(B_5) \, ,
	\label{eq2.87}\\
	{\cal P}_{(45)}^{q\to qg} &=  \frac{ 2\, n_0 L}{ B_4\,  B_5} 
		 \left({\cal S}(B_4) + {\cal S}(B_5) - {\cal S}(B_4- B_5)   \right) \, .
		 \label{eq2.88}
\end{align}
and for the virtual contributions, one finds
\begin{align}
	\bar{\cal P}_{(11)}^{q\to qg}  &= 
	\frac{n_0\, L}{B_1^2} \frac{\sin \left(B_1 L\right)}{B_1 L}
	\label{eq2.90} \\
	\bar{\cal P}_{(44)}^{q\to qg} &= \bar{\cal P}_{(55)}^{g\to c\bar{c}} = \frac{n_0\, L}{B_1^2} {\cal S}(B_1) \, ,
	\label{eq2.91} \\
	\bar{\cal P}_{(45)}^{q\to qg} &=  \frac{ 2\, n_0 L}{ B_1\,  \tilde{B}} 
	\left({\cal S}(B_1)  - {\cal S}(B_1- \tilde{B})   \right) \, .
	\label{eq2.92}
\end{align}

We close this section by taking a particularly simple rescattering limit of $\hat{P}_{q\to qg}^{(N=1)}$ in \eqref{eq2.71} that will be useful for our analysis of the medium-modified $q\to qc\bar{c}$ splitting function in section~\ref{sec5}. In this limit, the opacity $n_0 L \vert a({\bf q})\vert^2$ is kept fixed, while the transverse energies are taken to be parametrically large,
\begin{equation}
	n_0 L = \hbox{fixed}, \qquad \frac{1}{B_iL} \to 0\, .
	\label{eq2.93}
\end{equation}
Physically, taking all formation times $\propto 1/B_i$ to zero means that the $q \to qg$ splitting forms instantaneously. One thus expects that this splitting is not resolved by the medium, i.e., it proceeds as in vacuum, while the outgoing daughter partons undergo momentum broadening. Technically, $\lim_{1/B_iL \to 0} {\cal S}(B_i) = 1$ and, therefore, the longitudinal phase integrals \eqref{eq2.83}--\eqref{eq2.92} simplify to
\begin{align}
&\hbox{Values in the rescattering limit \eqref{eq2.93}:} \label{eq2.94} \\
 {\cal P}_{(11)}^{q\to qg} &= \frac{n_0\, L}{B_1^2}\, , \quad
	{\cal P}_{(44)}^{q\to qg} = \frac{2\, n_0\, L}{B_4^2} \, , \quad
	{\cal P}_{(55)}^{q\to qg} = \frac{2\, n_0\, L}{B_5^2}  \, , \nonumber \\
	{\cal P}_{(15)}^{q\to qg} &=   - \frac{2\, n_0\, L}{B_1\, B_5} \, ,\quad
	{\cal P}_{(45)}^{q\to qg} =  \frac{ 2\, n_0 L}{ B_4\,  B_5} \nonumber \\
	\bar{\cal P}_{(44)}^{q\to qg} &=  \frac{n_0\, L}{B_1^2}\, , \quad \bar{\cal P}_{(55)}^{g\to c\bar{c}} = \frac{n_0\, L}{B_1^2}\, . \nonumber
\end{align}
In this list, we have suppressed the term ${\cal P}_{(14)}^{q\to qg}$, as it multiplies a contribution that is subleading in $N_c$. 
\begin{align}
\hat{P}_{q\to qg,\, {\rm rescatt}}^{(N=1)} =& 2 n_0\, L\, \frac{N_c^2}{4} \, \int\frac{d{\bf q}}{(2\pi)^2} \vert a({\bf q}) \vert^2  \nonumber\\
&\times
\Bigg(  \left( \frac{D^{q\to qg}_{(44)}}{B_4^2} -   \frac{D^{q\to qg}_{\rm vac}}{B_1^2}\right) 
+ 2\, \left( \frac{D^{q\to qg}_{(55)}}{B_5^2} -   \frac{D^{q\to qg}_{\rm vac}}{B_1^2}\right) \nonumber \\
& \qquad - \left( \frac{D^{q\to qg}_{(15)}}{B_1\, B_5} -   \frac{D^{q\to qg}_{\rm vac}}{B_1^2}\right) 
- \left( \frac{D^{q\to qg}_{(45)}}{B_4\, B_5} -   \frac{D^{q\to qg}_{\rm vac}}{B_1^2}\right)  + {\cal O}\left( B_i^{-3} \right)
  \Bigg)\, .
 \label{eq2.95}
\end{align}
In this expression, each of the medium-modified products of Dirac terms and longitudinal phase factors $\tfrac{D^{q\to qg}_{(ij)}}{B_iB_j}$ may be rewritten as
\begin{align}
	 \frac{D^{q\to qg}_{(ij)}}{B_i\, B_j} &= 	
	  2 \zeta (1 - \zeta)  \frac{(1-\zeta)^2 + 1}{\zeta}\,  
	  \left( \frac{ {\bm{\upkappa}_1^{(i)} }   }{  \bm{\upkappa}_1^{(i)} \cdot  \bm{\upkappa}_1^{(i)} } \right) \cdot 
	  \left( \frac{ {\bm{\upkappa}_1^{(j)} }   }{  \bm{\upkappa}_1^{(j)} \cdot  \bm{\upkappa}_1^{(j)} } \right) \nonumber \\
	  &=  2 \zeta (1 - \zeta) \frac{(1-\zeta)^2 + 1}{\zeta}\,  
	  \left( \frac{ {\bm{\upkappa}_1 }   }{  (\bm{\upkappa}_1)^2  } \right)_{\bm{\upkappa}_1 \to \bm{\upkappa}_1^{(i)}} \cdot 
	  \left( \frac{ {\bm{\upkappa}_1^* }   }{ ( \bm{\upkappa}_1^*)^2 } \right)_{\bm{\upkappa}_1^* \to \bm{\upkappa}_1^{(j)}}  
	  \nonumber \\
	  &= 
	  "\left( \frac{D^{q\to qg}_{\rm vac}}{B_1^2} \right)_{\rm "shifted"}"\, .
	  \label{eq2.96}
\end{align}
The four contributions $\left( \tfrac{D^{q\to qg}_{(ij)}}{B_i\, B_j} - \tfrac{D^{q\to qg}_{\rm vac}}{B_1^2} \right)$ in \eqref{eq2.95} can therefore be interpreted as momentum-shifted vacuum contributions from which the unshifted vacuum baseline has been subtracted. This is the tell-tale sign of a rescattering contribution that redistributes yield from the vacuum distribution to a momentum-shifted one with a weight proportional to the opacity $n_0 L \vert a({\bf q}) \vert^2$. This rescattering, however, is more subtle than a probabilistic shift of the momentum of the outgoing quark (4) or gluon (5). Such classical momentum-shift contributions account for only the first two terms in \eqref{eq2.95}.

\section{The vacuum splitting $q \to q c \bar{c}$ in time-ordered perturbation theory}
\label{sec3}
In close analogy to the $1\to 2$ vacuum splitting function \eqref{eq2.13} analyzed in the previous section, we now write the $q \to q c \bar{c}$ vacuum splitting function with propagators in mixed representation
\begin{align}
	\hat{P}_{q \to q c \bar{c}}  =&   {\cal N}_{q \to q c \bar{c}} \, G^{\rho\mu}(p_2 + p_3)\,  {\rm Tr}\left[ \gamma_\mu \not{p}_{3} \gamma_\nu \not{p}_{2} \right]
	G^{\nu\lambda}(p_2+p_3) \nonumber \\
	& \times  \frac{1}{4 n\cdot p_1} {\rm Tr}\left[ \not{n}\, G^F(p_1)\, \gamma_\rho \not{p}_4 \gamma_\lambda G^F(p_1)\right]
	\nonumber \\
	=&  {\cal N}_{q \to q c \bar{c}}  \, \int dx^+\, dy^+\, d\bar{x}^+\, d\bar{y}^+\, e^{i (p_2^- + p_3^-)x^+}\, e^{i (p_2^- + p_3^- + p_4^-)y^+}\, 
	\nonumber \\
	& \qquad \qquad  \times e^{-i (p_2^- + p_3^-)\bar{x}^+}\,  e^{-i (p_2^- + p_3^- + p_4^-)\bar{y}^+} \nonumber \\
	& \qquad \qquad  \times G^{\rho\mu}(p_2 + p_3; x^+)\,  {\rm Tr}\left[ \gamma_\mu \not{p}_{3} \gamma_\nu \not{p}_{2} \right]
	G^{\nu\lambda}(p_2+p_3;\bar{x}^+)  \nonumber \\
	& \qquad \qquad  \times  \frac{1}{4 n\cdot p_1} {\rm Tr}\left[ \slashed{n}\, G^F(p_1;y^+)\, \gamma_\rho \slashed{p}_4 \gamma_\lambda G^F(p_1;\bar{y}^+)\right]\, .
	\label{eq3.1}
\end{align} 
The main aim of this section is to discuss the details of the calculation of \eqref{eq3.1} in time-ordered perturbation theory, before extending it in section~\ref{sec4} to include medium modifications. 

\subsection{Absence of instantaneous contributions to $q \to q c \bar{c}$}
\label{sec3.1}
As seen from \eqref{eq3.1}, this mixed representation complicates the calculation of vacuum diagrams. However, this complication is necessary in our approach, as it introduces the time coordinates that are central to embedding such splitting processes in a medium. One potential complication of the mixed representation is that the numerator of the fermion propagators becomes $i\slashed{p} \to i\slashed{p}  - i \gamma^+ p^- - \gamma^+ \tfrac{\partial}{\partial x^+}$ and also the gluon polarization tensor has a time derivative $\tfrac{\partial}{\partial x^+}$ in the $d^{--}$ component, as explained in \eqref{eq2.7} and \eqref{eq2.10}. Since these derivatives act on theta-functions, they generate contributions that are instantaneous in time. In this section, we explain why such contributions do not appear in our final result. 

For the fermion propagators, the absence of an instantaneous contribution is easily seen.  In \eqref{eq3.1}, the propagators $G^F(p;y^+)$ appear inside a trace adjacent to $\slashed{n} = \gamma^+$. Since the instantaneous part of $G^F(p;y^+)$ is proportional to $\gamma^+$, it does not contribute. One can therefore equivalently use the simpler numerator $i\slashed{p} $ for $G^F(p;y^+)$ in the mixed representation \eqref{eq3.1}. The same argument applies in the presence of medium-interactions. Since these enter fermion traces through vertices of the form $\gamma \cdot A = \gamma^+ A^-$ which are likewise proportional to $\gamma^+$, the instantaneous terms proportional to $\gamma^+ \tfrac{\partial}{\partial x^+}$ also vanish. 

For the gluon propagator, the difference between the gluon polarization tensors in the mixed representation and in the momentum representation can be written in terms of a simple correction factor:
\begin{equation}
	d^{\rho\mu}(p_2+p_3,x^+) = d^{\rho\mu}(p_2+p_3)  + \underbrace{\delta^{\rho -}\, \delta^{\mu -} \frac{2 \left(i \frac{\partial}{\partial x^+} - (p_2^- + p_3^-) \right)}{p_2^+ + p_3^+}}_{\equiv d^{\rho\mu}_{\rm correction}} \, .\label{eq3.2}
\end{equation}
To determine whether and how this correction $d^{\rho\mu}_{\rm correction}$ contributes, we first consider the term
\begin{align}
	 T^{\rho\mu} =& \int dx^+\, dy^+\,  e^{i (p_2^- + p_3^-)x^+}\, e^{i (p_2^- + p_3^- + p_4^-)y^+}\, 
		G^{\rho\mu}(p_2 + p_3; x^+)\,  G^F(p_1;y^+) \label{eq3.3} \\
	=& \int dx^+\, dy^+\,  e^{i (p_2^- + p_3^-)x^+}\, e^{i (p_2^- + p_3^- + p_4^-)y^+}\, d^{\rho\mu}(p_2 + p_3; x^+)\ d^F(p_1^+,0, {\bf p}_1) \, \nonumber \\
	& \times  \Theta(x^+)\, \frac{1}{2 p_1^+\, \zeta}	\exp\left[ - i \frac{({\bf p}_2 + {\bf p}_3)^2}{2p_1^+\, \zeta}  x^+ \right]\, 
	 \Theta(y^+)\, \frac{1}{2 p_1^+}
	\exp\left[ - i \frac{({\bf p}_2 + {\bf p}_3 + {\bf p}_4  )^2}{2p_1^+}  y^+ \right] \nonumber \, ,
\end{align}
which is one of the factors entering \eqref{eq3.1}. Explicit evaluation of the integrals in \eqref{eq3.3} shows that the correction term in \eqref{eq3.2} does not contribute to $T^{\rho\mu} $,
\begin{equation}
	T^{\rho\mu}_{\rm correction} = 0\, .
	\label{eq3.4}
\end{equation}
In summary, the term proportional to $\tfrac{\partial}{\partial x^+}$ does  contribute to \eqref{eq3.3} -- unlike in the case of the fermion propagator, where the spinor algebra alone guarantees the vanishing of the corresponding contribution. Nevertheless, $T^{\rho\mu} $ is invariant under $d^{\rho\mu}(p_2 + p_3; x^+) \to d^{\rho\mu}(p_2 + p_3)$, and the latter contains no instantaneous contribution. This makes it possible to formulate the vacuum splitting without instantaneous terms in mixed representation.

Also for gluon propagators, this observation generalizes to medium-induced interactions. As noted at the end of section~\ref{sec2.4.1}, contracting a three-gluon vertex with a medium $A^-$-field in the medium leads to a vertex $V^{\mu + \nu}$. Contracting this vertex with the correction term in \eqref{eq3.2} gives $V^{+ + \nu}$, which is non-vanishing only for $\nu = -$. However, the gluon propagator contracted with $V^{+ + \nu}$  would require a non-vanishing $+$-component, which does not exist. We therefore conclude that the correction term in \eqref{eq3.2} does not contribute to either vacuum or medium-induced processes for $q \to q c \bar{c}$. Consequently, in expressions written in the mixed representation, $d^{\rho\mu}(p_2+p_3,x^+)$ can be replaced by $d^{\rho\mu}(p_2+p_3)$ without affecting the result. This allows us to adopt a formulation free of instantaneous contributions. 

\subsection{The longitudinal phase integrals for the vacuum splitting }
\label{sec3.2}
Longitudinal phase space integrals appear when the $1\to 3$ splitting function  \eqref{eq3.1} is written in the mixed presentation. Since the Cauchy integral involved in transforming propagators from the momentum representation into the mixed representation amounts to trading the denominators of propagators for phases, it is clear that performing the resulting phase integrals should restore these denominators. To understand how this works explicitly, we start again from  \eqref{eq3.3}  which is one of the factors entering  \eqref{eq3.1}. Shifting the integration variable $x^+ \to x^+ + y^+$, we find
\begin{equation}
	T^{\rho\mu} =   \frac{d^{\rho\mu}(p_2 + p_3)\ d^F(p^+,0,{\bf p}) }{(2 p_1^+)\, (2 p_1^+\zeta)}
	\int_0^\infty  dy^+\, \int_{y^+}^\infty d\hat{x}^+   \exp\left[ i x^+ C_5 +i y^+ B_1\right] \, ,
\label{eq3.5}
\end{equation}
where 
\begin{align}
	B_1 &\equiv \left( \frac{({\bf p}_2 + {\bf p}_3)^2}{2p_1^+\, \zeta}  + p_4^- -  \frac{({\bf p}_2 + {\bf p}_3 + {\bf p}_4  )^2}{2p_1^+} \right) 
	=  \frac{ {\bm \kappa}_1^2 }{2 p_1^+ \zeta\, (1-\zeta)} \, ,  \label{eq3.6}\\
	C_5 &\equiv   \left( p_2^- + p_3^- - \frac{({\bf p}_2 + {\bf p}_3)^2}{2p_1^+\, \zeta} \right) = \frac{ {\bm \kappa}_2^2 }{2 p_1^+ \zeta\, z (1-z)} \, .\label{eq3.7}
\end{align}
The phase integral in \eqref{eq3.5} is understood as a regularized one,
\begin{equation}
\lim_{\epsilon \to 0} \lim_{\tau^+\to\infty} \int_0^{\tau^+}  dy^+\, \int_{y^+}^{\tau^+} d{x}^+ 
	 \exp\left[  i {x}^+   C_5  +   i y^+  B_1 - \epsilon x^+ - \epsilon y^+  \right]  = \frac{-1}{C_5 (B_1+C_5)}\, .\label{eq3.8}
\end{equation}
The shorthands $B_1$ and $C_5$ combine neatly into Lorentz-invariants
\begin{align}
	B_1+C_5 &=  \frac{ (p_2+p_3+p_4)^2}{2 n\cdot (p_2+p_3+p_4)} = \frac{\tilde{s}_{234} }{2 n\cdot p_1}\, , \label{eq3.9}\\
	C_5 &=   \frac{ (p_2+p_3)^2}{2 n\cdot (p_2+p_3)}  = \frac{\tilde{s}_{23} }{2 \zeta n\cdot p_1}\,  ,\label{eq3.10}
\end{align}
where $\tilde{s}_{23}$ is the squared invariant mass of the $c\bar{c}$ system, see \eqref{eq1.4} and the invariant mass of the parent quark is
\begin{align}
	\tilde{s}_{234} &= (p_2+p_3+p_4)^2 = \tilde{s}_{23} + \tilde{s}_{24} + \tilde{s}_{34}\, .	\label{eq3.11}
\end{align}
One therefore finds that the denominator of $T^{\rho\mu} =   \frac{d^{\rho\mu}(p_2 + p_3)\ d^F(p_1^+,0,{\bf p}_1) }{p_1^2\, (p_2+p_3)^2}$ is the product of the denominators of the gluon and quark propagator, as expected. In the following section~\ref{sec4}, we discuss the extension of these longitudinal phase integrals to in-medium interactions.

\subsection{The vacuum splitting $q \to q c \bar{c}$ in standard perturbation theory}
\label{sec3.3}
To facilitate comparison with the existing literature, we recall that  the massless triple-collinear splitting function  $q \to q c \bar{c}$ is typically expressed in terms of Lorentz invariants \cite{Catani:1999ss,Craft:2023aew,Dhani:2023uxu}  
\begin{align}
	\hat{P}_{q \to q c \bar{c}}  &= \frac{4}{\tilde{s}_{234}^2 }  C_F\, T_R\, \frac{\tilde{s}_{234} }{2  \tilde{s}_{23} }  \left[ 
	- \frac{\left[ \xi_2 ( \tilde{s}_{23} + 2 \tilde{s}_{34} ) - \xi_3 ( \tilde{s}_{23} + 2 \tilde{s}_{24} ) \right]^2 }{(\xi_2+\xi_3)^2  \tilde{s}_{23} \tilde{s}_{234} } 
	+  \frac{4\xi_4 + (\xi_2-\xi_3)^2 }{\xi_2+\xi_3}  \right. \nonumber \\
	& \qquad \qquad  \qquad \qquad   \qquad \qquad  \left.  
		+ 
		\left( \xi_2 + \xi_3 - \frac{ \tilde{s}_{23} }{\tilde{s}_{234} } \right)\right]\, ,
	\label{eq3.12}
\end{align} 

The presence of a medium rest frame breaks Lorentz invariance.  In the present work, medium-modifications to \eqref{eq3.12} arise from momentum ${\bf q}$ transferred by the medium transverse to the collinear direction. To track this medium-induced transverse momentum, it is useful therefore to start from a formulation of the vacuum kinematics in which the invariants  $\tilde{s}_{ij}$ and $\tilde{s}_{234} $ are expressed in terms of the boost-invariant
transverse momenta ${\bm \kappa}_i$ defined in \eqref{eq1.6}, \eqref{eq1.7}
\begin{align}
	\tilde{s}_{23} &= \frac{{\bm \kappa}_2\cdot {\bm \kappa}_2}{z (1-z)}\, , \label{eq3.13}\\
	\tilde{s}_{24} &= \frac{ (1-\zeta)^2 {\bm \kappa}_2\cdot {\bm \kappa}_2  - 2 z (1-\zeta) {\bm \kappa}_1\cdot {\bm \kappa}_2  + z^2 {\bm \kappa}_1\cdot {\bm \kappa}_1}{z \zeta (1-\zeta)}\, , \label{eq3.14}\\
	\tilde{s}_{34} &= \frac{ (1-z)^2 {\bm \kappa}_1\cdot {\bm \kappa}_1  + 2 (1-z) (1-\zeta) {\bm \kappa}_1\cdot {\bm \kappa}_2  + (1-\zeta)^2 {\bm \kappa}_2\cdot {\bm \kappa}_2}{(1-z) \zeta (1-\zeta)}\, , \label{eq3.15}\\
	\tilde{s}_{234} &=  \frac{{\bm \kappa}_1\cdot {\bm \kappa}_1}{\zeta (1-\zeta)}  + \frac{{\bm \kappa}_2\cdot {\bm \kappa}_2}{\zeta z (1-z)}\, .
	 \label{eq3.16}
\end{align}
Following the same logic as for $g\to c\bar{c}$ in \eqref{eq2.16} and $q\to q\, g$ in \eqref{eq2.74b}, we write \eqref{eq3.16} as the product 
of a color factor, a phase factor given by the square of the longitudinal phase integrals~\eqref{eq3.8} 
\begin{equation}
{\cal P}_{\rm vac} = \frac{1}{(B_1+C_5)^2\, C_5^2}\, .
\label{eq4.90}
\end{equation}
and a Dirac term $D_{\rm vac}^{q\to q c\bar{c}}$,
\begin{align}
	\hat{P}_{q \to q c \bar{c}}  &= C_F\, T_R\,  \frac{1}{C_5^2 \left( B_1+C_5\right)^2} \, D_{\rm vac}^{q\to q c\bar{c}}
	\label{eq3.17}
\end{align}
We parametrize the Dirac term in the form 
\begin{align}
	D_{\rm vac}^{q\to q c\bar{c}}
	 = & \frac{1}{2\left( p_1^+\right)^2} \frac{1}{2\left( p_5^+\right)^2}
	\left( d_{\rm vac}^{(11,22)}  \left( {\bm \kappa}_1\cdot {\bm \kappa}_1 \right)\, \left( {\bm \kappa}_2\cdot {\bm \kappa}_2 \right)  
	+  d_{\rm vac}^{(12,21)}  \left( {\bm \kappa}_1\cdot {\bm \kappa}_2 \right)\, \left( {\bm \kappa}_2\cdot {\bm \kappa}_1 \right) \right. 
	\nonumber\\
	& \qquad \qquad \qquad 
	\left. +\,  2\, d_{\rm vac}^{(12,22)}  \left( {\bm \kappa}_1\cdot {\bm \kappa}_2 \right)\, \left( {\bm \kappa}_2\cdot {\bm \kappa}_2 \right)  
	+  d_{\rm vac}^{(22,22)}  \left( {\bm \kappa}_2\cdot {\bm \kappa}_2 \right)\, \left( {\bm \kappa}_2\cdot {\bm \kappa}_2 \right) \right)\, ,
	\label{eq3.17b}
\end{align} 
where the prefactors $d_{\rm vac}^{(11,22)}$, $d_{\rm vac}^{(12,21)}$, $d_{\rm vac}^{(12,22)}$ and $d_{\rm vac}^{(22,22)}$  are somewhat involved rational functions of longitudinal momentum fractions
\begin{align}
	d_{\rm vac}^{(11,22)} &= \frac{ 2 \zeta^2 z^2 - 2 \zeta^2 z + \zeta^2 - 2\zeta + 2  }{(1-\zeta)\zeta^2(1-z)z}\, , \label{eq3.18} \\
	d_{\rm vac}^{(12,21)} &= - \frac{8}{\zeta^2}\, , \label{eq3.19} \\
	d_{\rm vac}^{(12,22)} &= - \frac{2 (2-\zeta)(2z-1)}{\zeta^2 (1-z) z}\, , \label{eq3.20} \\
	d_{\rm vac}^{(22,22)} &= \frac{8 (1-\zeta)}{\zeta^2 (1-z) z}\, . \label{eq3.21} 
\end{align}
To organize the various medium-induced contributions efficiently, it is convenient to decompose the gluon polarization tensor \eqref{eq2.5} into  transverse and longitudinal components,
\begin{align}
d_T^{\mu\nu}(q) &= - g^{\mu\nu} + \frac{q^\mu n^\nu + q^\nu n^\mu}{q\cdot n } - \frac{q^2\, n^\mu n^\nu}{(q\cdot n)^2  } \, , \label{eq3.22}\\
d_L^{\mu\nu}(q) &= \frac{q^2\, n^\mu n^\nu}{(q\cdot n)^2  }\, . \label{eq3.23}
\end{align}
This leads to the following decomposition of the Dirac term 
\begin{align}
	D_{\rm vac}^{q \to q c \bar{c}} &= \frac{1}{(2\zeta p_1^{+})^2}  \frac{1}{(2 p_1^{+})^2} \frac{1}{ 4 n\cdot p_1} d^{\rho\mu}(p_2 + p_3)\,  {\rm Tr}\left[ \gamma_\mu \not{p}_{3} \gamma_\nu \not{p}_{2} \right]
	d^{\nu\lambda}(p_2+p_3)  
	{\rm Tr}\left[ \not{n}\, \not{p}_1 \, \gamma_\rho \not{p}_4 \gamma_\lambda \not{p}_1 \right]
	\nonumber \\
	&= D_{\rm vac}^{(TT)} + D_{\rm vac}^{(TL)} +  D_{\rm vac}^{(LT)} +  D_{\rm vac}^{(LL)}\, .
	\label{eq3.24}
\end{align} 
where the first line gives the spinor structure and normalization of the Feynman diagram from which \eqref{eq3.17b}-\eqref{eq3.21} are obtained, while  the second line decomposes the result according to whether the intermediate gluon is transversely or longitudinally polarized gluon in the amplitude and the complex-conjugate amplitude, 
\begin{align}
	D_{\rm vac}^{(TT)} &=  d_{\rm vac}^{(11,22)}\, \frac{ {\bm \kappa}_1\cdot {\bm \kappa}_1 }{2(p_1^+)^2}\, 
	\frac{{\bm \kappa}_2\cdot {\bm \kappa}_2 }{2(p_5^+)^2}
				+ d_{\rm vac}^{(12,21)}  \frac{ {\bm \kappa}_1\cdot {\bm \kappa}_2}{2p_1^+ p_5^+}\, 
				\frac{{\bm \kappa}_2\cdot {\bm \kappa}_1 }{2p_1^+ p_5^+} \, , \label{eq3.25}\\
	D_{\rm vac}^{(TL)} &= 	 d_{\rm vac}^{(12,22)}  	 
	\frac{ {\bm \kappa}_1\cdot {\bm \kappa}_2 }{2(p_1^+)^2}\, \frac{ {\bm \kappa}_2\cdot {\bm \kappa}_2}{2(p_5^+)^2} \, ,	\label{eq3.26}\\
	D_{\rm vac}^{(LT)} &= D_{\rm vac}^{(TL)} \, , \label{eq3.27}\\
	D_{\rm vac}^{(LL)} &= 	 d_{\rm vac}^{(22,22)}  	 
	\frac{ {\bm \kappa}_2\cdot {\bm \kappa}_2}{2(p_1^+)^2}\, \frac{ {\bm \kappa}_2\cdot {\bm \kappa}_2 }{2(p_5^+)^2} \, . \label{eq3.28}
\end{align}

The decomposition \eqref{eq3.24} will be central for a compact formulation of medium-modifications in the $N=1$ opacity expansion.

%
\begin{figure}[t]
   \centering
   \newcommand{\diagraminclude}[1]{\includegraphics[width=0.30\textwidth]{#1}}
   \newcommand{\vacuumdiagraminclude}[1]{\raisebox{0.35cm}{\includegraphics[width=0.3\textwidth]{#1}}}
   \vacuumdiagraminclude{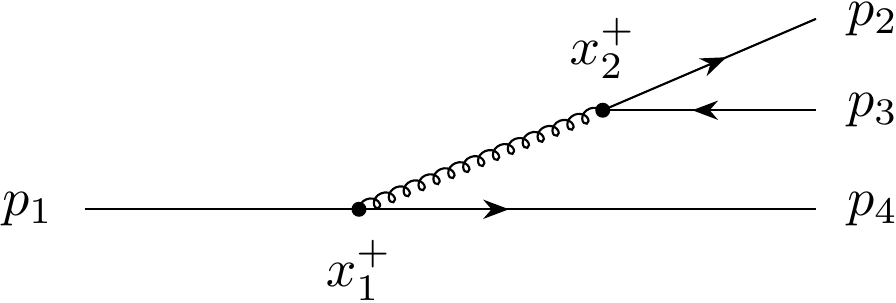}\hfill
   \diagraminclude{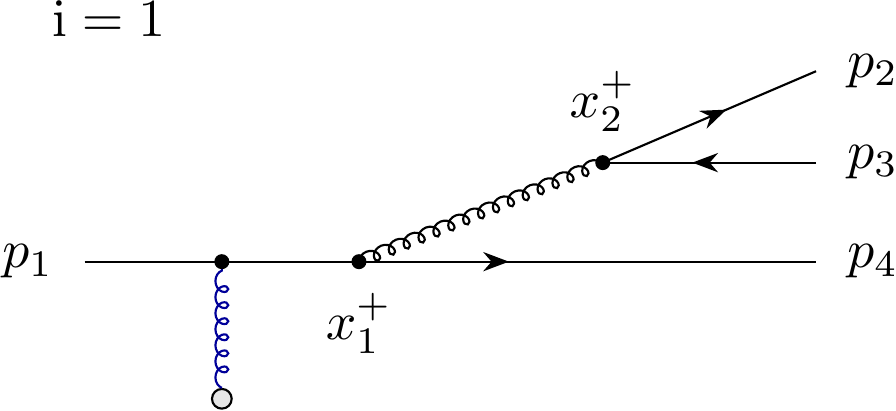}\hfill
   \diagraminclude{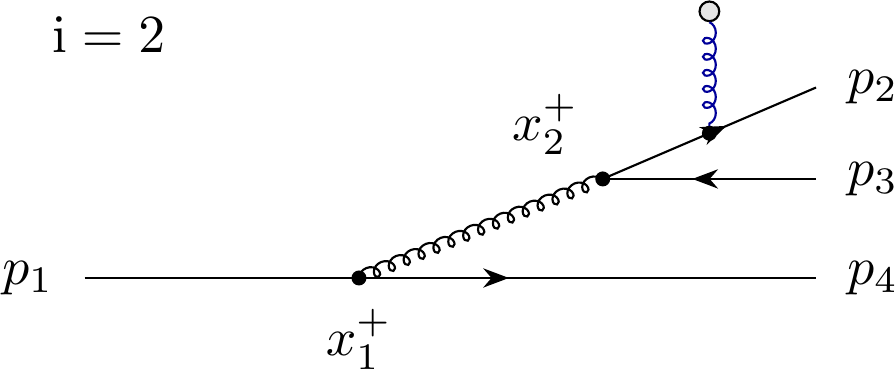}\hfill
   \diagraminclude{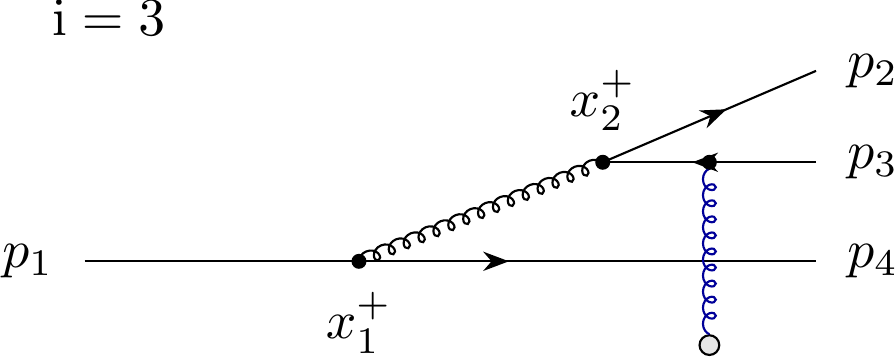}\hfill
   \diagraminclude{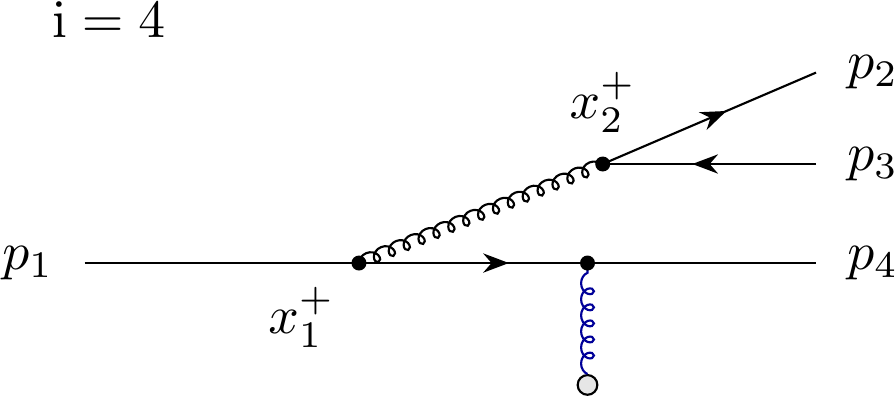}\hfill
   \diagraminclude{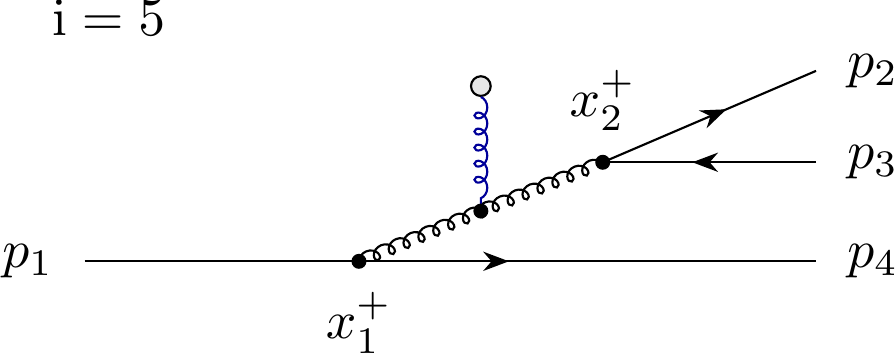}\\
   \caption{The vacuum diagram and the five real contributions to the medium-modification of the $q \to q c \bar{c}$ splitting function at order $N=1$ in opacity.}
   \label{fig4}
\end{figure}
%

%
\begin{figure}
   \centering
   \newcommand{\diagraminclude}[1]{\includegraphics[width=0.23\textwidth]{#1}}
   \diagraminclude{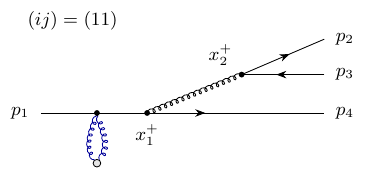}\hfill
   \diagraminclude{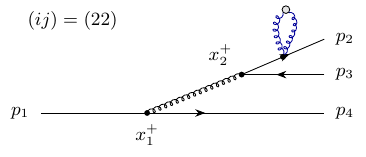}\hfill
   \diagraminclude{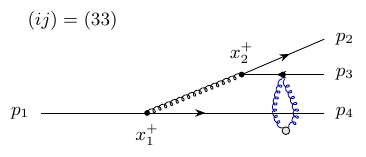}\hfill
   \diagraminclude{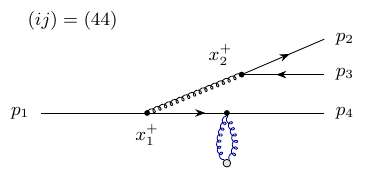}\\[0.2em]
   \diagraminclude{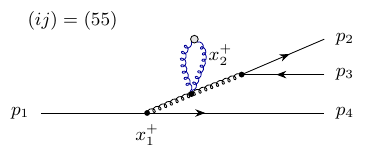}\hfill
   \diagraminclude{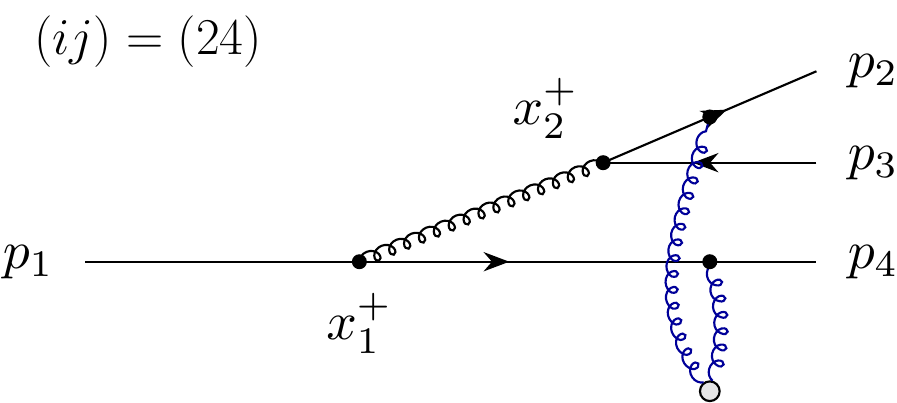}\hfill
   \diagraminclude{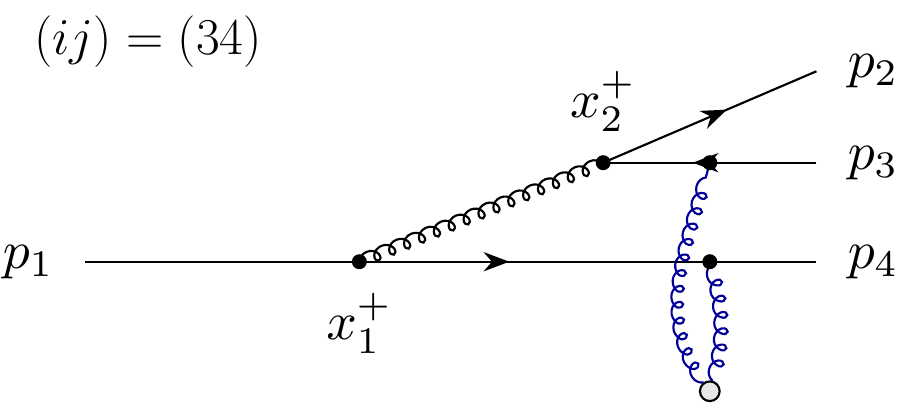}\hfill
   \diagraminclude{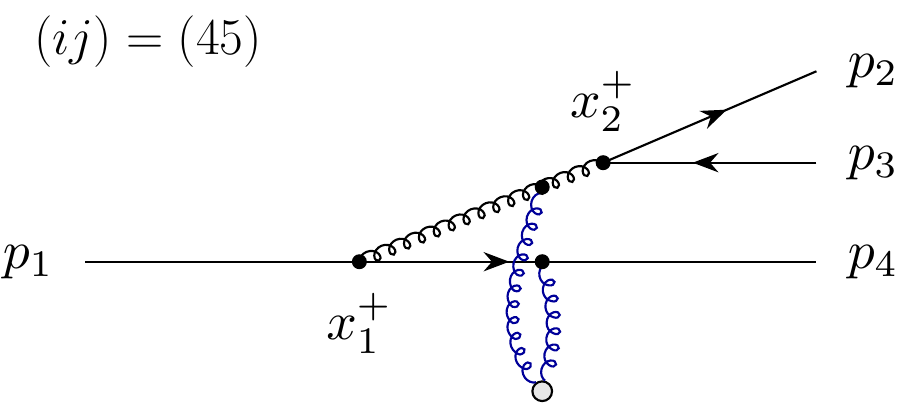}
   \caption{The five diagonal virtual diagrams $(ij)=(11)$, $(ij)=(22)$, $(ij)=(33)$, $(ij)=(44)$ and $(ij)=(55)$ and the three off-diagonal virtual diagrams $(ij)=(24)$, $(ij)=(34)$ and $(ij)=(45)$ that contribute to the medium-modification of the $q \to q c \bar{c}$ splitting function at order $N=1$ in opacity.}
   \label{fig5}
\end{figure}
%

\section{The $q \to q c \bar{c}$ splitting function to first order in opacity. }
\label{sec4}

At the amplitude level, a gluon can be exchanged with the $q \to q c \bar{c}$ splitting in five different ways: it can be attached to the incoming quark (1), the outgoing charm quark (2), the anti-charm quark (3) and outgoing quark of the same flavor as the parent quark (4), or to the intermediate gluon (5). These five contributions are depicted in Fig.~\ref{fig4}. 
To evaluate the medium modification of $q \to q c \bar{c}$ to first order in opacity, the sum of the five diagrams is contracted with its complex conjugate, yielding a sum over 15 symmetrized pairs $(ij)$ with $i \leq j$ and $i, j = 1, ..., 5$. Since, in these contributions, transverse momentum is transferred from the medium to the final state, we refer to them as  {\it real}. 

In addition, there are probability-conserving contributions. These are of same order in an expansion in opacity and arise from expanding the medium-modifications to second order in $A^-$ at the amplitude level and contracting with the vacuum splitting function. These contributions to the amplitude are depicted in Fig.~\ref{fig5}. By construction, in such contributions no momentum ${\bf q}$ is transferred to the final state: the two gluons exchanged at the amplitude level carry opposite transverse momentum $\pm {\bf q}$ and couple to the same scatterer in the medium. We refer to such contributions as 
 {\it virtual} and  denote them by barred quantities. The final result takes then the form \eqref{eq2.20}, \footnote{In eqs.~\eqref{eq2.20} and \eqref{eq2.71}, the color factors, Dirac terms and longitudinal phase integrals associated with the the medium-modification of $1 \to 2$ splitting functions are labeled by superscripts $g\to c\bar{c}$ and $q\to q g$, respectively. For notational simplicity, we do not introduce analogous superscripts in the definitions for the $q \to q c \bar{c}$ splitting function given in \eqref{eq4.1}. Throughout this work, quantities without superscript are understood to refer to this triple-collinear splitting function.}
\begin{equation}
\hat{P}_{q\to q c \bar{c}}^{(N=1)} = \int\frac{d{\bf q}}{(2\pi)^2} \vert a({\bf q}) \vert^2  
  \sum_{\substack{ (ij) \\ i\leq j} } \left(
 \underbrace{  C_{(ij)} D_{(ij)} {\cal P}_{(ij)} }_{\it real}  + 
 \underbrace{  \bar{C}_{(ij)} \bar{D}_{(ij)} \bar{\cal P}_{(ij)} }_{\it virtual}   \right)\, .
 \label{eq4.1}
\end{equation}
We now proceed to the calculation of the quantities entering \eqref{eq4.1}. 

\subsection{Color factors}
\label{sec4.1}
We determine the color factors associated with the different contributions $(ij)$ of $\hat{P}^{(N=1)} _{q \to q c \bar{c}}$ by averaging over the color of the incoming quark. The fifteen independent real contributions read
\begin{align}
	C_{(11)} &= C_{(22)} = C_{(33)} = C_{(44)}  \approx \frac{N_c^2}{8}\, , \label{eq4.2} \\
	C_{(55)} &\approx \frac{N_c^2}{4} \, ,\label{eq4.3} \\ 
 	C_{(12)} &=  C_{(13)}=C_{(15)}=C_{(25)}=C_{(35)} \approx \frac{N_c^2}{8} \, ,\label{eq4.4} \\ 
	C_{(24)} &=  C_{(34)}=C_{(45)} \approx -\frac{N_c^2}{8}\, , \label{eq4.5} \\
	C_{(14)} &=  C_{(23)} \approx 0\, . \label{eq4.6}
\end{align}
Here, the approximate signs indicate that the results are given to leading order in $N_c$. At this order, thirteen real terms must be retained in \eqref{eq4.1}, while the contributions $(14)$ and $(23)$ can be neglected. For the diagonal terms \eqref{eq4.2} and \eqref{eq4.3}, the values of the color factors can be readily understood: the vacuum contribution $q\to q c \bar{c}$ has a color factor $T_R C_F \approx \tfrac{N_c}{4}$, and interaction with a single scatterer yields one Casimir in the representation of the parton with which the medium interacts, namely $C_F \approx  \tfrac{N_c}{2}$ for \eqref{eq4.2}  and $C_A =  N_c$ for \eqref{eq4.3}.

At first order of opacity, both gluons are exchanged with the splitting at the same light-cone time. Therefore, probability-conserving contributions $(1j)$, with $j = 2, ..., 5$ and $(j5)$ with $j=2,3$ cannot arise since the corresponding partons do not propagate at the same light-cone time. The only non-vanishing contributions are
\begin{align}
	\bar{C}_{11} &= \bar{C}_{22} = \bar{C}_{33} = \bar{C}_{44}  \approx -\frac{N_c^2}{8}\, , \label{eq4.7} \\
	\bar{C}_{55} &\approx -\frac{N_c^2}{4} \, ,\label{eq4.8} \\ 
	\bar{C}_{24} &=  \bar{C}_{34}=\bar{C}_{45} \approx \frac{N_c^2}{8}\, , \label{eq4.9} \\
	\bar{C}_{23} & \approx 0\, . \label{eq4.10}
\end{align}
A technical comment is in order: All contributions to \eqref{eq4.1} are second order in the medium vertex $i \gamma^+ A^-$. If this vertex appears twice in the amplitude, its insertion yields a factor $i^2 = -1$, which differs by an overall sign from the case in which the vertex is inserted once in the amplitude and once in the complex conjugate amplitude. We account for this sign in the definition of the color factors, although it does not arise from the color algebra. 
For this reason, the expressions \eqref{eq4.7}--\eqref{eq4.10} differ by an overall sign from the corresponding expressions \eqref{eq4.2}--\eqref{eq4.6}.

\subsection{Longitudinal phase integrals}
\label{sec4.2}
In the following two subsections, we provide separate discussions of the longitudinal phase integrals ${\cal P}_{(ij)}$ that enter the real contributions in \eqref{eq4.1} and those that enter the virtual ones.

\subsubsection{Longitudinal phase integrals for real terms}
\label{sec4.2.1}
The longitudinal phase integrals ${\cal P}_{(ij)}$ for real contributions are of the form given in \eqref{eq2.47}, 
\begin{equation}
{\cal P}_{(ij)}  \equiv   \int_0^\infty d\check{y}^+ \, n(\check{y}^+)  
\frac{P_i(\check{y}^+) \,  P_j^*(\check{y}^+) +  P_j(\check{y}^+) \,  P_i^*(\check{y}^+) }{1+\delta_{ij}} \, , \qquad i, j = 1,...,5,  
\label{eq4.11}
\end{equation}
where $P_i(\check{y}^+) $ denotes the phase associated with the $i$-th diagram contributing to the amplitude in Fig.~\ref{fig4}. 
These phases read as follows:
\begin{align}
	P_1(\check{y}^+) &= 
	 \int dx^+_1\, dx^+_2  \theta\left(x_1^+ - \check{y}^+\right)\, 
	 \theta\left(x_2^+-x_1^+\right)\, \exp \big [ i \check{y}^+ 
	 \underbrace{\left( {\Gamma}_{234} -  \bar{\Gamma}_{234}\right)}_{\equiv A_1} \big ]
		\nonumber  \\
	& \qquad \times	
	\exp\big[  i x_1^+  \underbrace{\left( {\Gamma}_{4} +  {\Gamma}_{23} - {\Gamma}_{234}\right) }_{\equiv B_1}
	 + i x_2^+ \underbrace{\left( {\Gamma}_{2} +  {\Gamma}_{3} - {\Gamma}_{23} \right)}_{\equiv C_1} \big]\, ,
	 \label{eq4.12} \\
P_2(\check{y}^+) &=  
	 \int dx^+_1\, dx^+_2  \theta\left(\check{y}^+ - x_2^+\right)\, 
	 \theta\left(x_2^+-x_1^+\right)\, \exp\big[ i \check{y}^+ \underbrace{\left( \Gamma_2 -\bar{\Gamma}_2 \right)}_{\equiv A_2} \big]
	 	\nonumber \\
	& \qquad \times	
	\exp\big[ i x_1^+ \underbrace{\left( \Gamma_4 + \bar{\Gamma}_{23} -  \bar{\Gamma}_{234}\right) }_{\equiv B_2} 
	+i x_2^+ \underbrace{\left( \bar{\Gamma}_2 + \Gamma_3 - \bar{\Gamma}_{23} \right)}_{\equiv C_2} \big]\, ,
	 \label{eq4.13} \\
P_3(\check{y}^+) &=  
	 \int dx^+_1\, dx^+_2  \theta\left(\check{y}^+ - x_2^+\right)\, \theta\left(x_2^+-x_1^+\right)  
	 \exp\big[ i \check{y}^+ \underbrace{\left( \Gamma_3 -  \bar{\Gamma}_3 \right)}_{\equiv A_3} \big]
	 	\nonumber \\
	& \qquad \times	
	\exp\big[   i x_1^+ \underbrace{\left( \Gamma_4 + \bar{\Gamma}_{23} - \bar{\Gamma}_{234}\right)}_{\equiv B_3}  
	+ i x_2^+ \underbrace{\left( \Gamma_2 +  \bar{\Gamma}_3 - \bar{\Gamma}_{23} \right)}_{\equiv C_3} \big]\, ,
		\label{eq4.14} \\
P_4(\check{y}^+) &= 
	 \int dx^+_1\, dx^+_2  \theta\left(\check{y}^+ - x_1^+\right)\,  \theta\left(x_2^+-x_1^+\right) 
	 \exp\big[ i \check{y}^+ \underbrace{\left( \Gamma_4 - \bar{\Gamma}_4 \right)}_{\equiv A_4} \big]	
	 \nonumber \\
	& \qquad \times	
	\exp\big[  i x_1^+ \underbrace{\left( \bar{\Gamma}_4 +  \Gamma_{23} - \bar{\Gamma}_{234}\right) }_{\equiv B_4}
	 + i x_2^+ \underbrace{ \left( \Gamma_2 +  \Gamma_3 - \Gamma_{23} \right)}_{\equiv C_4} \big]\, ,
	 \label{eq4.15}  \\
P_5(\check{y}^+)  &= 
	 \int dx^+_1\, dx^+_2  \theta\left(\check{y}^+ - x_1^+\right)\,  \theta\left(x_2^+-\check{y}^+\right) 
	 \exp\big[ i \check{y}^+ \underbrace{\left( \Gamma_{23} - \bar{\Gamma}_{23} \right)}_{\equiv A_5} \big]
	 \nonumber	 \\
	& \qquad \times	
	\exp\big[  i x_1^+ \underbrace{\left( \Gamma_4 + \bar{\Gamma}_{23} -  \bar{\Gamma}_{234} \right)}_{\equiv B_5}  
	+ i x_2^+ \underbrace{\left( \Gamma_2 +  \Gamma_3 - \Gamma_{23} \right)}_{\equiv C_5} \big]\, .
	 \label{eq4.16} 
\end{align}
Here, the integrals are understood as being regularized, see \eqref{eq3.8}. The shorthands $\Gamma_{i ...k}$ denote transverse energies associated with the sum of momenta indicated by the indices, and the barred quantities $\bar{\Gamma}_{i ...k}$ correspond to the case in which one of these momenta is shifted by ${\bf q}$, for instance, 
\begin{equation}
	\Gamma_{2} \equiv \frac{ \left( {\bf p}_2 \right)^2 }{2 p^+_2}\, , \qquad
	\Gamma_{234} \equiv \frac{ \left( {\bf p}_2 + {\bf p}_3 + {\bf p}_4 \right)^2 }{2(p^+_2+p^+_3+p^+_4)}\, , \qquad
	\bar{\Gamma}_{234} \equiv \frac{ \left( {\bf p}_2 + {\bf p}_3 + {\bf p}_4 + {\bf q} \right)^2 }{2(p^+_2+p^+_3+p^+_4)}\, , \quad \hbox{etc.}
	\label{eq4.17}
\end{equation}
The phases \eqref{eq4.12}--\eqref{eq4.16} can then be inferred from Fig.~\ref{fig4} by simple diagrammatic rules: in each diagram, we associate the first branching $q \to q g$ with light-cone time $x_1^+$ and the second branching $g \to c \bar{c}$ with $x_2^+$.  The interaction with the medium-induced gluon is located at a light-cone time $\check{y}^+$ that is positioned with respect to  $x_1^+$ and $x_2^+$ as indicated by the $\theta$-functions. Each phase factor is then given by the sum of transverse energies leaving the branching vertex minus the transverse energy entering it. For instance, for diagram (5) of Fig.~\ref{fig4}, the momentum ${\bf q}$ is transferred to the intermediate gluon. In our convention, this amounts to the gluon and quark leaving the $q \to q g$  vertex at position $x_1^+$ carrying transverse momenta ${\bf p}_2 + {\bf p}_3 + {\bf q}$ and ${\bf p}_4$, respectively, while the incoming quark carries transverse momentum ${\bf p}_2 + {\bf p}_3 + {\bf p}_4 + {\bf q}$ into that vertex. The corresponding transverse energies entering and leaving this vertex are therefore $\bar{\Gamma}_{234}$, $ \bar{\Gamma}_{23}$ and $ \Gamma_4$, respectively, which explains the phase $i x_1^+ B_5$ in \eqref{eq4.16}.  

In equations \eqref{eq4.6}--\eqref{eq4.16}, the phases have been written in terms of ten different transverse energies, namely $\Gamma_2$, $\Gamma_3$, $\Gamma_4$, $\Gamma_{23}$, $\Gamma_{234}$ and the corresponding  barred quantities $\bar{\Gamma}_2$, $\bar{\Gamma}_3$, $\bar{\Gamma}_4$, $\bar{\Gamma}_{23}$, $\bar{\Gamma}_{234}$. Alternatively, we have expressed these ten transverse energies in terms of fifteen combinations $A_j$, $B_j$, $C_j$ with $j = 1,..., 5$ that are defined in eqs.~\eqref{eq4.12}-\eqref{eq4.16}.  Amongst these fifteen combinations, one can identify eight linear relations. We choose the following seven
\begin{equation}
	\hbox{linearly independent combinations:  $B_1$, $B_4$, $B_5$, $C_2$, $C_3$, $C_5$ and $A_1$.}
	\label{eq4.18} 
\end{equation}
This choice is informed by the fact that the $B_j$ and $C_j$ take particularly simple forms when written in terms of transverse momenta, while only six of them are linearly independent.  In terms of the seven independent combinations in \eqref{eq4.18}, the remaining eight linearly dependent ones can be written as 
\begin{align}
	B_2 &= B_5\, , \label{eq4.19}\\
	B_3 &= B_5\, , \label{eq4.20}\\
	C_1 &= C_5\, , \label{eq4.21}\\
	C_4 &= C_5\, , \label{eq4.22}\\
	A_4 &= A_1 + B_1 - B_4\, , \label{eq4.23}\\
	A_5 &= A_1 + B_1 - B_5\, , \label{eq4.24}\\
	A_2 &= A_5 + C_5 - C_2 = A_1 + B_1 - B_5 + C_5 - C_2\, ,\label{eq4.25}\\
	A_3 &= A_5 + C_5 - C_3 = A_1 + B_1 - B_5 + C_5 - C_3 \, .\label{eq4.26}
\end{align}
A further simplification arises from the observation that all $A_j$ can  be written as a term  $A_1$ plus sums of $B_j$'s and $C_j$'s. In the phase integrals \eqref{eq4.11},  all integrands are of the form $P_{i}(\check{z})\, P_{j}^*(\check{z}) \propto e^{i\check{z}(A_{i}-A_{j})}$;  the $A_1$-dependence cancels in these integrands because of the relations \eqref{eq4.23}--\eqref{eq4.26}. We therefore conclude that all phases ${\cal P}_{(ij)} $ depend only on $B_1$, $B_4$, $B_5$, $C_2$, $C_3$, $C_5$, which can be expressed as
\begin{align}
	B_1 &= \frac{\bm{\upkappa}_1^2}{2p^+\zeta (1-\zeta)}\, ,
	\label{eq4.27}  \\
	B_4 &= \frac{\left( \bm{\upkappa}_1 +  {\bf q} \zeta \right)^2}{2p^+\zeta (1-\zeta)}\, ,
	\label{eq4.28} \\
	B_5 &=  \frac{\left( \bm{\upkappa}_1 -  {\bf q}(1-\zeta) \right)^2}{2p^+\zeta (1-\zeta)}\, ,
	\label{eq4.29}  \\
	C_2 &= \frac{\left( \bm{\upkappa}_2 + {\bf q}(1-z) \right)^2}{2p^+\zeta z (1-z)}\, ,
	\label{eq4.30} \\
	C_3 &= \frac{\left( \bm{\upkappa}_2 - {\bf q} z \right)^2}{2p^+\zeta z (1-z)}\, ,
	\label{eq4.31} \\
	C_5 &= \frac{\bm{\upkappa}_2^2}{2p^+\zeta z (1-z)}\, .
	\label{eq4.32} 
\end{align}
One sees that $B_1$ ($C_5$) are the inverse formation times for an unperturbed first (second) splitting, as discussed in the text following \eqref{eq3.6}, \eqref{eq3.7}. The other expressions show the expected ${\bf q}$-dependence, which  can be inferred from shifting one of the outgoing transverse momenta ${\bf p}_j$ in Eqs.~\eqref{eq3.11} and \eqref{eq3.13}. 

To write the longitudinal phase integrals \eqref{eq4.11} in explicit analytic forms, we utilize again the interference factor ${\cal S}(C)$ introduced in \eqref{eq2.48}. For the homogeneous distribution \eqref{eq2.19} of scattering centers,  the longitudinal phase integrals ${\cal P}_{i,j}$ then take the following explicit form for the diagonal terms
\begin{align}	
	{\cal P}_{(11)} =& \frac{n_0 L}{C_5^2 \left( B_1+C_5\right)^2}\, ,
	\label{eq4.33} \\
	{\cal P}_{(22)} =& \frac{2\, n_0 L}{B_5^2 C_2 \left( B_5+C_2\right)}  {\cal S}(B_5)
	+ \frac{2\, n_0 L}{B_5 C_2^2 \left( B_5+C_2\right)} {\cal S}(C_2) \nonumber  \\
	& - \frac{2\, n_0 L}{B_5 C_2 \left( B_5+C_2\right)^2} {\cal S}(B_5+C_2)\, ,
	\label{eq4.34} \\
	{\cal P}_{(33)} =& \frac{2\, n_0 L}{B_5^2 C_3 \left( B_5+C_3\right)}  {\cal S}(B_5)
	+ \frac{2\, n_0 L}{B_5 C_3^2 \left( B_5+C_3\right)} {\cal S}(C_3) \nonumber\\
	& - \frac{2\, n_0 L}{B_5 C_3 \left( B_5+C_3\right)^2} {\cal S}(B_5+C_3)\, ,
	 \label{eq4.35}\\
	{\cal P}_{(44)} =& \frac{2\, n_0 L}{C_5^2 \left( B_4+C_5\right)^2}  {\cal S}(B_4+C_5)\, ,
	 \label{eq4.36}\\
	{\cal P}_{(55)} =& \frac{2\, n_0 L}{C_5^2 B_5^2}  {\cal S}(B_5)\, .
	\label{eq4.37}
\end{align}
The off-diagonal terms take the form
\begin{align}
	{\cal P}_{(12)}  = &  \frac{2\, n_0 L}{B_5 C_5 \left(B_1+C_5\right) C_2 }  {\cal S}(B_5)
	- \frac{2\, n_0 L}{C_5 \left(B_1+C_5\right) C_2  \left(B_5+ C_2\right)}  {\cal S}(B_5+C_2)\, ,
	 \label{eq4.38}\\
	{\cal P}_{(13)}  = & {\cal P}_{(12)}  \vert_{C_2 \to C_3} \, ,
	 \label{eq4.39}\\
	{\cal P}_{(15)} = & - \frac{2\, n_0 L}{B_5 C_5^2 \left(B_1+C_5\right) }  {\cal S}(B_5)\, ,
	\label{eq4.40} \\
	{\cal P}_{(24)} = & - \frac{2\, n_0 L}{B_5 C_5 \left(B_4+C_5\right) C_2}  {\cal S}(B_5)
	+ \frac{2\, n_0 L}{B_5 C_5 \left(B_4+C_5\right) C_2 }  {\cal S}(B_5-B_4-C_5)
	\nonumber \\
	& - \frac{2\, n_0 L}{C_5 \left(B_4+C_5\right) C_2 \left(B_5+C_2\right) }  {\cal S}(B_5-B_4-C_5+C_2)
	\nonumber \\
	& - \frac{2\, n_0 L}{B_5 C_5 \left(B_4+C_5\right) \left(B_5+C_2\right) }  {\cal S}(B_4+C_5)
	\nonumber \\
	&
	 + \frac{2\, n_0 L}{C_5 \left(B_4+C_5\right) C_2 \left(B_5+C_2\right) }  {\cal S}(B_5+C_2)\, ,
	 \label{eq4.41} \\
	{\cal P}_{(25)} =&  \frac{2\, n_0 L}{B_5 C_5 C_2 \left(B_5+C_2\right) }  {\cal S}(B_5+C_2)
	- \frac{2\, n_0 L}{B_5 C_5 C_2 \left(B_5+C_2\right)}  {\cal S}(C_2)
	\nonumber \\
	& - \frac{2\, n_0 L \left(B_5 + 2 C_2 \right) }{B_5^2 C_5 C_2 \left(B_5+C_2\right)  }  {\cal S}(B_5)\, ,
	 \label{eq4.42}\\ 
	 {\cal P}_{(34)} =& {\cal P}_{(24)}  \vert_{C_2 \to C_3} 
	  \label{eq4.43} \\
	{\cal P}_{(45)}  =&  \frac{2\, n_0 L}{B_5 C_5^2 \left(B_4+C_5\right) }  {\cal S}(B_4+C_5)
	+ \frac{2\, n_0 L}{B_5 C_5^2 \left(B_4+C_5\right) }  {\cal S}(B_5)
	\nonumber \\
	&
	- \frac{2\, n_0 L}{B_5 C_5^2 \left(B_4+C_5\right)}  {\cal S}(B_5-B_4-C_5)\, .
	\label{eq4.44}
\end{align}
Since we work to leading order in $N_c$, we do not write the off-diagonal terms ${\cal P}_{(14)}$ and ${\cal P}_{(23)}$, as they are multiplied by a color factor 
\eqref{eq4.6} that vanishes to leading order in $N_c$. 

\subsubsection{Longitudinal phase integrals for virtual terms}
\label{sec4.2.2}
The longitudinal phase integrals $\bar{\cal P}_{(ij)}$ for imaginary contributions are of the form \eqref{eq2.61},
\begin{equation}
\bar{\cal P}_{(ij)} \equiv   \int_0^\infty d\check{y}^+ \, n(\check{y}^+)  
\frac{ \bar{P}_{(ij)}(\check{y}^+) \,  P_{\rm vac}^* + P_{\rm vac}\,  \bar{P}^*_{(ij)}(\check{y}^+) }{1+\delta_{ij}} \, , \qquad i, j = 1,...,5\, .
\label{eq4.45}
\end{equation}
Here, the phase $P_{\rm vac}(\check{y}^+) $ associated with the vacuum amplitude of $q \to q c \bar{c}$ takes the form
\begin{equation}
	P_{\rm vac} = 
	 \int dx^+_1\, dx^+_2   \theta\left(x_2^+-x_1^+\right)\, 
	\exp\big[  i x_1^+  \underbrace{\left( {\Gamma}_{4} +  {\Gamma}_{23} -{\Gamma}_{234} \right) }_{\equiv B_1}
	 +i x_2^+ \underbrace{\left(  {\Gamma}_{2} +  {\Gamma}_{3} -  {\Gamma}_{23} \right)}_{\equiv C_5} \big]\, ,
	 \label{eq4.46} \\
\end{equation}
and $\bar{P}_{(ij)}(\check{y}^+) $ are phases associated with diagrams in which two gluons with equal and opposite momentum ${\bf q}$ are exchanged at light-cone time $\check{y}^+$ at the amplitude level. For diagonal virtual contributions, $\bar{P}_{(jj)}(\check{y}^+) $, the momentum flow remains unchanged compared to the vacuum amplitude and therefore the integrand of $\bar{P}_{(jj)}(\check{y}^+)$ equals the integrand of $P_{\rm vac}$ while the integrations over light-cone time are restricted by the position $\check{y}^+$ of the gluon exchange.
In addition, there are four off-diagonal virtual contributions
\begin{align}
\bar{P}_{(45)}(\check{y}^+)  &= 
	 \int dx^+_1\, dx^+_2  \theta\left(\check{y}^+ - x_1^+\right)\,  \theta\left(x_2^+-\check{y}^+\right) 
	 \exp\big[  i \check{y}^+ \underbrace{\left(  \Gamma_4 - \tilde{\Gamma}_4 + \Gamma_{23} - \bar{\Gamma}_{23} \right)}_{\equiv B_1 - \tilde{B}}  
	 \big]
	 \nonumber	 \\
	& \qquad \times	
	\exp\big[  i x_1^+ \underbrace{\left( \tilde{\Gamma}_4 + \bar{\Gamma}_{23} -\Gamma_{234} \right)}_{\equiv \tilde{B}}  
	+i x_2^+ \underbrace{\left( \Gamma_2 + \Gamma_3 - \Gamma_{23}\right)}_{\equiv C_5} \big] \, . 
	 \label{eq4.47} \\
\bar{P}_{(42)}(\check{y}^+)  &= 
	 \int dx^+_1\, dx^+_2  \theta\left(\check{y}^+ - x_1^+\right)\,  \theta\left(\check{y}^+ - x_2^+\right) 
	 \exp\big[  i \check{y}^+ \underbrace{\left( \Gamma_4 -  \tilde{\Gamma}_4 +  \Gamma_{2} - \bar{\Gamma}_{2}  \right)}_{\equiv \left(B_1 - \tilde{B} \right) + \left(C_5-C_2\right)}
	 \big]
	 \nonumber	 \\
	& \qquad \times	
	\exp\big[  i x_1^+ \underbrace{\left( \tilde{\Gamma}_4 + \bar{\Gamma}_{23} - \Gamma_{234}\right)}_{\equiv \tilde{B}}  
	+i x_2^+ \underbrace{\left( \bar{\Gamma}_2 + \Gamma_3 - \bar{\Gamma}_{23} \right)}_{\equiv C_2} \big] \, . 
	 \label{eq4.48} \\
\bar{P}_{(43)}(\check{y}^+)  &= 
	 \int dx^+_1\, dx^+_2  \theta\left(\check{y}^+ - x_1^+\right)\,  \theta\left(\check{y}^+ - x_2^+\right) 
	 \exp\big[ i \check{y}^+ \underbrace{\left( \Gamma_4 -  \tilde{\Gamma}_4 + \Gamma_{3} - \bar{\Gamma}_{3} \right)}_{\equiv \left(B_1 - \tilde{B} \right) + \left(C_5-C_3\right)}
	 \big]
	 \nonumber	 \\
	& \qquad \times	
	\exp\big[ i x_1^+ \underbrace{\left( \tilde{\Gamma}_4 + \bar{\Gamma}_{23} - \Gamma_{234}\right)}_{\equiv \tilde{B}}  
	+i x_2^+ \underbrace{\left( \Gamma_2 +  \bar{\Gamma}_3 -  \bar{\Gamma}_{23}\right)}_{\equiv C_3} \big] \, . 
	 \label{eq4.49} \\
\bar{P}_{(23)}(\check{y}^+)  &= 
	 \int dx^+_1\, dx^+_2  \theta\left(\check{y}^+ - x_1^+\right)\,  \theta\left(\check{y}^+ - x_2^+\right) 
	 \exp\big[  i \check{y}^+ \underbrace{\left( \Gamma_2 -  \bar{\Gamma}_2 + \Gamma_{3} - \bar{\Gamma}_{3}  \right)}_{\equiv \left(C_5 - \tilde{C} \right) }
	 \big]
	 \nonumber	 \\
	& \qquad \times	
	\exp\big[  i x_1^+ \underbrace{\left( \Gamma_4 + \Gamma_{23} - \Gamma_{234}\right)}_{\equiv B_1}  
	+i x_2^+ \underbrace{\left( \bar{\Gamma}_2 + \tilde{\Gamma}_3 - \Gamma_{23} \right)}_{\equiv \tilde{C}} \big]  \, . 
	 \label{eq4.50}
\end{align}
Here, we have distinguished between the transverse energies $\bar{\Gamma}_{i ... k}$ defined in \eqref{eq4.17}, and transverse energies $\tilde{\Gamma}_{i ... k}$ associated with  parton lines along which the momentum flows in the opposite direction:
\begin{equation}
	\tilde{\Gamma}_4 = \frac{\left( {\bf p}_4 - {\bf q} \right)^2}{2 p_4^+}\, ,\qquad \tilde{\Gamma}_3 = \frac{\left( {\bf p}_3 - {\bf q} \right)^2}{2 p_3^+}\, , \quad \dots \label{eq4.51}
\end{equation}
We encounter two additional combinations of transverse energies, 
\begin{align}
	\tilde{B} &=  \tilde{\Gamma}_4 + \bar{\Gamma}_{23} - \Gamma_{234}=  \frac{\left( \bm{\upkappa}_1 -  {\bf q}\right)^2}{2p^+\zeta (1-\zeta)}\, , 
	\label{eq4.52}\\
	\tilde{C} &= \bar{\Gamma}_{2}  + \tilde{\Gamma}_3- \Gamma_{23} =  \frac{\left( \bm{\upkappa}_2 +  {\bf q}\right)^2}{2p^+\zeta z (1-z)}\, .
	\label{eq4.53}
\end{align}
Since $\tilde{C}$ appears only in the term $(23)$ and since that term contributes only at subleading order in $N_c$, the final result  depends only on $\tilde{B}$.

The relevant phase space integrals in \eqref{eq4.1} can all be done analytically, yielding  
\begin{align}
\bar{\cal P}_{(11)} &=     \frac{n_0\, L}{C_5^2 \left(B_1+C_5\right)^2} -  
\frac{n_0\, L}{C_5^2 \left(B_1+C_5\right)^2}{\cal S}( B_1+C_5) \, ,
	 \label{eq4.54} \\
 \bar{\cal P}_{(22)} &=   \bar{\cal P}_{(33)}  =
 \frac{n_0\, L}{B_1\, C_5^2\, \left(B_1+C_5\right)} {\cal S}(C_5 ) - \frac{n_0\, L}{ B_1\, C_5\, \left(B_1+C_5\right)^2} {\cal S}( B_1+C_5)\, ,
	 \label{eq4.55} \\
\bar{\cal P}_{(44)}  &=   
 \frac{n_0\, L}{ C_5^2\, \left(B_1+C_5\right)^2} {\cal S}( B_1+C_5)\, ,
	 \label{eq4.56} \\
\bar{\cal P}_{(55)} &=   - \frac{n_0\, L}{ B_1 C_5^2\, \left(B_1+C_5\right)} {\cal S}( C_5 )
 + \frac{n_0\, L}{ B_1 C_5^2\, \left(B_1+C_5\right)} {\cal S}( B_1 +C_5 )\, ,
	 \label{eq4.57} \\
 \bar{\cal P}_{(45)}  &=  \frac{2 n_0\, L}{ \tilde{B} C_5^2\, \left( B_1+C_5\right)} {\cal S}( B_1 +C_5 ) -  \frac{2\, n_0\, L}{ \tilde{B} C_5^2\, \left(B_1+C_5\right)} {\cal S}( B_1-\tilde{B} + C_5 )\, ,
	 \label{eq4.58} \\
\bar{\cal P}_{(42)} &=  \frac{2\, n_0\, L}{ \tilde{B} C_5\, C_2\, \left( B_1+C_5\right)} {\cal S}( \tilde{B} - B_1 - C_5 ) 
- \frac{2\, n_0\, L}{ \tilde{B} C_5\, ( \tilde{B} + C_2 )\, \left( B_1+C_5\right)}  {\cal S}( B_1 + C_5 )\nonumber \\
& - \frac{2\, n_0\, L}{C_5\, C_2 ( \tilde{B} + C_2 )\, \left( B_1+C_5\right)}  {\cal S}( -B_1 + \tilde{B} - C_5 +C_2)\, ,
	 \label{eq4.59}\\ 
\bar{\cal P}_{(43)}  &= \bar{\cal P}_{(42)}  \Big\vert_{C_2 \to C_3}\, ,
	 \label{eq4.60}\\
	\bar{\cal P}_{(23)} &= \frac{2\, n_0\, L}{ B_1 C_5\, \tilde{C} \left( B_1+C_5\right)} {\cal S}( C_5 ) -  \frac{2\, n_0\, L}{ B_1 C_5 \left( B_1+C_5\right) (B_1+\tilde{C}) } {\cal S}( B_1+C_5) \nonumber \\
& - \frac{2\, n_0\, L}{C_5\, \tilde{C} ( B_1 + C_5 )\, ( B_1+\tilde{C})}  {\cal S}( C_5 - \tilde{C})\, .
	 \label{eq4.61}
\end{align}

\subsection{Dirac spinor structure}
\label{sec4.3}
We have associated the transverse boost-invariant momenta $\bm{\upkappa}_1$ in \eqref{eq3.11} and $\bm{\upkappa}_2$ in \eqref{eq3.13} to the first 
and second splitting of the vacuum amplitude $q\to q g \to q c \bar{c}$. For the corresponding five medium-modified splitting amplitudes in Fig.~\ref{fig4}, we introduce now 
shifted momenta $\bm{\upkappa}_1^{(i)}$, $\bm{\upkappa}_2^{(i)}$, $i = 1, ..., 5$. For instance, if transverse momentum ${\bf q}$ flows from the charm propagator to the medium (contribution 2), then $\bm{\upkappa}_n^{(2)} \equiv \bm{\upkappa}_n \vert_{{\bf p}_2 \to {\bf p}_2 + {\bf q}}$. The corresponding  
relative momenta associated to the medium-modified splitting amplitudes read
\begin{align}
	\bm{\upkappa}_1^{(1)} &= \bm{\upkappa}_1\, ,\qquad 
	\bm{\upkappa}_1^{(2)} = \bm{\upkappa}_1^{(3)} = \bm{\upkappa}_1^{(5)} = \bm{\upkappa}_1 - (1-\zeta){\bf q}\, ,\qquad 
	\bm{\upkappa}_1^{(4)} = \bm{\upkappa}_1 + \zeta {\bf q}\, ,\qquad 
	\label{eq4.62}\\
	\bm{\upkappa}_2^{(1)} &= \bm{\upkappa}_2^{(4)} = \bm{\upkappa}_2^{(5)} =\bm{\upkappa}_2\, ,\qquad 
	\bm{\upkappa}_2^{(2)} = \bm{\upkappa}_2 + (1-z){\bf q}\, ,\qquad \bm{\upkappa}_2^{(3)} = \bm{\upkappa}_2 -z {\bf q}\, .
	\label{eq4.63}	
\end{align}
In general, to first order in opacity, different relative momenta $\bm{\upkappa}_1^{(i)}$, $\bm{\upkappa}_2^{(i)}$ contribute in amplitude and complex conjugate amplitude to the spinor structure $D_{(i,j)}$. We denote relative momenta in the complex conjugate amplitude by $\bm{\upkappa}_1^{(i)*}$, $\bm{\upkappa}_2^{(i)*}$ -- since all relative momenta are real, no confusion should arise from this notation. The medium-modifications of the spinor structure 
are characteristically different depending on the polarization of the gluon. We now discuss the different cases that can arise.

\subsubsection{ The purely transverse contribution $D_{(ij)}^{(TT)}$  }
\label{sec4.3a}
The Dirac spinor structure of the terms $D_{(ij)}^{(TT)}$ can be computed straightforwardly using FeynCalc. However, the resulting kinematic expressions are lengthy -- O(10) lines -- and take the form of fourth-order polynomials in the transverse momenta, multiplied by prefactors that are rational functions of longitudinal momentum fractions. Without uncovering a simpler underlying structure, such results are of limited use. 

We therefore consider it a central results of this work that these lengthy expressions can be recast in a compact form, relying only on knowledge of the vacuum structure \eqref{eq3.25}. More precisely, we have verified through explicit calculation that, to first order in opacity, the purely transverse contributions to the spinor structure take the form
\begin{align}
	D_{(ij)}^{(TT)} &=  \alpha\, 
	\frac{ \bm{\upkappa}_1^{(i)}\cdot \bm{\upkappa}_1^{(j)*} }{2\left( p_1^+\right)^2}\,  
	\frac{ \bm{\upkappa}_2^{(i)}\cdot \bm{\upkappa}_2^{(j)*} }{2\left( p_5^+\right)^2}
	\nonumber\\
	&	+ \beta\, 
	\frac{ \bm{\upkappa}_1^{(i)}\cdot \bm{\upkappa}_2^{(j)} }{\left( 2 p_1^+ p_5^+\right)}\,  
	\frac{ \bm{\upkappa}_1^{(i)*}\cdot \bm{\upkappa}_2^{(j)*}}{\left( 2 p_1^+ p_5^+\right)}
	\nonumber \\
	&	+ \gamma\, 
	\frac{ \bm{\upkappa}_1^{(i)}\cdot \bm{\upkappa}_2^{(j)*}}{\left( 2 p_1^+ p_5^+\right)},  
	\frac{ \bm{\upkappa}_1^{(i)*}\cdot \bm{\upkappa}_2^{(j)} }{\left( 2 p_1^+ p_5^+\right)}\, ,
	\label{eq4.64}
\end{align}
where
\begin{align}
	\alpha &= d_{\rm vac}^{(11,22)} \, ,\label{eq4.65}\\
	\beta &= d_{\rm vac}^{(11,22)}  + d_{\rm vac}^{(12,21)} \, ,\label{eq4.66} \\
	\gamma &= - d_{\rm vac}^{(11,22)}  \, .\label{eq4.67}
\end{align}

The fact that the two relative momenta associated to a $1 \to 3$ splitting amplitude give rise, at the level of the squared amplitude, to three distinct combinations of scalar products was previously noted in the work of Arnold and Iqbal \cite{Arnold:2015qya}. Here, we identify this same structure in the $N=1$ opacity expansion. Our main observation is that knowledge of the simple momentum shifts \eqref{eq4.62}, \eqref{eq4.63} together with the corresponding vacuum coefficients \eqref{eq3.18}, \eqref{eq3.19},  is sufficient to reconstruct the full medium-modification of the vacuum splitting.

In the limit ${\bf q}\to 0$, the relative momentum dependence of the second and third term of \eqref{eq4.64} is identical, and 
\begin{equation}
	\lim_{{\bf q} \to 0} D_{(ij)}^{(TT)}  = 
	D_{\rm vac}^{(TT)} \, .
	\label{eq4.68}
\end{equation}

\subsubsection{ The purely longitudinal contribution $D_{(ij)}^{(LL)}$  }
\label{sec4.3b}
The longitudinal gluon polarization tensor \eqref{eq3.23} is propotional to the squared four-momentum of the propagating gluon. For the vacuum contribution,
this invariant mass is $M_{23}^2 =(p_2 + p_3)^2 = 2 p_2 \cdot p_3$, since both outgoing momenta are on-shell and massless,
\begin{equation}
		M_{23}^2  z (1-z) = \bm{\upkappa}_2\cdot \bm{\upkappa}_2\, .
		\label{eq4.69}
\end{equation} 
This explains why, in the decomposition of the vacuum spinor structure according to polarizations, the purely longitudinal contribution \eqref{eq3.28} is proportional to 
$\left(\bm{\upkappa}_2\cdot \bm{\upkappa}_2\right)^2$. Specifically, both the amplitude and its complex conjugate contribute one factor of 
the squared invariant mass of the charm - anticharm pair emerging from the second vertex. 

If the transverse momentum ${\bf q}$ flows from the charm or anti-charm propagator to the medium (contributions 2 and 3), one of the outgoing transverse momenta on the second vertex becomes off-shell, and the invariant mass of the gluon propagator entering this vertex is
\begin{equation}
	{M_{23}^2}^{(i)} \equiv (p_2 + p_3 + q)^2 = \frac{ \bm{\upkappa}_2\cdot \bm{\upkappa}_2 - z(1-z) \left(2 ({\bf p}_2+{\bf p}_3)\cdot {\bf q} + {\bf q}^2\right) }{z (1-z)}\, , \qquad \hbox{for $i=2,3$.}
	\label{eq4.70}
\end{equation}
If the transverse momentum ${\bf q}$ flows from the  light quark (contribution 1 and 4) to the medium,  the invariant mass of the charm- anticharm pair is still given by \eqref{eq4.69}. Accordingly, we define
\begin{equation}
	{M_{23}^2}^{(i)} \equiv M_{23}^2 \, , \qquad \hbox{for $i=1,4$.}
	\label{eq4.71}
\end{equation}
Using these shifted invariant masses, the purely longitudinal contribution to the spinor structure takes the form
\begin{equation}
	D_{(ij)}^{(LL)} =  d_{\rm vac}^{(22,22)} z^2\, (1-z)^2  \frac{ {M_{23}^2}^{(i)}}{2\left( p_1^+\right)^2} \, 
	\frac{ {M_{23}^2}^{(j)*} }{2\left( p_5^+\right)^2} \, , \qquad \hbox{for $i, j = 1,2,3,4$,}
	\label{eq4.72}
\end{equation}
where the star superscript indicates that ${M_{23}^2}^{(j)*}$ is the invariant mass of the $c\bar{c}$-pair emerging from the second vertex of the complex conjugate amplitude. 

On the other hand, longitudinally polarized gluons do not interact with the medium's color field $A^-$. Indeed, $V^{\mu\lambda\nu}A_\lambda  n_\nu \propto V^{\mu + +}$, and $V^{\mu + +}$ vanishes upon contraction with the longitudinal gluon polarization tensor $d^{\mu'\mu}_L$. Consequently, 
\begin{equation}
	D_{(ij)}^{(LL)} =  0 \, , \qquad \hbox{if $i = 5$ or $j = 5$.}
	\label{eq4.73}
\end{equation}
We have verified equations \eqref{eq4.72} and \eqref{eq4.73} by explicit calculation of all contributions in FeynCalc. In summary, the purely longitudinal medium-modifications of the vacuum spliting, to first order in opacity, can be expressed compactly using only the vacuum result \eqref{eq3.28} and the corresponding coefficient \eqref{eq3.21}, supplemented by the simple shifts in invariant mass \eqref{eq4.70}, \eqref{eq4.71} together with the fact that longitudinally polarized gluons do not interact with the medium.

\subsubsection{ The mixed transverse longitudinal contribution $D_{(ij)}^{(TL)}$  }
We have seen so far that to first order in opacity, purely transverse contributions $D_{(ij)}^{(TT)}$ can be written  compactly in terms of transverse momentum shifts \eqref{eq4.62}, \eqref{eq4.63} while purely longitudinal contributions $D_{(ij)}^{(LL)}$ should be understood in terms of shifted invariant masses \eqref{eq4.70}, \eqref{eq4.71}. The mixed contribution transverse longitudinal contribution can be understood in terms of the same two simple replacements. We have verified by explicit calculation that 
\begin{align}
	D_{(ij)}^{(TL)} &=  d_{\rm vac}^{(12,22)} \, z\, (1-z)\, 
	\frac{ {\bm \kappa}_1^{(i)} \cdot {\bm \kappa}_2^{(i)} }{2\left( p_1^+\right)^2} \, \frac{ {M_{23}^2}^{(j)*} }{2\left( p_5^+\right)^2}\, ,\qquad
	\hbox{for $j=1,2,3,4$,}
	\label{eq4.74}\\
	&= 0 \, , \qquad \qquad \qquad \qquad \qquad \qquad \qquad \qquad \qquad \hbox{for $j=5$.}
	\label{eq4.75}
\end{align}
Here, the last line follows again from the fact that longitudinally polarized gluons do not interact with the $A^-$ field of the medium. 


\subsubsection{Probability-conserving terms }
\label{sec4.3.4}
We finally consider terms in the expansion of \eqref{eq1.1} to second order in $A^-$, in which the amplitude is expanded to second order while its complex conjugate remains the vacuum one. In these contributions, the two gluons exchanged at the amplitude level carry opposite transverse momenta $\pm {\bf q}$ and couple to the same scatterer in the medium, i.e., both interactions with the medium occur at the same light-cone time. For this reason, such virtual contributions are either diagonal, of the form $(jj)$, or correspond to (45), (42), (43) and (23). They are depicted in Fig.~\ref{fig5}

For the diagional contributions, one finds
\begin{align}
	\bar{D}_{(jj)} &= D_{\rm vac}\, ,\qquad \hbox{for}\quad  j=1,\dots ,4 \, , \label{eq4.76} \\
	\bar{D}_{(55)} &= D_{\rm vac}^{(TT)} +  D_{\rm vac}^{(TL)} \, . \label{eq4.77}
\end{align}
We have verified these relations by explicit calculations with FeynCalc. Previous work on the opacity expansion has repeatedly shown that, if two gluons are exchanged with the same parton at the amplitude level, the result is proportional to the vacuum contribution.\footnote{We recall that the factor $-1$ that typically arises in such diagonal virtual terms has been absorbed in the present work in the color factor, see comment following \eqref{eq4.10}.} 
However, this simple relation between the Dirac structure of diagonal real and diagonal virtual terms holds only for quarks and for transversely polarized gluons. In the opacity expansion of $1 \to 2$ splitting functions, this does not pose a limitation, since incoming and outgoing gluons are transverse. In the present problem, however, the intermediate gluon has also a longitudinal component, and this component does not interact with the medium field $A^-$. Therefore,  in the amplitude expanded to second order in $A^-$, the intermediate gluon must be transverse, while the complex conjugate vacuum amplitude contains both transverse and a longitudinal terms. This leads to the combination $D_{\rm vac}^{(TT)} +  D_{\rm vac}^{(TL)}$ in \eqref{eq4.77}.

For  the off-diagonal contributions, we proceed by defining the transverse momenta associated to the medium-modified splittings of the amplitude expanded to second order in $A^-$,
\begin{align}
		\bm{\upkappa}_1^{(45)} &= \bm{\upkappa}_1 - {\bf q}\, ,\qquad  \bm{\upkappa}_2^{(45)} =\bm{\upkappa}_2 \, , \label{eq4.78}\\
		\bm{\upkappa}_1^{(42)} &= \bm{\upkappa}_1 - {\bf q}\, ,\qquad  \bm{\upkappa}_2^{(42)} =\bm{\upkappa}_2 + (1-z){\bf q}\, , \label{eq4.79}\\
		\bm{\upkappa}_1^{(43)} &= \bm{\upkappa}_1 - {\bf q}\, ,\qquad  \bm{\upkappa}_2^{(43)} =\bm{\upkappa}_2 -z {\bf q}\, , \label{eq4.80}\\
		\bm{\upkappa}_1^{(23)} &= \bm{\upkappa}_1 \, ,\qquad  \bm{\upkappa}_2^{(23)} =\bm{\upkappa}_2 + {\bf q}\, .\label{eq4.81}
\end{align}
For the virtual contribution (45), one finds then 
\begin{align}
	\bar{D}_{(45)}^{(TT)} &=  D_{(45,{\rm vac})}^{(TT)} \, ,\label{eq4.82}\\
	\bar{D}_{(45)}^{(TL)} &= D_{(45,{\rm vac})}^{(TL)} = d_{\rm vac}^{(12,22)} 
	\frac{ \bm{\upkappa}_1^{(45)} \cdot  \bm{\upkappa}_2^{(45)} }{2\left( p_1^+\right)^2}
	 \frac{ \bm{\upkappa}_2 \cdot  \bm{\upkappa}_2 }{2\left( p_5^+\right)^2} \, , \label{eq4.83}\\
	 \bar{D}_{(45)}^{(LT)} &= 0\, ,\label{eq4.84} \\
	 \bar{D}_{(45)}^{(LL)} & = 0\, . \label{eq4.85}
\end{align}
Here, the subscript $(45,{\rm vac})$ refers to the expression $D_{(45,{\rm vac})}^{(TT)}$ defined in  \eqref{eq4.64}, with $i\to (45)$ and the index $j$ set to the vacuum label $j = 1$. Analogously, \eqref{eq4.83} follows from \eqref{eq4.74}.
These results illustrate once more that contributions from longitudinally polarized gluons arise only from the vacuum amplitude, while such gluons do not interact with $A^-$. Therefore, the last two terms vanish. 

In a completely analogous way, one finds
\begin{align}
	\bar{D}_{(24)} &=  D_{(24,{\rm vac})} \, ,\label{eq4.86}\\
	\bar{D}_{(34)} &=  D_{(34,{\rm vac})} \, .\label{eq4.87}
\end{align}
Since in these terms the intermediate gluon does not interact with the medium field $A^-$, a decomposition into polarizations is not required and the results follow straightforwardly by associating the shifted transverse momenta \eqref{eq4.79}, \eqref{eq4.80} with the amplitude and the vacuum momenta $\bm{\upkappa}_1$, $\bm{\upkappa}_2$ with the complex conjugate amplitude. The contribution (23) is not needed in our study, as it contributes only at subleading order in $N_c$. 

\subsection{Summary of main result}
\label{sec4.4}
For clarity, we briefly summarize the main result of this section before discussing relevant limiting cases.
%
The massless triple-collinear LO vacuum splitting function $q \to q c\bar{c}$ of \eqref{eq3.12} can be written as
\begin{equation}
	\hat{P}_{q\to qc\bar{c}}^{\rm (vac)} = C^{q\to qc\bar{c}}_{\rm vac}\, D^{q\to qc\bar{c}}_{\rm vac}\, {\cal P}^{q\to qc\bar{c}}_{\rm vac}\, ,
	\label{eq4.88}
\end{equation}
where the color factor $ C_{\rm vac}^{q\to qc\bar{c}} = C_F\, T_R$
  is the product of the color factors associated with the $q\to qg$ and $g\to c\bar{c}$, and the longitudinal phase integral is defined in \eqref{eq4.90}, 
 ${\cal P}_{\rm vac}^{q\to qc\bar{c}} = \tfrac{1}{(B_1+C_5)^2\, C_5^2}$. The spinor structure $D^{q\to qc\bar{c}}_{\rm vac}$ was given in
 \eqref{eq3.17b} in terms of scalar products of the boost-invariant transverse momenta ${\bm \kappa}_i$ associated with the splittings $q\to qg$ and $g\to c\bar{c}$. 
 
The main result of this section is the medium-induced correction to this vacuum splitting function at first order in opacity and leading order in $N_c$. It is given by  \eqref{eq4.1}, 
\begin{equation}
\hat{P}_{q\to qc\bar{c}}^{(N=1)} = \int\frac{d{\bf q}}{(2\pi)^2} \vert a({\bf q}) \vert^2  
  \sum_{\substack{ (i,j) \\ i\leq j} }  \left(
 \underbrace{  C_{(ij)} D_{(ij)} {\cal P}_{(ij)} }_{\it real}  + 
 \underbrace{  \bar{C}_{(ij)} \bar{D}_{(ij)} \bar{\cal P}_{(ij)} }_{\it virtual}   \right)\, ,
 \label{eq4.92}
\end{equation}
where the sum runs over symmetrized diagrammatic contributions $(ij)$ in which the medium exchanges one gluon with the $i$-th and $j$-th parton of the vacuum branching process in the amplitude and/or the complex conjugate amplitude. 
\begin{enumerate}
	\item {\it The color factors $C_{(ij)}$, $\bar{C}_{(ij)}$, listed in \eqref{eq4.2}-\eqref{eq4.10}, are determined easily.  }\\
	They are naturally expressed in units of $\tfrac{N_c^2}{8}$ which, at leading order in $N_c$, is the product of the vacuum splitting color factor $C_{\rm vac}^{q\to qc\bar{c}} = C_F \, T_R$,  and the fundamental Casimir $C_F$ associated with the medium-induced gluon exchange off a quark. 
	\item {\it The longitudinal phase integrals ${\cal P}_{(ij)}$, $\bar{\cal P}_{(ij)}$ in \eqref{eq4.37}, \eqref{eq4.44}, \eqref{eq4.54}-\eqref{eq4.61} 
	encode the space-time information about how the splitting process is embedded in the medium.}\\
	They are expressed in terms of simple integrals and take the form of  interference factors that compare the medium length $L$ to formation times defined in terms of transverse energies $B_i$, $C_i$. 
	\item {\it The Dirac terms $D_{(ij)}$,  $\bar{D}_{(ij)}$ can be obtained from the vacuum Dirac structure $D^{q\to qc\bar{c}}_{\rm vac}$ through
	 simple replacement rules.}\\ 
	The replacement rules depend on the polarization of the intermediate gluon, as discussed in section~\ref{sec4.3}.  For  transversely polarized gluons, the relevant substitutions are 
	shifts of the transverse momenta ${\bm \kappa}_1$ and  ${\bm \kappa}_2$. For longitudinally polarized gluons, the shifted quantity is instead the gluon invariant mass; moreover, longitudinally polarized gluons do not interact with the medium's color field $A^-$. In summary, the medium-induced structures are obtained by: 
\begin{itemize}
\item replacing the purely transverse $(TT)$ contribution of $D_{\rm vac}$ by \eqref{eq4.64},
\item replacing the purely longitudinal $(LL)$ contribution by \eqref{eq4.72} and \eqref{eq4.73},
\item replacing the mixed $(LT)$ and $(TL)$ contributions by \eqref{eq4.74} and \eqref{eq4.75}.
\end{itemize}	
The probability-conserving virtual contributions $\bar{D}_{(ij)}$ are determined analogously, see subsection~\ref{sec4.3.4}.
\end{enumerate}

\section{Strongly ordered limits in vacuum and medium}
\label{sec5}

In the strongly ordered limit, in which the invariant mass of the $c\bar{c}$ pair is much smaller than all other invariant masses, 
the $q\to qc\bar{c}$ collinear splitting function is known to factorize into a product of consecutive $1\to 2$ splitting functions. Here, we aim to identify the limiting cases in which this factorization persists when the $1\to 3$ splitting process is embedded in the medium. 

Writing the medium modifications in the form \eqref{eq4.92} is particularly useful for studying in-medium factorization for two reasons. First, the Dirac terms $D_{(ij)}$ are obtained from $D^{q\to qc\bar{c}}_{\rm vac}$ by transverse momentum shifts. As discussed below, this allows $D_{(ij)}$, under certain conditions, to be expressed as tensor products of the Dirac terms for $1\to 2$ splitting functions, which is a key step for establishing factorization. Second, in the strongly ordered limits, all transverse contributions $D_{(ij)}^{(TT)}$ are parametrically leading, while the longitudinal phase integrals ${\cal P}_{(ij)}$ differ parametrically between $(ij)$. These interference factors encode how the $1\to 3$ splitting is embedded in the medium and thereby determine which scattering contributions $D_{(ij)}^{(TT)}$ are dominant or suppressed. The strongly ordered limits are therefore obtained by taking the appropriate limits of ${\cal P}_{(ij)}$ and then exploiting the known factorization properties of $D_{\rm vac}$.

To prepare for this analysis, we first recall how the $q \to qc\bar{c}$ vacuum splitting function factorizes in the strongly ordered limit. We then show that, once the splitting process is embedded in a medium,  several distinct strongly ordered limits arise, and we identify cases in which in-medium factorization persists. 

\subsection{Factorization of the Dirac term for $q \to qc\bar{c}$ }
\subsubsection{Factorization in the vacuum}
We are interested in the strongly ordered limit  in which the invariant mass $\tilde{s}_{23}$ of the $c\bar{c}$ pair is taken much smaller than the other invariant masses
\begin{equation}
	\tilde{s}_{23} \ll \tilde{s}_{24}\, , \, \tilde{s}_{34}\, ,\, \tilde{s}_{234} \, .
	\label{eq5.1}
\end{equation}
According to \eqref{eq3.13} -- \eqref{eq3.16}, the collinear limit can be reached also by requiring 
\begin{equation}
	{\bm \kappa}_2\cdot {\bm \kappa}_2 \ll {\bm \kappa}_1\cdot {\bm \kappa}_1\, ,
	\label{eq5.2}
\end{equation}
as long as the longitudinal momentum fractions $z$, $\zeta$ are ${\cal O}(1)$ and not soft.

This condition allows for ${\cal O} \left(  {\bm \kappa}_1^2  {\bm \kappa}_2^2\right) \sim {\cal O} \left(  ({\bm \kappa}_1\cdot{\bm \kappa}_2)^2\right)$, and therefore
\begin{equation}
	D_{\rm vac}^{q\to qc\bar{c}} \xrightarrow{ {\bm \kappa}_2\cdot {\bm \kappa}_2 \ll {\bm \kappa}_1\cdot {\bm \kappa}_1  }
	D_{\rm vac}^{(TT)} \, .
	\label{eq5.3}
\end{equation}
We use first the known decomposition properties~\cite{Craft:2023aew} of the vacuum splitting function $q \to qc\bar{c}$ to write a factorized expression for $D_{\rm vac}^{(TT)}$. To this end, we associate transverse tensors with the collinear $q \to qg$ and $g\to c\bar{c}$ splitting functions, respectively.
In particular, the spin-correlated LO $g\to c\bar{c}$ splitting tensor $\hat{P}_{\rm vac}^{g\to c \bar{c}, \mu\nu}$ can be written by replacing 
in \eqref{eq2.16} the Dirac term with the spin-correlated tensor
\begin{align}
	D_{\rm vac}^{g\to c\bar{c}, \, \mu\nu} 
	& = \frac{1}{\left(p_5^+ \right)^2} \frac{{\bm \kappa}_2\cdot {\bm \kappa}_2 }{ 2 z (1-z)} \Big( - g^{\mu\nu}   
	+ 4 z (1-z) \frac{ {\bm \kappa}_2^\mu {\bm \kappa}_2^{\nu} }{{\bm \kappa}_2\cdot {\bm \kappa}_2}  \Big)\, .
	\label{eq5.5}
\end{align}
The Dirac term  $D_{\rm vac}^{g\to c \bar{c}}$ follows from the spin-correlated tensor after performing the inclusive averages: first over the azimuthal orientation of the outgoing $c\bar{c}$ pair, 
$\langle \tfrac{ {\bm \kappa}_2^\mu {\bm \kappa}_2^{\nu} }{{\bm \kappa}_2\cdot {\bm \kappa}_2} \rangle = \tfrac{1}{2} g_\perp^{\mu\nu}$, and then over the helicity of the parent gluon by contracting with the polarization sum $\tfrac{1}{2} \sum_{\lambda=+,-} \epsilon_\lambda^\mu \epsilon_\lambda^{*\nu} = - \tfrac{1}{2} g_\perp^{\mu\nu}$. Likewise, the helicity-correlated tensor for $q \to q g$ can be written in the convention of Ref.~\cite{Craft:2023aew} as
\begin{align}
	D_{\rm vac}^{q\to qg,\, \mu\nu} 
	& =  \frac{1}{\left(p_1^+ \right)^2} 
	\frac{{\bm \kappa}_1\cdot{\bm \kappa}_1 }{ 2 \zeta (1-\zeta)}  \left[ \frac{1}{2} \zeta\, d^{\mu\nu}(p,n)\,  - 2 \frac{1-\zeta}{\zeta} 
	\frac{{\bm \kappa}_1^\mu {\bm \kappa}_1^\nu }{{\bm \kappa}_1\cdot{\bm \kappa}_1 } \right]\, , \label{eq5.4}
\end{align}
Contracting these tensors, we recover the purely transverse contribution \eqref{eq3.25} of the vacuum Dirac term for $q\to qc\bar{c}$. 
\begin{align}
	D_{\rm vac}^{(TT)} &= D_{\rm vac}^{g\to c\bar{c},\, \mu\nu} D_{\rm vac, \, \mu\nu}^{q\to qg} = \frac{1}{ \zeta(1-\zeta)z(1-z)} 
	\frac{1}{\left(2 p_1^+ \right)^2} \, \frac{1}{\left(2 p_5^+ \right)^2} 
	\nonumber \\
	&\quad \times 
	\left[   \left(\zeta + 2\frac{1-\zeta}{\zeta} - 2 \zeta z (1-z)\right) 
	 \left( {\bm \kappa}_1\cdot {\bm \kappa}_1\right)\, \left( {\bm \kappa}_2\cdot {\bm \kappa}_2\right) 
	  - 8 \frac{z(1-z)(1-\zeta)}{\zeta} \left( {\bm \kappa}_1 \cdot {\bm \kappa}_2\right)^2 \right] \nonumber \\
	 &= \frac{1}{\left(2 p_1^+ \right)^2} \, \frac{1}{\left(2 p_5^+ \right)^2} \left[ 
	 d_{\rm vac}^{(11,22)}\, \left( {\bm \kappa}_1\cdot {\bm \kappa}_1\right)\, \left( {\bm \kappa}_2\cdot {\bm \kappa}_2\right) 
						+ d_{\rm vac}^{(12,21)}  \left( {\bm \kappa}_1\cdot {\bm \kappa}_2\right)\, \left( {\bm \kappa}_2\cdot {\bm \kappa}_1\right) \right]\, .
	\label{eq5.8}
\end{align}
The factorization of the triple-collinear vacuum splitting function \eqref{eq3.17} into a tensor product of $1\to 2$ splitting functions,
\begin{equation}
	\mathscr{C}_{23} \hat{P}_{q\to qc\bar{c}} =  \hat{P}_{q\to qg}^{\mu\nu}\,  \hat{P}_{g\to c\bar{c},\, {\mu\nu}}\, ,
	\label{eq5.8b}
\end{equation}
can be understood from two simple observations: First, in the collinear limit denoted by the operator $\mathscr{C}_{23}$, where the invariant mass of the $c\bar{c}$ pair is much smaller than all other invariant masses, the longitudinal components of the Dirac term $D_{\rm vac}^{q\to qc\bar{c}}$ in \eqref{eq3.17b} become subleading, and only the purely transverse contribution survives. As shown in \eqref{eq5.8}, this surviving contribution factorizes.  Second, the $1\to 2$ splitting functions $\hat{P}_{g\to c\bar{c}}$ in \eqref{eq2.16} and $\hat{P}_{q\to qg}$ in \eqref{eq2.74b} are easily promoted to the two-dimensional tensors on the right hand side of \eqref{eq5.8b} by replacing the corresponding Dirac terms in their definition with \eqref{eq5.9} and \eqref{eq5.10}, respectively.

\subsubsection{Factorization of the Dirac terms to first order in opacity}
At first order in opacity, the purely transverse contributions to $D_{(ij)}$ are leading in the collinear limit \eqref{eq5.2}. However, the medium-modified splitting function $\hat{P}_{q\to qc\bar{c}}^{(N=1)}$ in \eqref{eq4.92} is obtained by summing over the different contributions $(ij)$, and its factorization depends on two conditions: For contributions $(ij)$ for which the corresponding color factors $C_{(ij)}$ survive at leading order in $N_c$, the corresponding interference terms ${\cal P}_{(ij)}$ must either factorize or vanish in the collinear limit, and the Dirac terms $D_{(ij)}$ must factorize. In the present subsection, we focus exclusively on the last of these conditions.

As we consider in-medium interactions to leading order in lightcone energy, they do not swap spin or helicity. Therefore, the spin- and helicity structure of the medium-modified Dirac terms $D_{(ij)}^{(TT)}$ are those of the vacuum contribution. This motivates  lifting the $1 \to 2$ vacuum splitting tensors \eqref{eq5.4} and \eqref{eq5.5} to medium-modified ones, 
\begin{align}
		D_{(ij)}^{q\to qg,\, \mu\nu} 
	& =   \frac{1}{\left(2 p_1^+ \right)^2} 
	 \frac{1}{\zeta (1-\zeta)}  \left[ \frac{1}{2} \zeta\, d^{\mu\nu}(p,n)\, {\bm \kappa}_1^{(i)}\cdot{\bm \kappa}_1^{(j)*}  - 2 \frac{1-\zeta}{\zeta} 
	{\bm \kappa}_1^{(i)\, \mu} {\bm \kappa}_1^{(j)*\, \nu}  \right]\, ,\label{eq5.9} \\
D_{(ij)}^{g\to c\bar{c}, \, \mu\nu} 
	& =  \frac{1}{\left(2 p_5^+ \right)^2}   \frac{1}{ z (1-z)} \Big( - g^{\mu\nu} {\bm \kappa}_2^{(i)}\cdot {\bm \kappa}_2^{(j)*}   
	+ 4 z (1-z) {\bm \kappa}_2^{(i)\, \mu} {\bm \kappa}_2^{(j)*\, \nu}  \Big)\, .
		\label{eq5.10}	
\end{align}
We note that the medium-modified Dirac terms $D_{(ij)}^{(TT)}$ in \eqref{eq4.64} have different prefactors 
$\beta = d_{\rm vac}^{(11,22)} + d_{\rm vac}^{(12,21)} $ and $\gamma = - d_{\rm vac}^{(11,22)}$
multiplying the contribution $ \left( \bm{\upkappa}_1^{(i)}\cdot \bm{\upkappa}_2^{(j)}\right)\,  \left( \bm{\upkappa}_1^{(i)*}\cdot \bm{\upkappa}_2^{(j)*}\right)$ and 
$\left( \bm{\upkappa}_1^{(i)}\cdot \bm{\upkappa}_2^{(j)*}\right)\,  \left( \bm{\upkappa}_1^{(i)*}\cdot \bm{\upkappa}_2^{(j)}\right)$, respectively. In contrast, in $D_{\rm vac}^{(TT)}$, the prefactor of 
$\left( \bm{\upkappa}_1\cdot \bm{\upkappa}_2 \right)^2$ fixes only the sum 
$\beta + \gamma$. Therefore, in general, the terms $D_{(ij)}^{(TT)} $ cannot be written in terms of tensor products of the form $D_{(ij)\, \mu\nu}^{q\to qg} D_{(ij)}^{g\to c\bar{c}\, \mu\nu} $. However, it is possible 
to write
\begin{align}
	D_{(ij)}^{(TT)} = D_{(ij)}^{g\to c\bar{c},\, \mu\nu} D_{(ij), \, \mu\nu}^{q\to qg} \quad \hbox{if}\quad
	\bm{\upkappa}_1^{(i)} = \bm{\upkappa}_1^{(j)*}\quad \hbox{or} \quad \bm{\upkappa}_2^{(i)} = \bm{\upkappa}_2^{(j)*}\, .
	\label{eq5.11}
\end{align}
In the following, this factorization property \eqref{eq5.11} of the Dirac terms will serve as a building block in the discussion of the structure 
of  the medium-modified splitting function $\hat{P}_{q\to qc\bar{c}}^{(N=1)}$ in \eqref{eq4.92}, although it does not by itself imply its factorization.  

Inspecting the explicit expressions for the shifted transverse momenta $\bm{\upkappa}_1^{(i)}$, $\bm{\upkappa}_2^{(i)}$ in \eqref{eq4.62}, \eqref{eq4.63}, one finds that the factorization property \eqref{eq5.11} holds for all diagonal contributions $(11)$, $(22)$, $(33)$, $(44)$, $(55)$ as well as for the off-diagonal contributions $(14)$, $(15)$, $(23)$, $(25)$, $(35)$ and $(45)$ whereas it fails for $(12)$, $(13)$, $(24)$ and $(34)$. 
A simple physical picture helps understand the cases for which the factorization property \eqref{eq5.11} is violated: 

The contributions $(12)$ and $(13)$ are precisely those for which the same location in the medium couples, in the amplitude and complex conjugate amplitude, to both the parent "1" and the granddaughters "2" and "3" of the $1 \to 3$ splitting process. Similarly, the terms $(24)$ and $(34)$ are precisely those for which, in the amplitude and complex conjugate amplitude, the medium couples to partonic fragments that, in a factorized picture, would be attributed exclusively to the first $1\to 2$ splitting (namely "4") or to the second $1\to 2$ splitting (namely "2" or "3"). Such medium-induced cross talk between the two branchings likewise obstructs factorization on physical grounds. In  contrast, all diagonal and off-diagonal contributions for which \eqref{eq5.11} holds can be associated, within a factorized picture, entirely with either the first or the second $1\to 2$ splitting. 

\subsection{Strongly ordered collinear limits in medium}
For the medium-modified splitting function $\hat{P}_{q\to qc\bar{c}}^{(N=1)}$ to factorize in a collinear limit, the interference terms ${\cal P}_{(ij)}$ contributing to the sum in \eqref{eq4.92} must either factorize or vanish in that limit. We therefore consider interference terms in the kinematic regime in which the longitudinal momentum fractions $\zeta$,  $z$ are not soft but ${\cal O}(1)$, and 
\begin{equation}
{\bm \kappa}_2^{(j)} \cdot {\bm \kappa}_2^{(j)}  \ll {\bm \kappa}_1^{(i)} \cdot {\bm \kappa}_1^{(i)} \, .
\label{eq5.12}
\end{equation}
This is a natural extension of the strong ordering \eqref{eq5.2} to medium interactions. It assumes that the momentum transfer from the medium is sufficiently small so that the strong transverse momentum ordering of the vacuum splitting is preserved. It directly translates to 
 \begin{equation}
 	B_i \gg C_j\, . 
	\label{eq5.13}
 \end{equation}
 
 In a space-time picture, $B_i \gg C_j$ also implies that the first splitting is much closer to the production point of the parent quark than the second one. In a medium of length $L$, the relevant dimensionless quantities are $B_i L$ and  $C_j L$, and there are several physically distinct strongly ordered limits of the form \eqref{eq5.12}, 
 \begin{enumerate}
	\item $\mathscr{C}_{C_jL}$:   {\it $B_i \gtrsim \tfrac{1}{L} \gg C_j$: Strong ordering without resolving the second splitting in medium } 
	\item $\mathscr{C}_{\frac{1}{B_iL}}$:   {\it $B_i  \gg C_j \gtrsim \tfrac{1}{L}$: Instantaneous embedding of first splitting in the medium}
	 \item $\mathscr{C}_{B_iL}$:  {\it $\tfrac{1}{L} \gg B_i \gg C_j$: Totally coherent (vacuum) limit }
 \end{enumerate}
 In analogy with the notation $\mathscr{C}_{23}$ introduced in \eqref{eq5.8b}, we denote the corresponding limiting operations by $\mathscr{C}_{C_jL}$, $\mathscr{C}_{\frac{1}{B_iL}}$ and $\mathscr{C}_{B_iL}$, where the subscript indicates the small parameter in the expansion.   Figure~\ref{figlast} illustrates the different space--time embeddings of the $1\to 3$ splitting in each of these three limits. 
  
 \subsubsection{Strongly ordered limit without resolving the second splitting in medium}
To consider the hierarchy $B_i \gtrsim \tfrac{1}{L} \gg C_j$, we study here the limit 
 \begin{equation}
 	C_j L \to 0\, ,
	\label{eq5.14}
\end{equation} 
in which the formation time of the second splitting becomes arbitrarily large, that means, the $g \to c\bar{c}$ lies outside the medium. $B_i$ can then be either commensurate with or much larger than $\tfrac{1}{L}$, as long as \eqref{eq5.13} holds. On physical grounds, one expects that the $1 \to 3$ splitting function can be written in this limit as the tensor product of the medium-modified $q \to qg$ splitting function times the vacuum splitting function for $g \to c\bar{c}$. 

To demonstrate how this expectation is realized, we consider first the longitudinal phase integrals ${\cal P}_{(ij)}$ that enter \eqref{eq4.92}. In the limit \eqref{eq5.14}, the following real terms are of leading order ${\cal O}( C_i^{-2})$
\begin{align}
	{\cal P}_{(11)} &=   \frac{n_0\, L}{C_5^2\, B_1^2} + {\cal O}( C_i^{-1})\, , 	\label{eq5.15}\\
	{\cal P}_{(44)} &=    \frac{2\, n_0\, L}{C_5^2\, B_4^2}  {\cal S}(B_4) 
				+ {\cal O}( C_i^{-1})\, ,   	\label{eq5.16} \\
	{\cal P}_{(55)} &=    \frac{2\, n_0\, L}{C_5^2\, B_5^2}  {\cal S}(B_5)  
				 + {\cal O}( C_i^{-1})\, ,   	\label{eq5.17} \\
	{\cal P}_{(15)} &=  -  \frac{2\, n_0\, L}{C_5^2\, B_1\, B_5}  {\cal S}(B_5) 
				+ {\cal O}( C_i^{-1})\, ,    	\label{eq5.18} \\
	{\cal P}_{(45)} &=   \frac{2\, n_0\, L}{C_5^2\, B_5 B_4} \left(  {\cal S}(B_5) +  {\cal S}(B_4) -  {\cal S}(B_5-B_4) \right)
	 + {\cal O}( C_i^{-1})\, , 	\label{eq5.19}
\end{align} 
with all other real terms ${\cal P}_{(ij)}$ vanishing up to ${\cal O}( C_i^{-1})$. Likewise, one finds for the virtual contributions
\begin{align}
	\bar{\cal P}_{(11)} &=   \frac{n_0\, L}{C_5^2\, B_1^2}   -  \frac{n_0\, L}{C_5^2\, B_1^2} {\cal S}(B_1) + {\cal O}( C_i^{-1})\, , 	\label{eq5.20} \\
	\bar{\cal P}_{(44)} &=    \frac{n_0\, L}{C_5^2\, B_1^2}  {\cal S}(B_1) 
				+ {\cal O}( C_i^{-1})\, ,   \label{eq5.21}  \\
	\bar{\cal P}_{(55)} &=    \frac{n_0\, L}{C_5^2\, B_1^2}  {\cal S}(B_1)   
				 + {\cal O}( C_i^{-1})\, ,    \label{eq5.22}  \\
	\bar{\cal P}_{(45)} &=   \frac{2\, n_0\, L}{C_5^2\, \tilde{B} B_1} \left(  {\cal S}(B_1)  -  {\cal S}(\tilde{B} -B_1) \right) 
	+ {\cal O}( C_i^{-1})\, , \label{eq5.23} 
\end{align} 
with all other virtual contributions vanishing up to ${\cal O}( C_i^{-1})$. The longitudinal phase integrals that survive in this limit are precisely those associated with a medium-modified $q(1)\to q(4)g(5)$ splitting function at first order in opacity, multiplied by the longitudinal phase integral 
${\cal P}_{\rm vac}^{g \to c\bar{c}} = \tfrac{1}{C_5^2}$ associated with a vacuum $g \to c\bar{c}$ splitting. This yields the factorization 
\begin{align}
	\mathscr{C}_{C_jL} \, {\cal P}_{(ij)} &= {\cal P}_{\rm vac}^{g \to c\bar{c}}\, {\cal P}_{(ij)}^{q \to qg} 
	\, , \quad \hbox{for $i,j = 1,4,5$}\, , \label{eq5.24} \\
	\mathscr{C}_{C_jL} \,  {\cal P}_{(ij)} &= 0 
	\, , \quad \hbox{for $i=2,3$ or $j=2,3$.} \label{eq5.25} 
\end{align}
An analogous relation holds for virtual contributions. In particular, in the limit \eqref{eq5.14}, no interaction  of the charm or anti-charm quark with the medium survives  since the corresponding longitudinal phase integrals vanish to leading order ${\cal O}( C_i^{-2})$. As a direct consequence, see Eqs.~\eqref{eq4.63} and \eqref{eq4.78}--\eqref{eq4.81}, the surviving terms satisfy $ \bm{\upkappa}_2^{(i)} =  \bm{\upkappa}_2$, which allows us to invoke the factorization relation \eqref{eq5.11},
\begin{align}
	\mathscr{C}_{C_jL} \, D_{(ij)}^{(TT)} =& \frac{1}{2\left( p_1^+\right)^2}
	\frac{1}{(1-\zeta)\, \zeta} \left[ \frac{1}{2} \zeta\, d^{\mu\nu}(p,n)  \left({\bm \kappa}_1^{(i)}\cdot {\bm \kappa}_1^{(i)}\right)  - 
	 2 \frac{1-\zeta}{\zeta} {\bm \kappa}_1^{(i)\, \mu} {\bm \kappa}_1^{(j)\, \nu\, *}  \right] \nonumber \\
	& \times \frac{1}{2\left( p_5^+\right)^2} \frac{1}{(1-z)\, z}  
	\left[- g^{\mu\nu} \, {\bm \kappa}_2^2 + 4 z (1-z) {\bm \kappa}_2^\mu {\bm \kappa}_2^\nu \right]  \nonumber \\
	=& D_{(ij), \, \mu\nu}^{q\to qg}    D_{\rm vac}^{g\to c\bar{c},\, \mu\nu} \, , \quad \hbox{for $i,j = 1,4,5$}\, .
	 \label{eq5.26} 
\end{align}
Also the virtual terms and the color factors show the relevant factorization properties,
\begin{align}
	\mathscr{C}_{C_jL} \, C_{(ij)} &= T_R\, C^{q\to qg}_{(ij)} \, , \quad &\hbox{for $i,j = 1,4,5$},  \label{eq5.27} \\
	\mathscr{C}_{C_jL} \, \bar{C}_{(ij)} &= T_R\, \bar{C}^{q\to qg}_{(ij)} \, ,\quad &\hbox{for $i,j = 1,4,5$},  \label{eq5.28}  \\
	\mathscr{C}_{C_jL} \, \bar{\cal P}_{(ij)} &= {\cal P}_{\rm vac}^{g \to c\bar{c}}\, \bar{\cal P}_{(ij)}^{q \to qg} 
	\quad &\hbox{for $i,j = 1,4,5$},  \label{eq5.29}  \\
	\mathscr{C}_{C_jL} \, \bar{D}_{(ij)}^{(TT)} &=  D_{\rm vac}^{g\to c\bar{c},\, \mu\nu} \, \bar{D}_{(ij), \, \mu\nu}^{q\to qg}    \, , \quad &\hbox{for $i,j = 1,4,5$}.  \label{eq5.30} 
\end{align}
This establishes factorization of the medium-modified splitting function \eqref{eq4.92} in the strongly ordered limit defined by \eqref{eq5.13}, \eqref{eq5.14} in which the formation time of the second splitting becomes so long that the splitting occurs outside the medium
\begin{equation}
\mathscr{C}_{C_jL} \, \hat{P}_{q\to qc\bar{c}}^{(N=1)} 
= \hat{P}_{q\to qg,\, \mu\nu}^{(N=1)} \,  \hat{P}_{\rm vac}^{g\to c\bar{c},\, \mu\nu}\, ,
\label{eq5.31}
\end{equation}
where 
\begin{align}
\hat{P}_{q\to qg}^{(N=1)\, \mu\nu} &= \int\frac{d{\bf q}}{(2\pi)^2} \vert a({\bf q}) \vert^2  
  \sum_{\substack{ (i,j),\,  i\leq j   \\  i,j \in 1,4,5} }  \left(
  C^{q\to qg}_{(ij)} D^{q\to qg,\, \mu\nu}_{(ij)} {\cal P}^{q\to qg}_{(ij)}  + 
  \bar{C}^{q\to qg}_{(ij)} \bar{D}^{q\to qg,\, \mu\nu}_{(ij)} \bar{\cal P}^{q\to qg}_{(ij)}  \right)\, .
 \label{eq5.32}
\end{align}
and 
\begin{equation}
	\hat{P}_{\rm vac}^{g\to c\bar{c},\, \mu\nu} = T_R\, \frac{1}{C_5^2}\, D_{\rm vac}^{g\to c\bar{c}, \, \mu\nu}\, .
	\label{eq5.33}
\end{equation}
To summarize this technical discussion: in the limit $C_i L \to 0$, where the $g\to c\bar{c}$ vertex lies outside the medium, all contributions in which the charm- or anti-charm quark interact with the medium vanish. This follows from the property
$\mathscr{C}_{C_jL}  {\cal P}_{(ij)} = 0$ for $i, j = 2,3$. Consequently, every non-vanishing contribution to the sum in \eqref{eq4.92} factorizes into a $g\to c\bar{c}$ part and a $q\to qg$ part. The factors associated with the $g\to c\bar{c}$ splitting are precisely the vacuum ones:  the color factor $T_R$ for color, the longitudinal phase integral ${\cal P}_{\rm vac}^{g\to c\bar{c}} = \tfrac{1}{C_5^2}$, and the 
helicity-dependent Dirac term $D_{\rm vac}^{g\to c\bar{c}, \, \mu\nu}$ defined in \eqref{eq5.5}. Since these factors are common to all contributions, they can be factored out of the sum in \eqref{eq4.92}. The remaining terms are then precisely those of the medium-modified $q \to qg$ splitting function \eqref{eq2.71}, with the corresponding Dirac terms replaced by the spin-dependent tensors \eqref{eq5.26}. This  establishes the factorization property \eqref{eq5.31}.

 \subsubsection{Strongly ordered limit with first splitting embedded instantaneously in the medium}
The hierarchy $B_i  \gg C_j \gtrsim \tfrac{1}{L}$ also satisfies the generalized strong ordering condition \eqref{eq5.13}. In contrast to the previous case, however, we no longer require the formation time $\propto 1/C_i$ associated with the second splitting to be large compared to the medium length $L$. Instead, we allow $C_i L$ to be order one or even larger. Strong ordering in the sense of \eqref{eq5.13} is then ensured by imposing 
 \begin{equation}
 	\frac{1}{B_i\, L} \to 0\, .
	\label{eq5.34}
 \end{equation}
Increasing the scale $B_i$ shortens the length scale over which the first splitting forms. In the limit $B_i L \to \infty$, when the splitting occurs effectively instantaneously, one expects on physical grounds that this first splitting process is not resolved by the medium. In what follows, we clarify in which sense this limit admits a probabilistic interpretation in terms of: (i) a $q\to qg$ splitting that is unresolved by the medium and proceeds as in vacuum, (ii) medium-induced momentum broadening of the daughters produced in the $q\to qg$ splitting, and (iii) a medium-modified $g\to c\bar{c}$ splitting governed by $\hat{P}_{g\to c\bar{c}}^{(N=1)}$. 

Similar to the discussion in the previous section, we begin by collecting the longitudinal phase integrals that dominate in the limit \eqref{eq5.34}. Retaining only terms up to ${\cal O}(B_i^{-2})$, we find the real contributions
\begin{align}
	{\cal P}_{(11)} &=   \frac{n_0\, L}{C_5^2\, B_1^2} + {\cal O}\left( B_i^{-3}\right)\, ,  \label{eq5.35} \\
	{\cal P}_{(22)} &=    \frac{2\, n_0\, L}{C_2^2\, B_5^2} {\cal S}(C_2) + {\cal O}\left( B_i^{-3}\right)\, ,  \label{eq5.36}\\
	{\cal P}_{(33)} &=    \frac{2\, n_0\, L}{C_3^2\, B_5^2} {\cal S}(C_3) + {\cal O}\left( B_i^{-3}\right)\, ,  \label{eq5.37}\\
	{\cal P}_{(44)} &=   \frac{2\, n_0\, L}{C_5^2\, B_4^2} + {\cal O}\left( B_i^{-3}\right)\, , \label{eq5.38} \\
	{\cal P}_{(55)} &=    \frac{2\, n_0\, L}{C_5^2\, B_5^2} + {\cal O}\left( B_i^{-3}\right)\, ,  \label{eq5.39}\\
	{\cal P}_{(15)} &=    - \frac{2\, n_0\, L}{C_5^2\, B_1\, B_5} + {\cal O}\left( B_i^{-3}\right)\, , \label{eq5.40} \\
	{\cal P}_{(25)} &=    - \frac{2\, n_0\, L}{C_5\, C_2\, B_5^2} {\cal S}(C_2) + {\cal O}\left( B_i^{-3}\right)\, , \label{eq5.41} \\
	{\cal P}_{(35)} &=    - \frac{2\, n_0\, L}{C_5\, C_3\, B_5^2} {\cal S}(C_3) + {\cal O}\left( B_i^{-3}\right)\, ,  \label{eq5.42}\\
	{\cal P}_{(45)} &=    \frac{2\, n_0\, L}{C_5^2\, B_5\, B_4} + {\cal O}\left( B_i^{-3}\right)\, . \label{eq5.43}
\end{align} 
In the same strongly ordered limit \eqref{eq5.34}, the longitudinal phase integrals of virtual contributions read, up to ${\cal O}(B_i^{-2})$,
\begin{align}
	\bar{\cal P}_{(22)} &=    \frac{n_0\, L}{B_1^2\, C_5^2} {\cal S}(C_5) + {\cal O}\left( B_i^{-3}\right)\, , \label{eq5.44} \\
	\bar{\cal P}_{(33)} &=    \frac{n_0\, L}{B_1^2\, C_5^2} {\cal S}(C_5) + {\cal O}\left( B_i^{-3}\right)\, , \label{eq5.45} \\
	\bar{\cal P}_{(44)} &=   \frac{n_0\, L}{B_1^2\, C_5^2} + {\cal O}\left( B_i^{-3}\right)\, , \label{eq5.46}  \\
	\bar{\cal P}_{(55)} &=    \frac{n_0\, L}{B_1^2\, C_5^2} \frac{\sin\left( C_5 L\right)}{C_5 L} + {\cal O}\left( B_i^{-3}\right)\, . \label{eq5.47}
\end{align} 
We recall that the only off-diagonal virtual contributions are $(45)$, $(24)$, $(34)$ and $(23)$. Among these, the first three are subleading, of order ${\cal O}\left( B_i^{-3}\right)$. The fourth contribution, $(23)$, is leading at ${\cal O}\left( B_i^{-2}\right)$, but is multiplied by a color factor that is subleading in $N_c$. For this reason, it is  not listed here. 

To understand how these longitudinal phase integrals impact the medium-modified splitting function \eqref{eq4.1} in this strongly ordered limit, we first note that to leading order in $N_c$, the contributions that are proportional to the vacuum structure $\tfrac{1}{B_1^2 C_5^2} D_{\rm vac}$ sum up to 
\begin{equation}
 \mathscr{C}_{\frac{1}{B_iL}} \left( C_{(11)} D_{(11)} {\cal P}_{(11)}  + 
  \sum_{(ij)} \left( \bar{C}_{(ij)} \bar{D}_{(ij)} \bar{\cal P}_{(ij)} \right) \right) = - \frac{N_c^2}{8} \frac{2\, n_0\, L}{C_5^2\, B_1^2} D_{\rm vac} \, .
  \label{eq5.48}
\end{equation}
We note in particular that the term $\bar{\cal P}_{(55)}$ cancels the pieces proportional to $\tfrac{\sin\left( C_5 L\right)}{C_5 L}$ in $\bar{\cal P}_{(22)}$
and  $\bar{\cal P}_{(33)}$. This is so since the color factor multiplying $\bar{\cal P}_{(55)}$ is twice as large as that multiplying $\bar{\cal P}_{(22)}$
and  $\bar{\cal P}_{(33)}$. Using \eqref{eq5.48} and inserting all factors into equation \eqref{eq4.1}, we find
\begin{align}
\mathscr{C}_{\frac{1}{B_iL}}
\hat{P}_{q\to qc\bar{c}}^{(N=1)} =& 2 n_0\, L\, \frac{N_c^2}{8} \, \int\frac{d{\bf q}}{(2\pi)^2} \vert a({\bf q}) \vert^2  \nonumber\\
&\times
\Bigg(  
\frac{1}{C_5^2} \Bigg\lbrace
\left( \frac{D_{(44)}}{B_4^2} -   \frac{D_{\rm vac}}{B_1^2}\right) + 2\, \left( \frac{D_{(55)}}{B_5^2} -   \frac{D_{\rm vac}}{B_1^2}\right) \nonumber \\
& \qquad \qquad \qquad 
- \left( \frac{D_{(15)}}{B_1\, B_5} -   \frac{D_{\rm vac}}{B_1^2}\right) - \left( \frac{D_{(45)}}{B_5\, B_4} -   \frac{D_{\rm vac}}{B_1^2}\right) 
\Bigg\rbrace \nonumber \\
& \qquad +  \frac{1}{B_5^2} \Bigg\lbrace \left( \frac{D_{(22)}}{C_2^2} -  \frac{D_{(25)}}{C_2\, C_5} \right)  {\cal S}(C_2) + 
\left( \frac{D_{(33)}}{C_3^2} - \frac{D_{(35)}}{C_3\, C_5} \right) {\cal S}(C_3) \Bigg\rbrace  
  \Bigg)\, .
 \label{eq5.49}
\end{align}
Here, the factor $2$ in front of the $D_{(55)}$ contribution arises from the fact that the color factor of the $(55)$ contribution is $C_A=N_c$ while all other color factors are $N_c/2$. 

We observe that the terms in the first curly bracket $\lbrace \dots \rbrace$ of  \eqref{eq5.49}  have the same structure as the medium-modified $q \to qg$ splitting function in the rescattering limit \eqref{eq2.95}. Moreover, all medium-modified Dirac terms in this first curly bracket can be written in the form
$ D_{(ij)} = D_{(ij)}^{q\to qg,\, \mu\nu} D_{{\rm vac},\, \mu\nu}^{g\to c\bar{c}}$. This allows us to write the contribution from the first curly bracket as a tensor product $\hat{P}_{q\to qg,\, {\rm rescatt},\, \mu\nu}^{(N=1)} \hat{P}_{\rm vac}^{g\to c\bar{c}, \mu\nu }$, where $\hat{P}_{q\to qg,\, {\rm rescatt},\, \mu\nu}^{(N=1)}$ is defined by lifting the Dirac terms in \eqref{eq2.95} to the corresponding tensors \eqref{eq5.4} and \eqref{eq5.9}.

The terms in the second curly bracket $\lbrace \dots \rbrace$ of  \eqref{eq5.49} have the same structure as the medium-modified $g\to c\bar{c}$ splitting function in \eqref{eq2.67}. Moreover,  if the medium transfers transverse momentum ${\bf q}$ to the $g \to c\bar{c}$ splitting, then the internal gluon propagator carries a shifted transverse momentum ${\bf p}_5 \to {\bf p}_5+{\bf q}$. Therefore, this last term can be written as the tensor product of a medium-modified $g\to c\bar{c}$ splitting contracted with a vacuum $q\to q g$ splitting in which the outgoing gluon was shifted in transverse momentum, $ \hat{P}^{q\to qg}_{{\rm vac},\, \mu\nu} \Big\vert_{{\bf p}_5 \to {\bf p}_5+{\bf q}}  \hat{P}^{(N=1), \, \mu\nu }_{g\to c\bar{c} }$.

Combining these observations, we obtain 
 \begin{equation}
 \mathscr{C}_{\frac{1}{B_iL}} \hat{P}_{q\to qc\bar{c}}^{(N=1)} =  \hat{P}_{q\to qg,\, {\rm rescatt},\, \mu\nu}^{(N=1)} \hat{P}_{\rm vac}^{g\to c\bar{c}, \mu\nu }
 +
 \hat{P}^{q\to qg}_{{\rm vac},\, \mu\nu} \Big\vert_{{\bf p}_5 \to {\bf p}_5+{\bf q}}  \hat{P}^{(N=1), \, \mu\nu }_{g\to c\bar{c} }
 \label{eq5.55}
 \end{equation}
 The collinear limit \eqref{eq5.55} thus admits a simple probabilistic interpretation as the iteration of (i) an unresolved $q\to q g$ splitting that subsequently rescatters, and (ii) a gluon emerging from the unresolved vacuum-like $q\to q g$ splitting that undergoes medium-modified $g\to c\bar{c}$ splitting. Since we work only to first order in opacity, these two possibilities contribute separately to the sum in \eqref{eq5.55}. Either the unresolved $q\to q g$ rescatters, in which case the subsequent $g\to c\bar{c}$ proceeds as in vacuum because no second gluon exchange is available, or the unresolved $q\to q g$ splitting evolves as in vacuum, in which case the $g\to c\bar{c}$ splitting receives the medium modification. At second order in opacity, one would expect not only contributions in which the unresolved $q\to q g$ splitting rescatters twice or the $g\to c\bar{c}$ splitting is modified by two gluon exchanges, but also mixed contributions in which both the unresolved but rescattering $q\to q g$ and the subsequent $g\to c\bar{c}$ splitting receive medium modifications to first order in opacity. 

\subsubsection{The totally coherent (vacuum) limit} 
Also the hierarchy $\tfrac{1}{L} \gg B_i \gg C_j$ satisfies the generalized  strong ordering condition \eqref{eq5.13}. This condition may be written as
\begin{equation}
	B_j L \to 0\, ,\quad \hbox{and} \quad C_j L \to 0\, ,\quad \hbox{and} \quad B_j \gg C_j\, .
	\label{eq5.56}
\end{equation}
it describes situations in which the formation times of both splittings are so long, that both splittings take place outside the medium. Equivalently, this limit can be reached by requiring that $1/L$ sets parametrically the largest transverse momentum in the problem
\begin{equation}
  n_0\, L = {\rm const}\, ,\qquad \hbox{and}\quad L \to 0\, .
  \label{eq5.57}
\end{equation}  
One readily verifies that in this limit, the only surviving longitudinal phase integrals in  Eqs. \eqref{eq4.33}--\eqref{eq4.44} and \eqref{eq4.54}--\eqref{eq4.61} are ${\cal P}_{(11)}$ and the first term of $\bar{\cal P}_{(11)}$. This indicates that only the parent quark can interact with the medium, since any splitting occurs at a finite distance and therefore outside the medium. However, the $(11)$ contribution is independent of the momentum transfer ${\bf q}$, i.e. it is vacuum-like, and is therefore exactly canceled by the probability-conserving virtual contribution,
\begin{equation}
\mathscr{C}_{B_iL}
\hat{P}_{q\to qc\bar{c}}^{(N=1)} = 0 \, . \label{eq5.58}
\end{equation}
The medium-modification therefore vanishes in the totally coherent limit \eqref{eq5.57}. In this limit, the 
$1 \to 3$ splitting function is the vacuum splitting function that factorizes according to \eqref{eq5.8b}.

%
\begin{figure}[!htbp]
    \centering
      \includegraphics[width=0.7\textwidth]{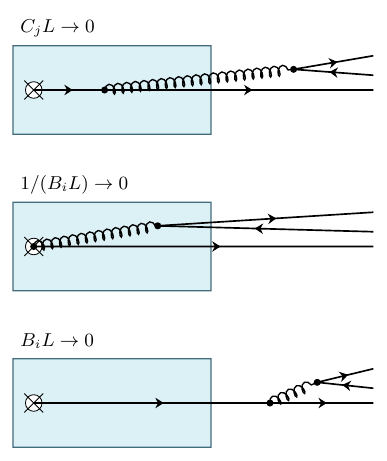}
    \vspace{-0.5em}
    \caption{Schematic summary of the three limits in which the medium-modified $q \to q c\bar{c}$ splitting function factorizes into a tensor product of $1\to 2$ splitting functions. In all three strongly ordered limits, the formation time associated with the first branching ($1/B_i$) is parametrically shorter than that of the second branching ($1/C_j$), implying $B_iL \gg C_jL$. i) For sufficiently small invariant mass of the $c\bar{c}$ pair, the second splitting forms so late that it occurs outside the medium. In this limit, the factorization  is given by the product of a medium-modified $q\to q g$ splitting and a vacuum $g\to c\bar{c}$ splitting, see \eqref{eq5.31}. ii) In the limit $B_iL \to \infty$, the $q\to q g$ splitting forms so rapidtly that it cannot be resolved by the medium. The factorization is then given by a vacuum $q\to qg$ splitting that subsequently rescatters in the medium, multiplied by a medium-modified $g\to c \bar{c}$ splitting, see \eqref{eq5.55}. iii) If the first splitting is delayed such that it occurs outside the medium, the medium-modification vanishes, see \eqref{eq5.58}. The resulting factorization therefore reduces to that of the vacuum splitting function, \eqref{eq5.8b}.}
\label{figlast}
\end{figure}

\section{Conclusions and Outlook}
\label{sec6}
In the present work, we have developed a variant of the opacity expansion for calculating medium modifications of the $1\to n$ collinear splitting functions defined through the dispersive parts of the quark and gluon two-point functions \eqref{eq1.1}, \eqref{eq1.2}. The formalism factorizes, to all orders in opacity, the space-time embedding of the splitting process from the partonic kinematics encoded in Dirac spinor terms. The space-time dependence is captured by simple longitudinal phase integrals that can be evaluated analytically, whereas the Dirac terms become increasingly intricate  with the order of opacity and the number of final-state partons. By analyzing the instantaneous contributions of propagators, we have shown that the gluon and fermion polarization tensors, which in time-ordered perturbation theory contain time derivatives, can be recast in a form free of such derivatives, making them amenable to symbolic algebra packages such as FeynCalc. This proved instrumental not only in simplifying the Dirac terms but also in demonstrating that they can be expressed in terms of vacuum contributions evaluated with specific momentum shifts. 

In section~\ref{sec4}, we have computed medium modifications to the triple-collinear splitting function $q \to qc\bar{c}$, where $c$ is treated as a massless quark of different flavor than $q$. This splitting is simpler than other triple-collinear splittings because only one vacuum branching history contributes, and thus only five real amplitude contributions arise at first order in opacity. In contrast, the $q \to q q\bar{q}$ splitting function has two vacuum branching histories, while the $q \to q g g$ and $g \to c\bar{c}g$ and $g \to g g g$ each have three. Accordingly, these cases require 10 and 15 distinct real amplitude contributions, respectively, at first order in opacity, all of which must then be squared. Although these calculations are therefore more involved than the present one, we expect that the corresponding medium-modified triple-collinear splitting functions can be computed using the same techniques. In particular, we already know that their space-time dependence is governed by simple analytic longitudinal phase integrals, and that the Dirac terms can be evaluated and simplified using symbolic algebra packages. 

In section~\ref{sec5}, we have demonstrated that the medium-modified $q\to qc\bar{c}$ splitting function factorizes in three distinct strongly ordered collinear limits. This is the first proof that a medium-modified $1\to 3$ splitting function factorizes in a strongly ordered limit and is the main physics result of the present work. It also provides a theoretical foundation for describing  a $1\to 3$ splitting as a sequence of successive Markovian $1\to 2$ processes.  A schematic summary is provided in Fig.~\ref{figlast}. Our proof was greatly simplified by the fact that taking collinear limits of the longitudinal phase diagrams is straightforward. Identifying which of the longitudinal phase diagrams vanish in a given strongly ordered limit  and which survive provided clear guidance as to whether factorization should be expected and what form it would take. We expect that establishing factorization of other triple-collinear splitting functions, such as $q \to q g g$, $g \to c\bar{c}g$ and $g \to g g g$, will benefit from a similar strategy. As a first step, it would be particularly interesting to determine whether phase factors associated with contributions in which the amplitude and complex conjugate amplitude follow different vacuum branching histories survive in the relevant strongly ordered limits. On physical grounds, one may expect that many, if not all, of these contributions vanish, in which case the problem would become more similar to the one considered here. In summary, we expect that both the calculation of the medium-modified $q\to q c \bar{c}$ splitting function in section~\ref{sec4} and its physics analysis in strongly ordered collinear limits given in section~\ref{sec5} can be extended to all other triple-collinear splitting functions using the techniques developed in this work. 

A further set of questions concerns  higher orders in opacity and the color flow between medium and splitting process. At the simplest level, we have presented results only at leading order in $N_c$, raising the question of whether additional structures appear at subleading order and whether factorization of medium-modified $1\to 3$ splittings continues to hold. Understanding  the nature of color decorrelations that arise at higher orders in opacity is likely to be more challenging. One expects that resumming the opacity expansion yields an expression in terms of medium-averages over eikonal Wilson lines. However, collecting sufficient information at higher orders in opacity to carry out such a resummation may prove difficult.

The technical advances reported in this work -- namely, an efficient and compact organization of the computation of $1\to 3$ splitting functions in a form where space-time information factorizes -- and the conceptual insights into the factorization of $1\to 3$ splitting functions in medium, may not only be extended to the above-mentioned class of related problems, but may also stimulate further developments in the near future. In particular: 

One important direction for future research is the further improvement of jet quenching parton showers. Their construction relies on the factorization of the branching process into  a Markovian chain of successive $1\to 2$ splitting functions and establishing
factorization of medium-modified $1\to 3$ splitting functions provides a systematic way of defining the correspoinding medium-modified splitting kernels. For instance,  the recently developed leading-logarithmic jet-quenching Monte Carlo JetMED~\cite{Caucal:2020xad,Caucal:2020uic} accounts for the possibility that an initial branching occurs at sufficiently high scale in the medium to remain unresolved by the medium, while the daughter partons subsequently rescatter. In this sense, the physical picture underlying the factorization formula \eqref{eq5.55} derived here is already implemented phenomenologically in JedMed. However, the rescattering prescription employed in JetMed is simpler than the factor that we have computed explicitly in \eqref{eq2.95}. As the medium modifications and factorization properties of the complete set of $1\to 3$ splitting functions are computed, they may provide the basis for more refined building blocks in jet-quenching parton showers. 

Another important open question concerns the region of overlapping formation times. The factorized, strongly ordered limits studied in section~\ref{sec5} correspond to negligible overlap between the formation times of  successive splittings.  Away from these limits, the results of the opacity expansion derived here can be evaluated numerically to determine the kinematic region over which the factorized description in terms of successive $1\to 2$ splittings remain quantitatively reliable. 

 Another promising direction concerns the unique opportunities offered by boosted $c\bar{c}$-pairs inside jets to probe the microscopic dynamics of jet-medium interactions~\cite{Attems:2022otp}, and the role of parton formation times~\cite{Brewer:2025wol}. One challenge for phenomenological studies is the medium-induced gluon radiation associated with the $g\to c\bar{c}$ splitting. Extending our analysis to $g\to c\bar{c} g$ could provide the basis for a more accurate modelling of these effects. From a conceptual perspective, it would be particularly interesting to investigate the soft-gluon limit of this process. In this limit, the energetic partons are expected to be describable by effective currents. For the $g\to c\bar{c} g$ splitting, one could then explore possible connections between the soft limit and  an antenna description formulated in terms of a pair of such currents. 
 
 As another promising direction, we highlight the recent proposal to use collinear spin correlations in dense QCD matter as a novel probe of jet-medium modifications~\cite{Silva:2025dan}. These collinear azimuthal correlations among three fragmenting partons are physically distinct from the azimuthal correlations of two outgoing partons with respect to a global even orientation that arise in anisotropic media~\cite{Hauksson:2023tze,Barata:2024bqp}.
 Our calculation of the medium-modified $q\to q c\bar{c}$ splitting provides QCD input needed for phenomenological studies of such collinear spin correlations. Extending our calculations to the complete set of triple-collinear splitting functions would further strengthen this input. At this stage, our results support the expected physical picture:  spin-flip contributions are absent in the high-energy collinear limit, medium-induced changes in spin correlations arise entirely from kinematic effects, i.e. the momentum transfer ${bf q}$ from the medium tilts the azimuthal orientation of the parton pair. This effect becomes sizeable when  $\vert{\bf q}\vert$  becomes comparable  to the smaller of the two boost-invariant relative momentum, $\bm{\upkappa}_2$ which. Medium-induced spin-decorrelations are therefore expected in a kinematic window  where the smallest invariant mass in a triple collinear splitting is large enough for the second splitting to occur inside the medium, yet small enough that $\bm{\upkappa}_2$ remains comparable to the characteristic transverse momentum transfer. 
 
These and other physics questions  (such as questions about the QGP-induced modifications to multi-point energy correlators~\cite{Bossi:2024qho,Barata:2025fzd}) on jet propagation in dense QCD matter would benefit from further computations of medium-modified triple-collinear splitting functions, their detailed analysis, and their implementation and phenomenological validation in jet-quenching Monte Carlo generators.

\acknowledgments We are greatly indebted to Peter Arnold for several very helpful discussions over the past two years, and for drawing our attention to the importance of understanding instantaneous contributions in time-ordered perturbation theory.  We thank Pier Monni for discussions about the Catani-Grazzini representation at an early stage of this work. We also acknowledge helpful comments and discussions with  Joao Barata, Ian Moult and Krishna Rajagopal,

\bibliography{jhepSU.bib}
  \end{document}